\documentclass[twocolumn,iop]{emulateapj}
\usepackage{apjfonts}
\usepackage{color}

\usepackage{amsmath,graphicx,longtable,hyperref}
\usepackage{natbib,threeparttable,rotating}
\usepackage{ulem}
\newcommand{\etal}{et al.}
\newcommand{\hbeta}{H{$\beta$}}

\newcommand{\lya}{Ly\,$\alpha$}
\newcommand{\CIV}{C\,{\sevenrm IV}}

\newcommand{\SiIV}{Si\,{\sevenrm IV}}
\newcommand{\CIII}{C\,{\sevenrm III]}}

\newcommand{\AlIII}{Al\,{\sevenrm III}}

\newcommand{\SiIII}{Si\,{\sevenrm III]}}

\def\FeII{Fe\,{\sc ii}}
\def\MgII{Mg\,{\sc ii}}
\def\HeII{He\,{\sc ii}}
\def\O3{O\,{\sc iii}]}

\newcommand{\NeV}{[Ne\,{\sevenrm\,V}]}

\def \CII {[C\,{\sc ii}]}

\newcommand{\OIV}{O\,{\sevenrm IV]}}

   \font\sevenrm=cmr7 scaled 1000
\newcommand{\comments}[1]{}

\def\kms{{\rm km\,s^{-1}}}

\def\ergs{${\rm erg\,s^{-1}}$}

\begin{document}

\title{Gemini GNIRS near-infrared spectroscopy of 50 Quasars at $z\gtrsim 5.7$ }

\author{Yue Shen$^{1,2,*}$, Jin Wu$^{3,4}$, Linhua Jiang$^4$, Eduardo Ba{\~n}ados$^5$, Xiaohui Fan$^6$, Luis C. Ho$^{3,4}$, Dominik A. Riechers$^7$, Michael A. Strauss$^8$, Bram Venemans$^9$, Marianne Vestergaard$^{10,6}$, Fabian Walter$^9$, Feige Wang$^{11}$, Chris Willott$^{12}$, Xue-Bing Wu$^{3,4}$, Jinyi Yang$^6$} 

\altaffiltext{1}{Department of Astronomy, University of Illinois at Urbana-Champaign, Urbana, IL 61801, USA; shenyue@illinois.edu}
\altaffiltext{2}{National Center for Supercomputing Applications, University of Illinois at Urbana-Champaign, Urbana, IL 61801, USA}
\altaffiltext{3}{Department of Astronomy, School of Physics, Peking University, Beijing 100871, China}
\altaffiltext{4}{Kavli Institute for Astronomy and Astrophysics, Peking University, Beijing 100871, China}
\altaffiltext{5}{The Observatories of the Carnegie Institution for Science, 813 Santa Barbara Street, Pasadena, CA 91101, USA}
\altaffiltext{6}{Steward Observatory, University of Arizona, 933 North Cherry Avenue, Tucson, AZ 85721-0065, USA}
\altaffiltext{7}{Cornell University, Space Sciences Building, Ithaca, NY 14853, USA}
\altaffiltext{8}{Princeton University Observatory, Peyton Hall, Princeton, NJ 08544, USA}
\altaffiltext{9}{Max Planck Institut f\"ur Astronomie, K\"onigstuhl 17, D-69117, Heidelberg, Germany}
\altaffiltext{10}{The Niels Bohr Institute at University of Copenhagen, Juliane Maries Vej 30, 2100 Copenhagen, Denmark}
\altaffiltext{11}{Department of Physics, Broida Hall, University of California, Santa Barbara, CA 93106, USA}
\altaffiltext{12}{NRC Herzberg, 5071 West Saanich Road, Victoria, BC V9E 2E7, Canada}
\altaffiltext{$^*$}{Alfred P. Sloan Research Fellow.}

\shorttitle{Near-IR Spectroscopy of $z\gtrsim 5.7$ Quasars}
\shortauthors{Shen \etal}

\begin{abstract}

We report initial results from a large Gemini program to observe $z\gtrsim 5.7$ quasars with GNIRS near-IR spectroscopy. Our sample includes 50 quasars with simultaneous $\sim 0.85-2.5$\,\micron\ spectra covering the rest-frame ultraviolet and major broad emission lines from \lya\ to \MgII. We present spectral measurements for these quasars and compare to their lower-redshift counterparts at $z=1.5-2.3$. We find that when quasar luminosity is matched, there are no significant differences between the rest-UV spectra of $z\gtrsim 5.7$ quasars and the low-$z$ comparison sample. High-$z$ quasars have similar continuum and emission line properties and occupy the same region in the black hole mass and luminosity space as the comparison sample, accreting at an average Eddington ratio of $\sim 0.3$. There is no evidence for super-Eddington accretion or hypermassive ($>10^{10}\,M_\odot$) black holes within our sample. We find a mild excess of quasars with weak \CIV\ lines relative to the control sample. Our results, corroborating earlier studies but with better statistics, demonstrate that these high-$z$ quasars are already mature systems of accreting supermassive black holes operating with the same physical mechanisms as those at lower redshifts. 
\end{abstract}
% \lya, on the other hand, is more heavily absorbed in the high-$z$ sample compared to the control sample, which may enhance the fraction of weak-line quasars at $z\gtrsim 5.7$ based on \lya\ alone with optical spectroscopy. 

\section{Introduction}\label{sec:intro}

The combination of recent progress in deep imaging sky surveys and optical/near-IR spectroscopic capabilities on large-aperture telescopes has revolutionized the study of high-redshift ($z\gtrsim 5.7$) quasars \citep[e.g.,][]{Fan_etal_2001,Fan_etal_2003,Fan_etal_2004,Fan_etal_2006araa, Cool_etal_2006, Goto_2006,McGreer_etal_2006,Jiang_etal_2008, Jiang_etal_2009, Jiang_etal_2015,Jiang_etal_2016,Venemans_etal_2007,Venemans_etal_2015,Mortlock_etal_2009,Mortlock_etal_2011,Willott_etal_2007,Willott_etal_2009,Willott_etal_2010a,Morganson_etal_2012,Banados_etal_2014,Banados_etal_2016,Matsuoka_etal_2016,Matsuoka_etal_2018a,Matsuoka_etal_2018,Tang_etal_2017,Reed_etal_2017}. These earliest quasars, powered by supermassive black holes (SMBHs), not only constrain the physics of black hole accretion and host galaxy assembly at cosmic dawn, but also provide critical information on the physical conditions of the intergalactic medium (IGM) in the early Universe \citep[e.g.,][]{Fan_etal_2006araa}. Since the first successful identification of $z\gtrsim 6$ quasars \citep{Fan_etal_2001}, the searches for these high-$z$ quasars have matured and dramatically increased the inventory of these rare objects. There are now more than $250$ quasars known at $z>5.6$, with the most distant quasar reaching $z\sim 7.5$ \citep{Banados_etal_2018}. These impressive statistics now enable a transition from individual case studies to ensemble studies of high-$z$ quasars. 

The vast majority of $z\gtrsim 6$ quasars are selected from wide-field optical$+$IR imaging with the drop-out technique and are confirmed with optical spectroscopy \citep[e.g.,][]{Fan_etal_2001}. The confirmation optical spectroscopy provides limited information about these high-$z$ quasars themselves, and to probe the physical properties of these systems (e.g., BH mass, spectral properties, broad and narrow absorption lines, etc.), near-IR spectroscopy is necessary to cover the rest-frame UV from \CIV\,1549\AA\ to \MgII\,2798\AA. Some earlier near-IR spectra of $z\gtrsim 6$ quasars already provided a glimpse of their physical properties, such as BH masses and Eddington ratios \citep[e.g.,][]{Kurk_etal_2007,Jiang_etal_2007,Willott_etal_2010b,DeRosa_etal_2014,Mazzucchelli_etal_2017,Kim_etal_2018}, and chemical abundances in the broad-line region \citep[e.g.,][]{Barth_etal_2003,Kurk_etal_2007,DeRosa_etal_2011}. These measurements of BH properties also facilitate the study of the relations between BH growth and host galaxy assembly in the early universe, where the host galaxy properties are mostly inferred from the molecular gas emission in the millimeter regime \citep[e.g.,][]{Walter_etal_2003,Walter_etal_2009,Riechers_etal_2009,Wang_etal_2013,Venemans_etal_2016,Willott_etal_2017,Decarli_etal_2017,Decarli_etal_2018}. However, near-IR spectroscopy of $z\gtrsim 6$ quasars is expensive and thus only a small fraction of them have existing near-IR spectra, some of which are also of low quality (e.g., insufficient spectral coverage, signal-to-noise ratio, and spectral resolution). Given the importance of understanding the physical properties and growth of these high-$z$ SMBHs, we have assembled a large sample of high-quality near-IR spectra for these objects, as statistics are the key to addressing most of the relevant questions pertaining to the early growth and evolution of these SMBHs and their hosts.  

We have conducted a Gemini-GNIRS survey (Gemini Large and Long Program LLP-7) to obtain near-IR $YJHK$ simultaneous spectroscopy for a sample of $50$ quasars at $z\gtrsim 5.7$. The primary goal is to compile a large statistical sample of these objects with near-IR spectroscopic coverage to understand their rest-frame UV properties, and derive physical properties such as BH mass and Eddington ratios with better spectral quality and sample statistics. This sample, when supplemented with existing near-IR spectroscopic data for high-$z$ quasars, can be used to measure the demography of these SMBHs in terms of their mass function (after the complicated selection function is properly quantified), as well as statistical studies such as chemical abundance in the quasar broad-line region, the prevalence of broad and narrow absorption-line systems, etc. The GNIRS sample presented here is substantially larger than earlier near-IR samples and enables a uniform analysis with the same data format and spectral fitting tools. 

In this paper we describe the details of our program (\S\ref{sec:data}), and present the spectral fitting results on the GNIRS sample (\S\ref{sec:spe_ana}). As one application of our data, we compare the rest-frame UV spectral properties of our sample to their lower-redshift counterparts (\S\ref{sec:results}) and discuss the implications in \S\ref{sec:dis}. Additional applications of our data will be presented in successive work. Throughout this paper we adopt a flat $\Lambda$CDM cosmology with $\Omega_\Lambda=1-\Omega_0=0.7$ and $H_0=70\,\kms {\rm Mpc}^{-1}$.

\section{Data}\label{sec:data}

\begin{table*}
\caption{Sample Summary}\label{tab:sample}
\centering
\scalebox{0.95}{
\begin{tabular}{lcccccccccccccc}
\hline\hline
ObjID         & R. A. (J2000)   & Decl. (J2000)  & $z_{\rm old}$     &  $z_{\rm sys}$ & $z_{\rm sys, err}$ & $J$ & $J$ err  &    $H$  & $H$ err   &   $K_s$ & $K_s$ err & $J$ ref  & dis. ref & comment\\
(1) & (2) & (3) & (4) & (5) & (6) & (7) & (8) & (9) & (10) & (11) & (12) & (13) & (14) & (15) \\
\hline
P000+26      & 00:01:21.63 & $+$26:50:09.2 & 5.75  & 5.733  & 0.007  & 18.36 & 0.08  &        &        &       &      & UHS            &  1 &    \\   
J0002+2550   & 00:02:39.39 & $+$25:50:34.9 & 5.82  & 5.818  & 0.007  & 18.46 & 0.09  &        &        &       &      & UHS            &  2 &    \\   
J0008$-$0626 & 00:08:25.77 & $-$06:26:04.6 & 5.93  & 5.929  & 0.006  & 19.43 & 0.13  &        &        &       &      & Jiang2016      &  3 &   BAL \\   
J0028+0457   & 00:28:06.56 & $+$04:57:25.7 & 5.99  & 5.982  & 0.012  & 19.16 & 0.12  & 19.05  & 0.20   & 18.32 & 0.18 & UKIDSS LAS     &  4 &    \\   
J0033$-$0125 & 00:33:11.40 & $-$01:25:24.9 & 6.13  & 5.978  & 0.010  & 20.64 & 0.20  &        &        &       &      & Willott2007    &  5 &    \\   
J0050+3445   & 00:50:06.67 & $+$34:45:22.6 & 6.25  & 6.251  & 0.006  & 19.05 & 0.15  &        &        &       &      & UHS            &  6 &    \\   
J0055+0146   & 00:55:02.91 & $+$01:46:18.3 & 6.02  & 6.017  & 0.054  & 20.93 & 0.15  &        &        &       &      & Willott2009    &  7 &    \\   
J0136+0226   & 01:36:03.17 & $+$02:26:05.7 & 6.21  & 6.206  & 0.009  & 21.15 & 0.22  &        &        &       &      & Willott2010    &  6 &    \\   
J0203+0012   & 02:03:32.38 & $+$00:12:29.2 & 5.86  & 5.709  & 0.010  & 19.05 & 0.10  & 17.75  & 0.07   & 17.32 & 0.09 & UKIDSS LAS     &  8 &  BAL  \\   
J0221$-$0802 & 02:21:22.71 & $-$08:02:51.5 & 6.16  & 6.161  & 0.054  & 21.09 & 0.14  &        &        &       &      & Willott2010    &  6 &    \\   
J0227$-$0605 & 02:27:43.29 & $-$06:05:30.2 & 6.2   & 6.212  & 0.007  & 21.07 & 0.22  &        &        &       &      & VIK            &  7 &    \\   
J0300$-$2232 & 03:00:44.18 & $-$22:32:27.2 & 5.7   & 5.684  & 0.008  & 18.71 & 0.08  & 19.42  & 0.08   &       &      & Banados2014    &  4 &    \\   
J0353+0104   & 03:53:49.72 & $+$01:04:04.4 & 6.049 & 6.057  & 0.005  & 19.45 & 0.16  & 18.53  & 0.17   & 18.16 & 0.22 & UKIDSS LAS     &  9 &   BAL  \\   
P060+24      & 04:02:12.69 & $+$24:51:24.4 & 6.18  & 6.170  & 0.006  & 19.05 & 0.10  & 18.69  & 0.16   & 17.81 & 0.14 & UKIDSS GCS     &  1 &    \\   
J0810+5105   & 08:10:54.32 & $+$51:05:40.1 & 5.80  & 5.805  & 0.010  & 18.77 & 0.06  &        &        &       &      & Jiang2016      & 10 &    \\   
J0835+3217   & 08:35:25.76 & $+$32:17:52.6 & 5.89  & 5.902  & 0.009  & 20.50 & 0.20  &        &        &       &      & Jiang2016      & 10 &    \\   
J0836+0054   & 08:36:43.85 & $+$00:54:53.3 & 5.82  & 5.834  & 0.007  & 17.70 & 0.03  & 17.02  & 0.03   & 16.18 & 0.03 & UKIDSS GCS     & 11 & radio-loud    \\   
J0840+5624   & 08:40:35.09 & $+$56:24:19.9 & 5.85  & 5.816  & 0.010  & 19.00 & 0.01  &        &        &       &      & Jiang2016      & 12 &    \\   
J0841+2905   & 08:41:19.52 & $+$29:05:04.4 & 5.96  & 5.954  & 0.005  & 19.17 & 0.09  & 18.62  & 0.18   & 17.84 & 0.15 & UKIDSS LAS     & 13 &  BAL  \\   
J0842+1218   & 08:42:29.43 & $+$12:18:50.5 & 6.055 & 6.069  & 0.009  & 18.78 & 0.11  &        &        &       &      & UHS            & 3  &    \\   
J0850+3246   & 08:50:48.25 & $+$32:46:47.9 & 5.867 & 5.730  & 0.013  & 18.74 & 0.08  &        &        &       &      & UHS            & 3  &    \\   
J0927+2001   & 09:27:21.82 & $+$20:01:23.7 & 5.79  & 5.770  & 0.013  & 19.12 & 0.17  &        &        &       &      & UHS            & 12 &    \\   
J1044$-$0125 & 10:44:33.04 & $-$01:25:02.2 & 5.8   & 5.780  & 0.007  & 18.31 & 0.05  & 17.92  & 0.12   & 17.03 & 0.07 & UKIDSS LAS     & 14 &  BAL  \\   
%J1059-0906  & 10:59:28.61 & $-$09:06:20.4 & 5.92  &        &        & 19.50 & 0.18  &        &        &       &      & VHS            &  6 &    \\       
J1137+3549   & 11:37:17.73 & $+$35:49:56.9 & 6.01  & 6.009  & 0.010  & 18.48 & 0.07  &        &        &       &      & UHS            & 12 &    \\   
J1143+3808   & 11:43:38.33 & $+$38:08:28.7 & 5.81  & 5.800  & 0.010  & 19.00 & 0.14  &        &        &       &      & UHS            & 10 &    \\   
J1148+0702   & 11:48:03.29 & $+$07:02:08.3 & 6.339  & 6.344  & 0.006  & 19.36 & 0.11  & 18.39  & 0.12   & 17.51 & 0.11 & UKIDSS LAS     & 10 &    \\   
J1148+5251   & 11:48:16.64 & $+$52:51:50.3 & 6.43  & 6.416  & 0.006  & 18.17 & 0.06  &        &        &       &      & UHS            & 15 &    \\   
J1207+0630   & 12:07:37.43 & $+$06:30:10.1 & 6.04  & 6.028  & 0.013  & 19.35 & 0.14  &        &        & 17.50 & 0.12 & UKIDSS LAS     & 3  &    \\   
J1243+2529   & 12:43:40.81 & $+$25:29:23.9 & 5.85  & 5.842  & 0.006  & 19.21 & 0.13  & 18.29  & 0.10   & 17.54 & 0.10 & UKIDSS LAS     & 10 &    \\   
J1250+3130   & 12:50:51.93 & $+$31:30:22.9 & 6.13  & 6.138  & 0.005  & 19.22 & 0.12  & 18.17  & 0.14   & 17.40 & 0.09 & UKIDSS LAS     & 12 &  BAL  \\   
J1257+6349   & 12:57:57.47 & $+$63:49:37.2 & 6.02  & 5.992  & 0.010  & 19.78 & 0.08  &        &        &       &      & Jiang2016      & 3  &    \\   
J1335+3533   & 13:35:50.81 & $+$35:33:15.8 & 5.93  & 5.870  & 0.020  & 18.90 & 0.13  &        &        & 17.61 & 0.14 & UKIDSS LAS     & 12 &    \\   
P210+27      & 14:01:47.34 & $+$27:49:35.0 & 6.14  & 6.166  & 0.007  & 19.56 & 0.21  &        &        &       &      & UHS            & 1  &    \\   
J1403+0902   & 14:03:19.13 & $+$09:02:50.9 & 5.86  & 5.787  & 0.013  & 19.17 & 0.10  & 18.59  & 0.10   & 17.93 & 0.11 & UKIDSS LAS     & 3  &    \\   
J1425+3254   & 14:25:16.34 & $+$32:54:09.6 & 5.85  & 5.862  & 0.006  & 19.22 & 0.17  &        &        &       &      & UHS            & 16 &    \\   
J1427+3312   & 14:27:38.59 & $+$33:12:41.0 & 6.12  & 6.118  & 0.005  & 19.68 & 0.05  &        &        &       &      & McGreer2006    & 17 &  radio-loud; BAL  \\   
J1429+5447   & 14:29:52.17 & $+$54:47:17.7 & 6.21  & 6.119  & 0.008  & 19.70 & 0.07  &        &        &       &      & Willott2010    & 6  &  radio-loud  \\   
J1436+5007   & 14:36:11.74 & $+$50:07:06.9 & 5.83  & 5.809  & 0.010  & 18.99 & 0.14  &        &        &       &      & UHS            & 12 &    \\   
P228+21      & 15:14:44.91 & $+$21:14:19.8 & 5.92  & 5.893  & 0.015  & 19.00 & 0.02  &        &        &       &      & Banados2016    & 1  &    \\   
J1545+6028   & 15:45:52.08 & $+$60:28:24.0 & 5.78  & 5.794  & 0.007  &       &       &        &        &       &      &                & 18 &    \\   
J1602+4228   & 16:02:53.98 & $+$42:28:24.9 & 6.07  & 6.083  & 0.005  & 18.82 & 0.12  &        &        &       &      & UHS            & 2  &    \\   
J1609+3041   & 16:09:37.27 & $+$30:41:47.6 & 6.16  & 6.146  & 0.006  & 19.39 & 0.15  & 18.72  & 0.18   & 18.15 & 0.22 & UKIDSS LAS     & 10 & radio-loud   \\   
J1621+5155   & 16:21:00.92 & $+$51:55:48.9 & 5.71  & 5.637  & 0.008  & 18.52 & 0.10  &        &        &       &      & UHS            & 10 &    \\   
J1623+3112   & 16:23:31.81 & $+$31:12:00.5 & 6.22  & 6.254  & 0.006  & 19.16 & 0.11  & 18.45  & 0.12   & 17.86 & 0.13 & UKIDSS LAS     & 2  &    \\   
J1630+4012   & 16:30:33.90 & $+$40:12:09.6 & 6.05  & 6.066  & 0.007  & 19.73 & 0.22  &        &        &       &      & UHS            & 15 &    \\   
P333+26      & 22:15:56.63 & $+$26:06:29.4 & 6.03  & 6.027  & 0.007  & 19.52 & 0.14  &        &        &       &      & UHS            & 1  &    \\   
J2307+0031   & 23:07:35.35 & $+$00:31:49.4 & 5.87  & 5.900  & 0.010  & 20.43 & 0.11  &        &        &       &      & VHS            & 9  &    \\   
J2310+1855   & 23:10:38.88 & $+$18:55:19.7 & 6.04  & 5.956  & 0.011  & 17.94 & 0.05  &        &        &       &      & UHS            & 10 &    \\   
J2329$-$0403 & 23:29:14.46 & $-$04:03:24.1 & 5.90  & 5.883  & 0.007  & 21.06 & 0.19  &        &        &       &      & Willott2009    & 7  &  BAL  \\   
J2356+0023   & 23:56:51.58 & $+$00:23:33.3 & 6     & 5.987  & 0.014  & 21.18 & 0.07  &        &        &       &      & Jiang2016      & 9  &    \\ 
\hline
\hline\\
\end{tabular}
}
\begin{tablenotes}
      \small
      \item NOTE. --- Column (1) lists the abbreviated object IDs we assigned to each object in our GNIRS sample, which are adopted throughout this work. The original redshifts from the discovery paper for each object are compiled in Column (4) and our improved systemic redshifts (see \S\ref{sec:spe_ana}) are compiled in Column (5). Note that some of these original redshifts may have been updated in other works. $JHK$ magnitudes are in the Vega system. The $J$ magnitudes are used to normalize the spectra, which are taken from different references and converted from AB magnitude (if needed) using $J_{\rm Vega}=J_{\rm AB} - 0.94$. Reference keys for the $J$ magnitudes (and $H,K_s$ magnitudes if available):  UHS -- \citet{Dye_etal_2018}; VIK -- VIKING DR4 \citep{Edge_etal_2013}; VHS -- \citet{McMahon_et_al_2013}; UKIDSS GCS -- \citep{Lawrence_etal_2007}; UKIDSS LAS -- UKIDSS LAS DR10 \citep{Lawrence_etal_2007}; Jiang2016 -- \citet{Jiang_etal_2016}; Banados2016: \citet{Banados_etal_2016}; Banados2014 -- \citet{Banados_etal_2014}; Willott2009 -- \citet{Willott_etal_2009}; Willott2007 -- \citet{Willott_etal_2007}; McGreer2006 -- \citet{McGreer_etal_2006}. Reference keys for the discovery papers in Column (14) are: 1. \citet{Banados_etal_2016}; 2. \citet{Fan_etal_2004}; 3. \citet{Jiang_etal_2015}; 4. \citet{Banados_etal_2014}; 5. \citet{Willott_etal_2007}; 6. \citet{Willott_etal_2010a}; 7. \citet{Willott_etal_2009}; 8. \citet{Venemans_etal_2007}; 9. \citet{Jiang_etal_2008}; 10. \citet{Jiang_etal_2016}; 11. \citet{Fan_etal_2001}; 12. \citet{Fan_etal_2006b}; 13. \citet{Goto_2006}; 14. \citet{Fan_etal_2000}; 15. \citet{Fan_etal_2003}; 16. \citet{Cool_etal_2006}; 17. \citet{McGreer_etal_2006}; 18. \citet{Wang_etal_2016}. Radio-loud identification is either from the discovery paper or from \citet{Banados_etal_2015}.
\end{tablenotes}
\end{table*}

Our target pool includes all known quasars at $z\gtrsim 5.7$ compiled from the literature and unpublished works for which we are collecting near-IR spectroscopy from different sources. The majority of these quasars were discovered from dedicated high-$z$ quasar searches targeted to different depths from wide-area imaging surveys. During the 15B-17A semesters we observed a total of 51 quasars from our target pool with GNIRS on Gemini-North during Band 2 allocation, although for some targets the observations were only partially executed. Most of the observations were carried out in queue mode by Gemini staff. We preferentially excluded objects with existing near-IR spectra with reasonably good quality and $JHK$ wavelength coverage. Bright targets were assigned a higher priority over faint targets given the total time allocation. The selection of targets also depend on their visibility in each observing semester. Our GNIRS sample is by no means a complete flux-limited sample, but it covers a broad range of luminosities and samples the diversity of quasar properties at $z\gtrsim 5.7$. As described in \S\ref{sec:results}, we create a comparison sample at lower-$z$ that matches the luminosity distribution of the GNIRS sample for a fair comparison. The incompleteness in our near-IR spectroscopic follow-up will be taken into account in our future work that requires the detailed selection function (such as the black hole mass function).

We used the cross-dispersion (XD) mode (32 l/mm) on GNIRS with the short blue camera and a slit width of 0.675" to balance the need for spectral resolution and throughput. This configuration provides simultaneous spectral coverage of $\sim 0.85-2.5\,\mu m$ at a spectral resolution $R \sim 650$ with a pixel scale of 0.15"/pix, sufficient to resolve the broad emission lines. We used a fixed position angle (PA$=90^{\circ}$ east of north) to minimize the effect of differential flexure, as recommended on the GNIRS instrument page\footnote{https://www.gemini.edu/sciops/instruments/gnirs/spectroscopy/observing-strategies}. Given the typical low airmass of our observations ($<1.2$), the atmospheric differential refraction introduces positional shifts of $\lesssim 0.06$\arcsec\ across JHK bands, which is less important than differential flexure and will not affect our relative flux calibration much. Each target was observed for a period of 30 minutes to 5 hrs depending on the target brightness (with a typical single-exposure time of $300$\,s), in order to reach an accumulative continuum SNR of $\sim 5$ per pixel averaged over $H$ band. The observations were executed in the standard ABBA sequence (with an offset of 3\arcsec\ along the slit between A and B positions), and we observed one A0 star for telluric correction and flux calibration immediately preceding or following the science observation. For absolute flux calibration we use the photometry of the targets compiled from the literature, as described in \S\ref{sec:reduc}.

%We process the GNIRS data with a custom reduction pipeline that we improved upon existing GNIRS pipelines ({\bf cite refs, Jin should write this up}). We flux calibrate the spectra using the photometric magnitudes in $JHK$ bands and we ignore the potential variability between the GNIRS and earlier photometric observations. More technical details regarding the reduction pipeline will be presented in a follow-up paper. 

\subsection{GNIRS data reduction}\label{sec:reduc}
\label{2ddatareduction}

%The spectra were obtained using the cross-dispersed (XD) mode of the Gemini Near-IR Spectrograph on the 8.1 m Gemini North telescope with the ``short blue'' camera, 32 mm grating and 0.675" slit. This mode gives simultaneous spectral coverage from $0.85 \sim 2.5\mu m$ at $R \sim 650$ with a pixel scale of 0.15"/pix. 

Our data reduction pipeline is a combination of two existing pipelines for GNIRS. The first one is the PyRAF-based XDGNIRS ~\citep{Mason2015XDGNIRS}; the other one is the IDL-based  XIDL package \footnote{http:\slash \slash www.ucolick.org\slash \ensuremath{\sim}xavier\slash IDL\slash }. XDGNIRS is commonly used for bright nearby galaxy targets; it uses the standard ABBA method to perform sky subtraction and spectrum combination with 2D images before final 1D extraction, while XIDL does an extra spline fitting after the A minus B step and extracts the 1D spectrum for each single subtraction image. Both methods have their pros and cons, and we chose to combine these two methods to make our final results more accurate and robust. In brief, our data reduction consists of 3 steps, (1) preprocessing; (2) wavelength and S-distortion mapping; (3) 1D spectrum extraction and combination. The data are processed in ABBA sequence groups. All final 1D spectra are calibrated and stored in vacuum wavelengths. 

In the preprocessing step, we first clean the large-scale pattern noise by fitting a periodic function to the image and subtracting it using existing routines in the XIDL package. Cosmic rays are masked using the LAcosmic method ~\citep{Dokkum2001LACosmic}. The images are then flatfielded using dome flats. We split the image into different echelle orders and then apply the A-B method to do sky subtraction for each ABBA sequence. The typical exposure time for a single observation is 300 seconds. During this time, the sky emission may have large variations in the infrared and the A-B method will lead to large residuals near sky emission lines in such cases. After this step, the XDGNIRS pipeline directly combines the sky-subtracted frames, which is reasonable for bright objects but not ideal for our faint targets. We therefore apply an extra correction to suppress these residuals.

\begin{figure*} %[htbp]
\centering
\includegraphics[keepaspectratio,width=\textwidth,height=0.75\textheight]{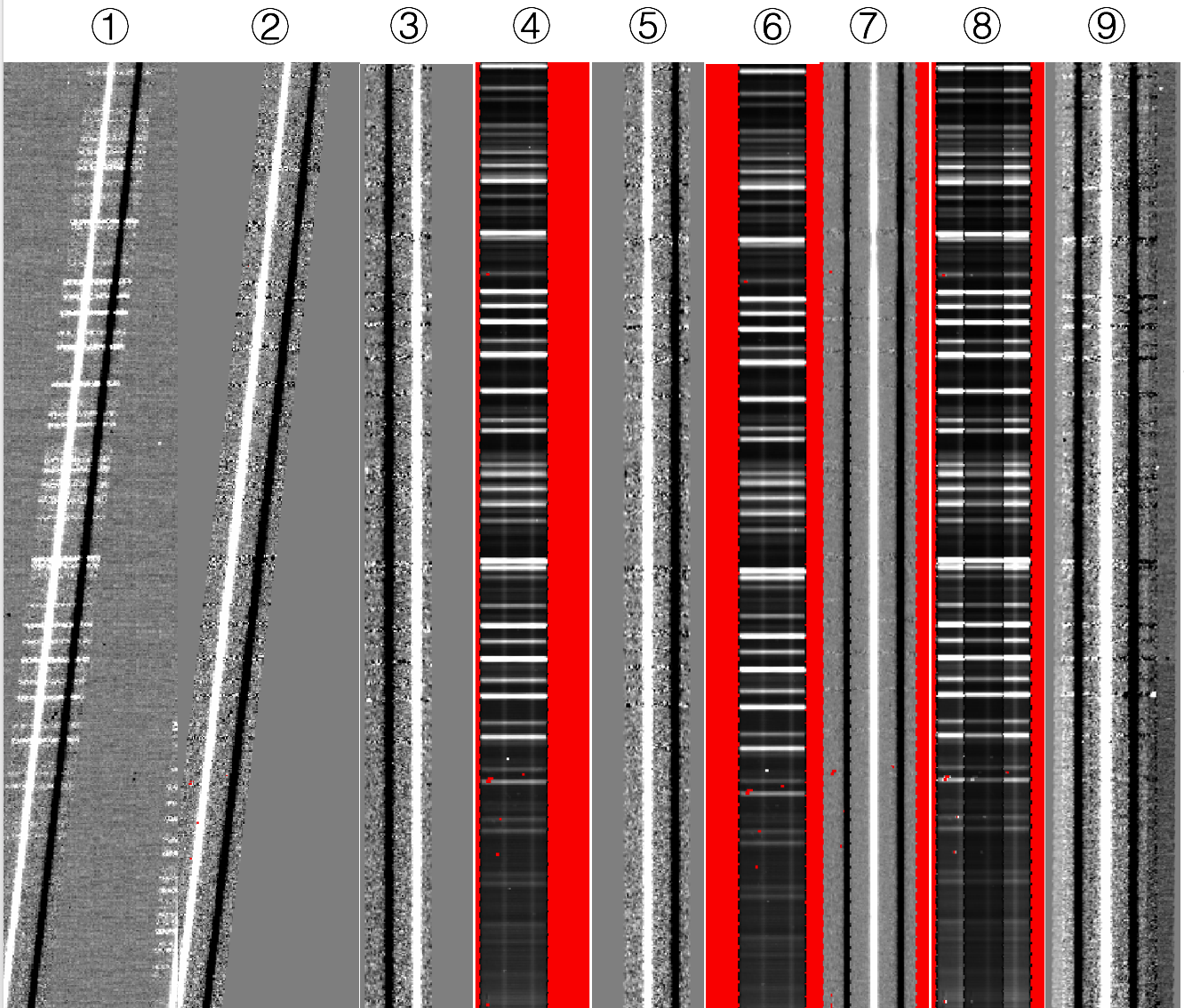}
\caption{Sky subtraction and 2D combination results for a single order with b-spline correction by our pipeline. The 9 images are, from left to right: 1. A-B; 2. A-B corrected by b-spline; 3. Align B position into the center; 4. variance image of 3; 5. Align A position into the center; 6. variance image of 5; 7. combined result of multiple aligned A positions and B positions; 8. variance image of 7; 9. the same combined result without b-spline correction from XDGNIRS for the same order. The difference between images 7 and 9 demonstrates the improvement of our hybrid approach over using XDGNIRS alone without the b-spline correction.}
\label{fig:our_skysub}
\end{figure*}

In the wavelength and S-distorting mapping step, we use arc observations to fit the 2D y-wavelength relation and use the pinhole observations to get the 2D x-slit mapping. Having these mappings, we are able to correct the skyline residuals using the b-spline method ~\citep{Kelson2003BSpline}. This correction improves the data reduction in the wavelength range with numerous sky lines. Fig.\ \ref{fig:our_skysub} demonstrates the improvement using the b-spline method.

After that, a common rectifying grid is determined for each spectrum order in each ABBA image sequence. We used our custom remapping function, which guarantees the conservation of total flux. Using the mappings above and our remapping function, all the b-spline corrected data are rectified onto the same grid. The XIDL pipeline makes 1D extraction directly from the non-rectified data and then combines the spectrum. Instead, we decided to first combine all 2D images and then use a single extraction to get the final spectrum. The main reason is that we will sometimes have undetected bad pixels after the pattern noise removal and cosmic ray rejection process, which can be identified using sigma rejection method in the 2D combine process. It is more robust to reject these bad pixels in the 2D image than in the extracted 1D spectrum. All the A-B images were visually checked before they were combined to make sure that all the features in the final spectrum are genuine and were not due to instrumental defects.

In the last 1D spectrum extraction and combination step, we use the boxsum (i.e., with a boxcar aperture) method to extract 1D spectra with an aperture size of 5 pixels. We also tested optimal extraction \citep{Horne_1986} and found nearly identical results. We obtain both the quasar raw spectrum and standard star raw spectrum as well as their error spectrum. We perform absorption line removal for the standard star by fitting Voigt profiles at the position of hydrogen lines. After that, most of the absorption features in the standard star spectrum come from atmospheric absorption. We then correct for telluric absorption in our object spectrum by dividing it by the spectrum of the standard star. We always chose the standard star that has the smallest difference in air mass and observation time for a given science target. After telluric correction, we obtain the relative flux ratio between the science object and the standard star. We then use a blackbody spectrum to model the standard star with its effective temperature. Multiplying the relative flux ratio by the model star spectrum gives a flux-calibrated spectrum. Then different orders are combined using the flux in their overlap region, and we combine all the 1D spectra for one object in different ABBA sequences with 5-sigma clipping to obtain the final 1D spectrum. 

As the last step, we rescale the flux (density) using the available $J$ band magnitude of the specific quasar target (see Table \ref{tab:sample}). We ignore the possibility of quasar variability between the $J$ observation and the GNIRS spectrum, which is less important than the sky and seeing variations between the quasar and standard star observations in our GNIRS program. Since these high-$z$ quasars are point sources in rest-frame UV, this last flux rescaling step effectively corrects for slit losses. There is one object (J1545$+$6028) for which we do not have available $J$ band magnitudes, and we used the standard star for absolute flux calibration as well. 

Excluding one object where no signal was received at all during exposures under poor observing conditions, our final GNIRS sample includes 50 quasars. The basic target information is summarized in Table \ref{tab:sample}.

\section{Spectral Analysis}\label{sec:spe_ana}

\begin{table}
\caption{Line Fitting Parameters}\label{tab:linefit}
\centering
\scalebox{1.0}{
\begin{tabular}{cccc}
\hline\hline
Line Name & Vacuum Rest Wavelength & $n_{\rm gauss}$ & Complex \\
& [\AA] & & \\
(1) & (2) & (3) & (4)  \\
\hline
\MgII & 2798.75 & 2 & \MgII   \\
\CIII & 1908.73 & 2 & \CIII  \\
\SiIII & 1892.03 & 1 & \CIII    \\
\AlIII & 1857.40 & 1 & \CIII   \\
\CIV & 1549.06 & 2  & \CIV   \\
\HeII & 1640.42 & 1  & \CIV   \\
\O3  &  1663.48 & 1  & \CIV  \\
\SiIV & 1399.41$^{*}$ & 2 & \SiIV/\OIV  \\
\hline
\hline\\
\end{tabular}
}
\begin{tablenotes}
      \small
      \item NOTE. --- Fitting parameters for the lines considered in this work. The third column lists the total number of Gaussians used for each line. Multiple lines in the same line complex as specified by name are fit simultaneously. The rest wavelengths of the lines are taken from \citet{Vandenberk_etal_2001}.  
     
      $^*$The wavelength of the \SiIV\ line is taken as the arithmetic mean of the central wavelengths of \SiIV\ and \OIV\ (1396.76\,\AA\ and 1402.06\,\AA).
\end{tablenotes}
\end{table}

%\clearpage 
%\begin{longtable*}{llll}
%\caption[notes]{FITS Catalog Format}\label{tab:format}\\
%\hline \hline
%   \multicolumn{1}{l}{\textbf{Column}} &
%   \multicolumn{1}{l}{\textbf{Format}} &
%   \multicolumn{1}{l}{\textbf{Units}} &
%   \multicolumn{1}{l}{\textbf{Description}} \\[0.5ex] \hline
%   \\ [0.5ex]
%\endfirsthead
%%This is the header for the remaining page(s) of the table...
%\multicolumn{4}{l}{{\tablename} \thetable{} -- Continued} \\[0.5ex]
%  \hline \hline \\[-2ex]
%  \multicolumn{1}{l}{\textbf{Column}} &
%  \multicolumn{1}{l}{\textbf{Format}} &
%  \multicolumn{1}{l}{\textbf{Units}} &
%  \multicolumn{1}{l}{\textbf{Description}} \\[0.5ex] \hline \\ [1ex]
%\endhead
%%This is the footer for all pages except the last page of the table...
%  \hline
%  \multicolumn{4}{l}{{Continued on Next Page\ldots}} \\
%\endfoot
%%This is the footer for the last page of the table...
%  \\  \hline \hline \\ [1ex]
%\endlastfoot
%%Now the data...
\begin{table*}
\caption{FITS Catalog Format}\label{tab:format}
\centering
\scalebox{1.0}{
\begin{tabular}{llll}
\hline\hline
Column & Format & Units & Description \\
\hline
OBJID	&	STRING		&			& Object ID of the GNIRS sample \\                              
%RA		&	DOUBLE	&	degree	& J2000 R.A. \\                                                
%DEC		& DOUBLE 	& degree	& J2000 Decl.  \\ 
ZOLD         & DOUBLE       &  & Original redshift from the discovery paper \\                              
ZSYS	& DOUBLE	& 		& Improved systemic redshift \\ 
ZSYS\_ERR & DOUBLE &            & Uncertainty in ZSYS \\       
ZMGII       &   DOUBLE &  &  Redshift based on broad \MgII \\
ZMGII\_ERR      &   DOUBLE &  &  Uncertainty in ZMGII \\
ZCIII       &   DOUBLE &  &  Redshift based on the \CIII\ complex \\
ZCIII\_ERR      &   DOUBLE &  &  Uncertainty in ZCIII \\
ZCIV       &   DOUBLE &  &  Redshift based on broad \CIV \\
ZCIV\_ERR      &   DOUBLE &  &  Uncertainty in ZCIV \\
ZSIIV       &   DOUBLE &  &  Redshift based on broad \SiIV/\OIV \\
ZSIIV\_ERR      &   DOUBLE &  &  Uncertainty in ZSIIV \\
LOGL1350       & DOUBLE        &  [\ergs] & Continuum luminosity at restframe 1350\,\AA  \\                                                
LOGL1350\_ERR   & DOUBLE        & [\ergs]  &  Uncertainty in LOGL1350  \\                                               
LOGL1700       & DOUBLE        & [\ergs]  &  Continuum luminosity at restframe 1700\,\AA \\                                                
LOGL1700\_ERR   & DOUBLE        & [\ergs]  & Uncertainty in LOGL1700  \\                                               
LOGL3000       & DOUBLE        & [\ergs]  & Continuum luminosity at restframe 3000\,\AA  \\                                                
LOGL3000\_ERR   & DOUBLE        & [\ergs]  &  Uncertainty in LOGL3000 \\    
LOGLBOL       & DOUBLE        & [\ergs]  & Bolometric luminosity  \\                                                
LOGLBOL\_ERR   & DOUBLE        & [\ergs]  &  Uncertainty in LOGLBOL \\     
MGII         & DOUBLE[5]  & \AA, $\kms$, [\ergs], \AA, \AA  & peak wavelength, FWHM, $\log L_{\rm line}$, restframe EW, top 50\% flux centroid \\ 
CIII\_ALL        & DOUBLE[5]  & ...  & For the entire \CIII\ complex (\CIII, \SiIII, \AlIII)  \\
CIV             & DOUBLE[5]  & ...   & For the entire \CIV\ line  \\     
SIIV\_OIV        & DOUBLE[5]  & ...  &  For the 1400\,\AA\ complex \\                                                  
MGII\_ERR     & DOUBLE[5]  & \AA, $\kms$, [\ergs], \AA, \AA  & Measurement errors in MGII \\      
CIII\_ALL\_ERR    & DOUBLE[5]  & ...  & Measurement errors in CIII\_ALL  \\                                                 
CIV\_ERR         & DOUBLE[5]  & ...  & Measurement errors in CIV \\                                                  
SIIV\_OIV\_ERR    & DOUBLE[5]  & ...  & Measurement errors in SIIV\_OIV \\           
LOGBH\_CIV\_VP06  & DOUBLE     &  [$M_\odot$] & Single-epoch virial BH mass based on \CIV\ \citep{Vestergaard_Peterson_2006} \\                                                 
LOGBH\_CIV\_VP06\_ERR & DOUBLE  & ...  & Measurement errors in LOGBH\_CIV\_VP06 \\                                                 
LOGBH\_MGII\_S11  & DOUBLE     & [$M_\odot$]  & Single-epoch virial BH mass based on \MgII\ \citep{Shen_etal_2011} \\                                                 
LOGBH\_MGII\_S11\_ERR & DOUBLE  & ...  & Measurement errors in LOGBH\_MGII\_S11 \\
LOGBH  & DOUBLE     & [$M_\odot$]  & Adopted fiducial BH mass; based on \MgII\ if available, otherwise based on \CIV\ \\                                                 
LOGBH\_ERR & DOUBLE  & ...  & Measurement errors in LOGBH \\
LOGEDD\_RATIO   &  DOUBLE & ... & Eddington ratio based on the fiducial BH mass \\
LOGEDD\_RATIO\_ERR   &  DOUBLE & ... & Measurement errors in LOGEDD\_RATIO \\ 
\hline
\hline\\
\end{tabular}
}
\begin{tablenotes}
      \small
      \item NOTE. --- (1) Bolometric luminosities were computed using a bolometric correction of 5.15
\citep{Richards_etal_2006b} using the $3000$\AA\ monochromatic luminosities; (2) Uncertainties are measurement errors only; (3) Null value (indicating unmeasurable) is zero for a quantity and $-1$ for its associated error, except for LOGEDD\_RATIO where the null value is -99. 
\end{tablenotes}
\end{table*}      

\begin{figure}
\centering
    \includegraphics[width=0.48\textwidth]{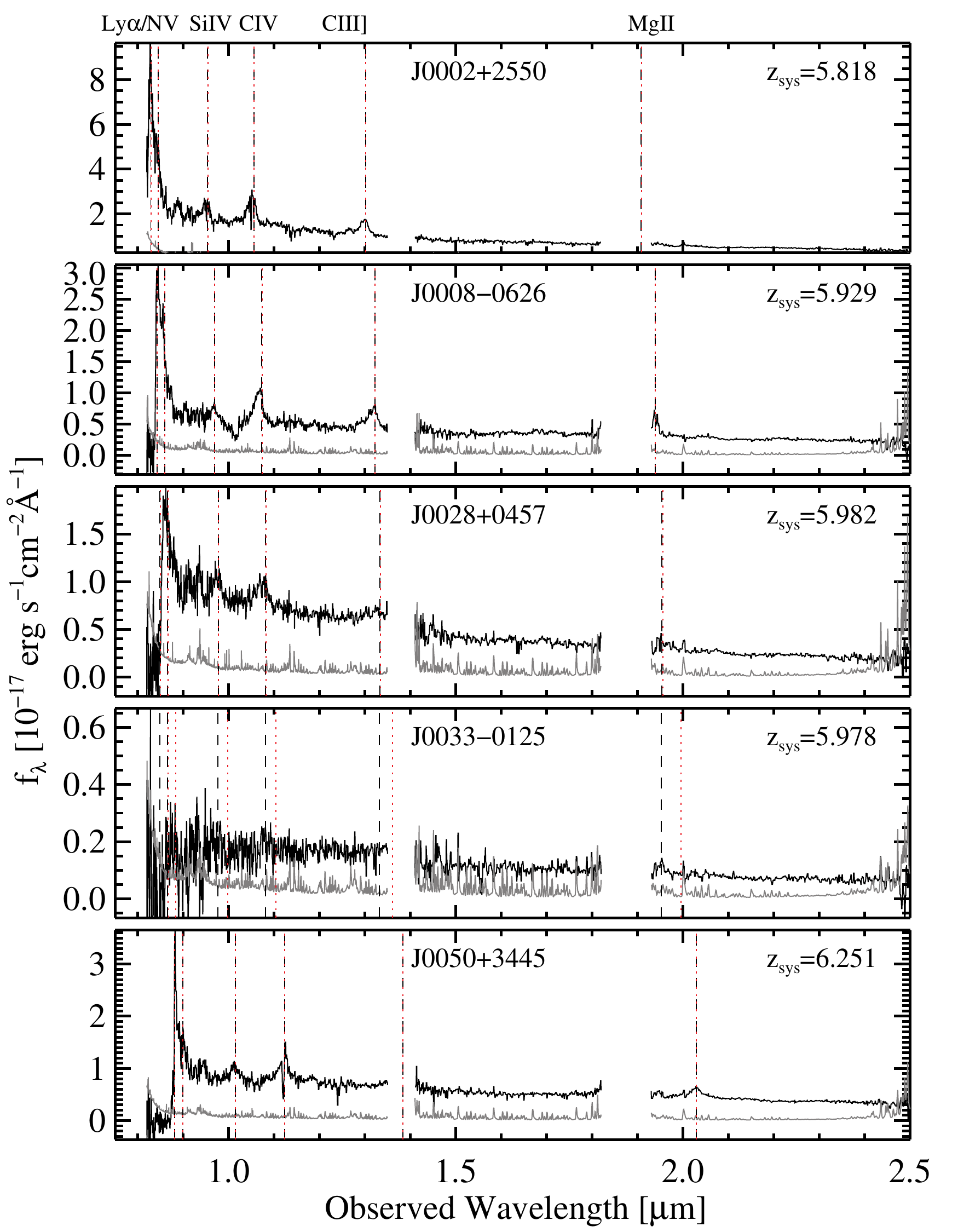}
     \caption{GNIRS spectra of our high-$z$ quasar sample. The spectrum (black; error in gray) has been smoothed with a $5$-pixel median filter. The vertical black dashed lines indicate the major emission lines using our best estimate of the systemic redshifts $z_{\rm sys}$ (see \S\ref{sec:spe_ana} for details), and the red dotted lines indicate those using the original redshifts. The full set of spectra is included in the online version of the paper. }
    \label{fig:spec}
\end{figure}

\begin{figure}
\centering
    \includegraphics[angle=-90,width=0.48\textwidth]{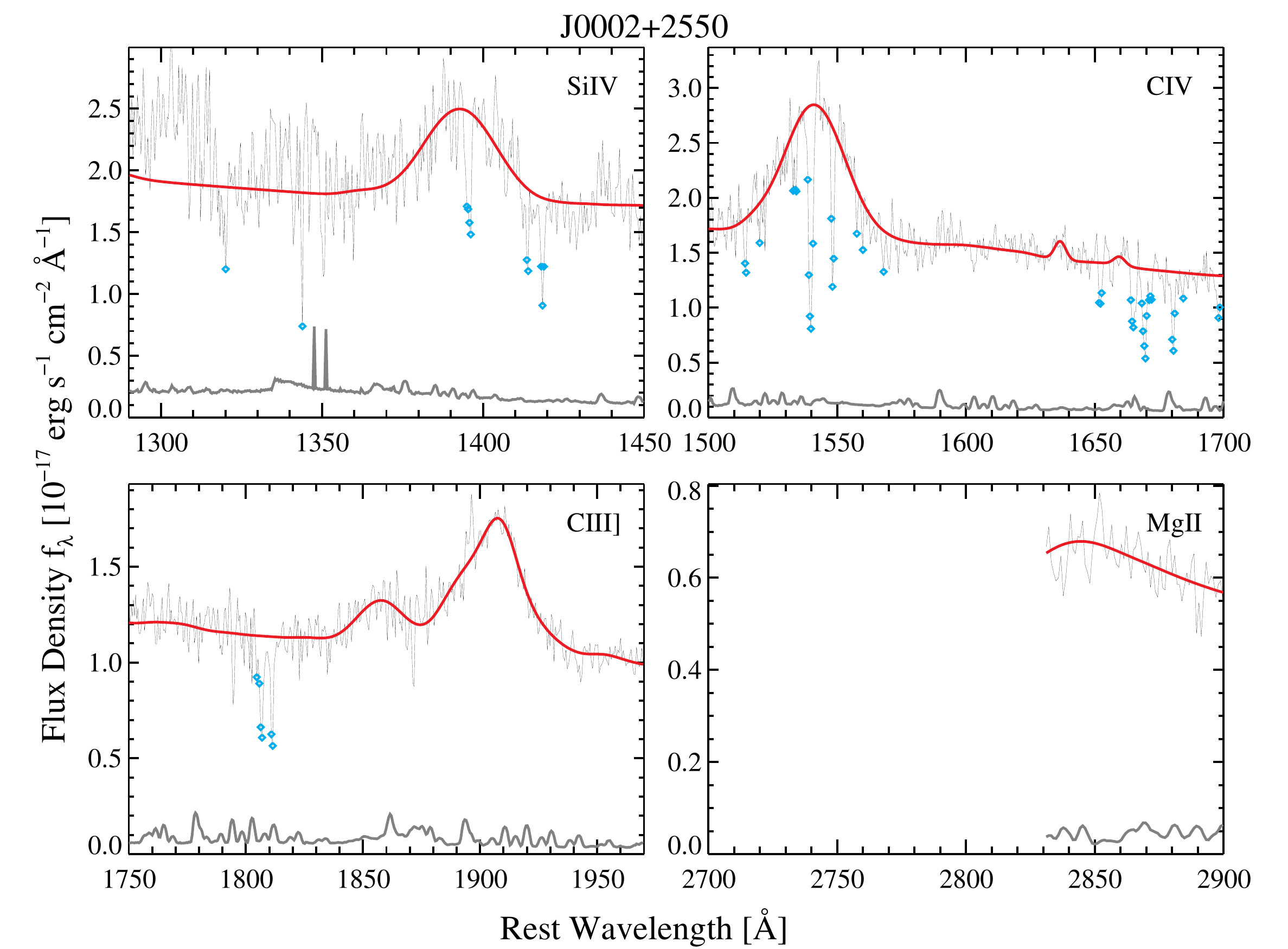}
     \caption{An example of our spectral fits around several major broad lines. The original spectrum and flux errors are shown in black and gray, respectively. The red lines are our model. The cyan points are masked absorption pixels in our fits. The full set of the fitting quality assessment plots is provided in an online figure set.}
    \label{fig:spec_qa}
\end{figure}

We fit the GNIRS spectrum following the approach detailed in, e.g., \citet[][]{Shen_Liu_2012}. In short, we shift the spectrum to restframe using the initial redshift, and fit the spectrum with a global continuum$+$emission line model \footnote{The full technical details of the spectral fitting are described in \citet{Shen_etal_2018b} and the associated code is made public along with that paper.}. This differs slightly from our earlier work which used a local continuum$+$line fit around each broad line \citep[e.g.,][]{Shen_etal_2008,Shen_etal_2011}. Several wavelength windows free of major emission lines (except for the broad-band \FeII\ emission) are used to fit the global continuum as a first step. The global continuum is modeled by a power-law plus a 3rd-order polynomial, and UV \FeII\ emission is modeled using empirical templates from the literature \citep[e.g.,][]{Vestergaard_Wilkes_2001} that are scaled and broadened to match our spectrum. The additional 3rd-order polynomial component is introduced to account for any peculiar continuum curvature in the rest-frame UV that may be caused by internal reddening, as observed in a small fraction of quasars \citep[e.g.,][]{Shen_etal_2018b}. The continuum and the \FeII\ emission form a pseudo-continuum, which is subtracted from the spectrum, leaving a line-only spectrum for which we model the emission lines with multiple Gaussians. We fit the broad emission lines in individual line complexes specified in Table \ref{tab:linefit}, where the main line and adjacent weak lines are fit simultaneously. We found that the number of Gaussians we use for each line is sufficient to reproduce the line profile, and using more Gaussians is unnecessary given the medium spectral quality of our sample. Fig.\ \ref{fig:spec_qa} compares our model and the data around several major broad lines in one of our objects, and the full set of fitting results is provided as an online figure set. This figure set can also be used to reject certain line fits, e.g., if the line largely falls within one of the telluric gaps in the spectrum. 

In some cases, the original redshift from the discovery paper is inaccurate, leading to small coverages of the broad emission lines in the de-redshifted near-IR spectrum and potential biases in the emission line measurements. Therefore we perform a second fit to the spectrum with the updated redshfit from the first fit and update the fitting results for all objects in our sample. One such iteration is sufficient for the fit to converge.

We measure the continuum and emission-line properties from the best-fit models. To estimate the uncertainties of these spectral measurements, we use a Monte Carlo approach \citep[e.g.,][]{Shen_etal_2011}: for each original spectrum, we create a trial sample of 50 mock spectra, {each is generated by randomly shuffling the flux (density) values in the original spectrum by adding a Gaussian random deviate at each pixel with its dispersion equal to the spectral error at that pixel}; the same fitting approach was applied to the mock spectra and the measured quantities recorded; the nominal uncertainties of the measured spectral properties are then estimated as the semi-quantile of the range enclosing the 16th and 84th percentiles of the distribution. Adding flux perturbations to the original spectrum instead of our model spectrum preserves details in the spectral features that are not captured or well-fit by our model (such as absorption lines). On the other hand, the original spectrum is already a perturbed version of the noise-free true spectrum, hence the mock spectra are slightly noisier than the original spectrum, and therefore our approach will produce overly conservative measurement errors in the spectral quantities. We have found that this approach produces very reasonable estimation of the measurement uncertainties \citep[e.g.,][]{Shen_etal_2011}. \citet{Shen_etal_2016b} and \citet{Shen_etal_2018b} further studied the dependence of measured spectral quantities and their uncertainties on the SNR of the original spectrum and demonstrated that this overall fitting approach and error estimation are quite robust and do not depend on the SNR of the spectrum.

Once we have fitted the emission lines, we use the peak wavelength of the major broad emission lines measured from the best-fit model to improve the systemic redshift estimate of the quasar. Available optical spectra of these high-$z$ quasars often only cover the heavily absorbed \lya\ line, and the resulting systemic redshift derived from \lya\ is highly uncertain. Our near-IR spectroscopy covers additional broad emission lines such as \SiIV, \CIV, \CIII\ and \MgII, which provide more accurate systemic redshifts.   

However, it is well known that quasar emission lines are often shifted from the systemic velocity due to various dynamical and/or radiative processes \citep[e.g.,][]{Shen_etal_2016b}. We follow the approach detailed in \citet{Shen_etal_2016b} to derive the best systemic redshift estimates based on a series of lines that takes into account the velocity shifts of each line relative to systemic as a function of quasar continuum luminosity. While it is difficult to measure the redshifts of these high-$z$ quasars to better than $\sim 200\,\kms$ with broad lines only \citep[e.g.,][]{Shen_2016,Shen_etal_2016b}, these new systemic redshifts are an improvement over some of the previous redshift estimates based on optical spectroscopy alone. In the catalog described in Table \ref{tab:format} (also in Table \ref{tab:sample}) we provide the improved systemic redshifts and their uncertainties including both measurement errors and systematic errors from intrinsic line velocity shifts as quantified in \citet{Shen_etal_2016b}. The median redshift uncertainty of our GNIRS sample based on the broad emission lines is $\sim 330\,{\rm km\,s^{-1}}$. One broad-absorption-line quasar, J0203$+$0012, has a systemic redshift $z_{\rm sys}=5.777\pm0.011$ determined from \CIV\ that is significantly lower than the redshift ($z=5.86$) based on optical spectroscopy reported in the discovery paper \citep{Venemans_etal_2007}. If we adopted the discovery redshift, then \CIV\ and \SiIV\ would be blueshifted by $\sim 5000\,\kms$, which would be extreme but still possible. Later near-IR spectroscopy of this object confirmed its BAL nature and derived revised redshifts of $5.70<z<5.74$ \citep{Mortlock_etal_2009} and $z=5.706$ \citep{Ryan-Weber_etal_2009} largely based on \CIV. Our derived redshift is slightly larger than the latter two redshift estimates because we took into account the typical blueshift of \CIV. In any case, we are less confident about the systemic redshift determined for this object given its BAL nature and the lack of \MgII\ coverage.

\begin{figure}
\centering
    \includegraphics[width=0.48\textwidth]{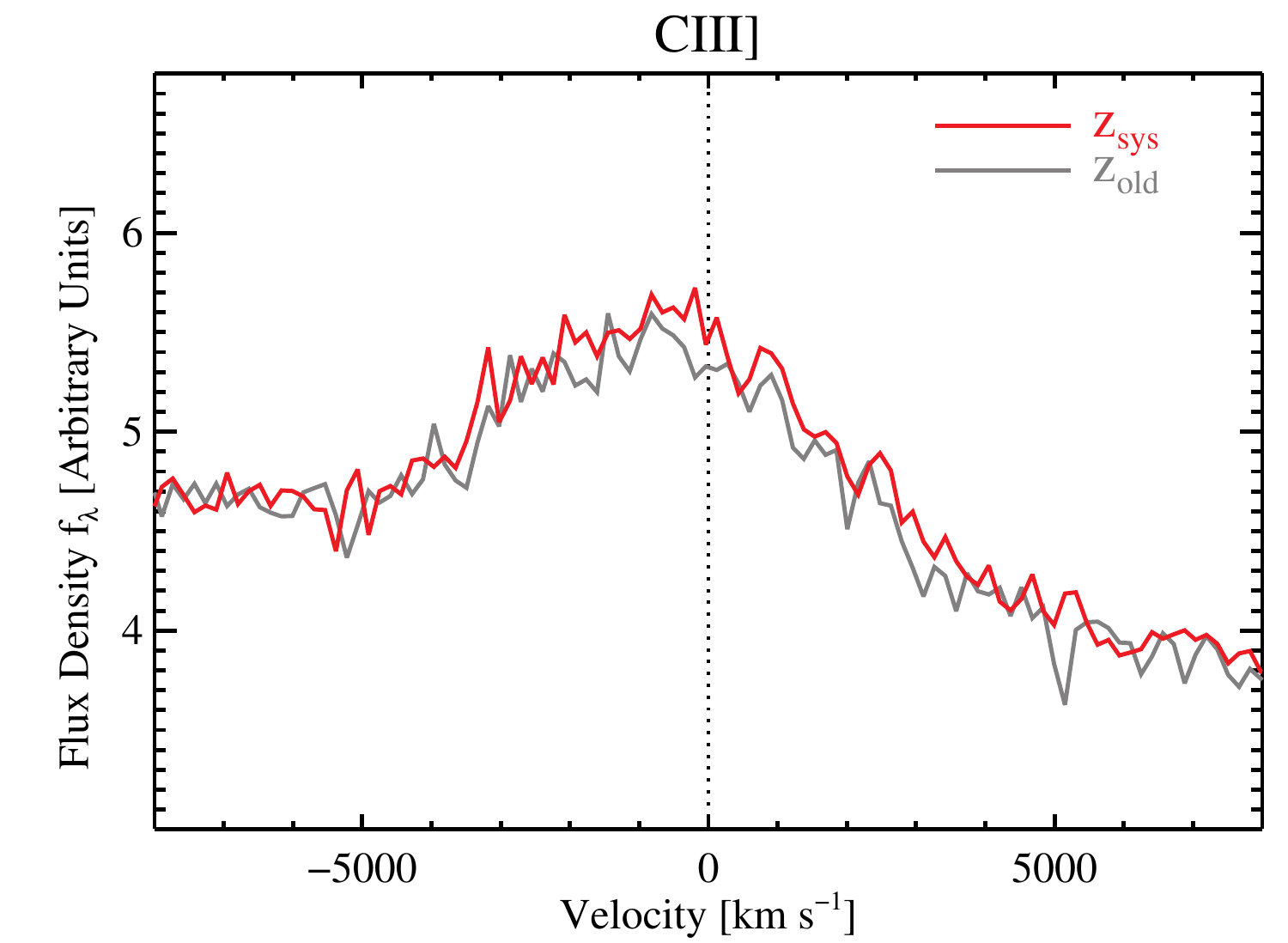}
     \caption{Composite spectrum for our GNIRS sample around the \CIII\ line. The red line shows the result using the improved redshifts and the gray line shows the result using the old redshifts. The peak of the \CIII\ complex is expected to be within $\sim 200\,\kms$ of systemic \citep[e.g.,][]{Shen_etal_2016b}. Therefore our redshift estimates are on average better than previous estimates. }
    \label{fig:CIII}
\end{figure}

To evaluate the overall improvement of our redshift estimation over previous results, we plot the median composite spectrum (see \S\ref{sec:results}) around the \CIII\ line in Fig.\ \ref{fig:CIII}. Unlike \CIV, the peak of the \CIII\ complex is known to have a modest average velocity shift of $\sim -220\,\kms$ from systemic, with negligible luminosity dependence \citep{Richards_etal_2011,Shen_etal_2016b}. The composite \CIII\ profile using our improved redshifts is aligned with the expected location, while the composite line using previous redshifts shows a much larger blueshift. Thus we conclude that our improved redshifts are on average better than previous estimates.

\begin{figure}
\centering
    \includegraphics[width=0.48\textwidth]{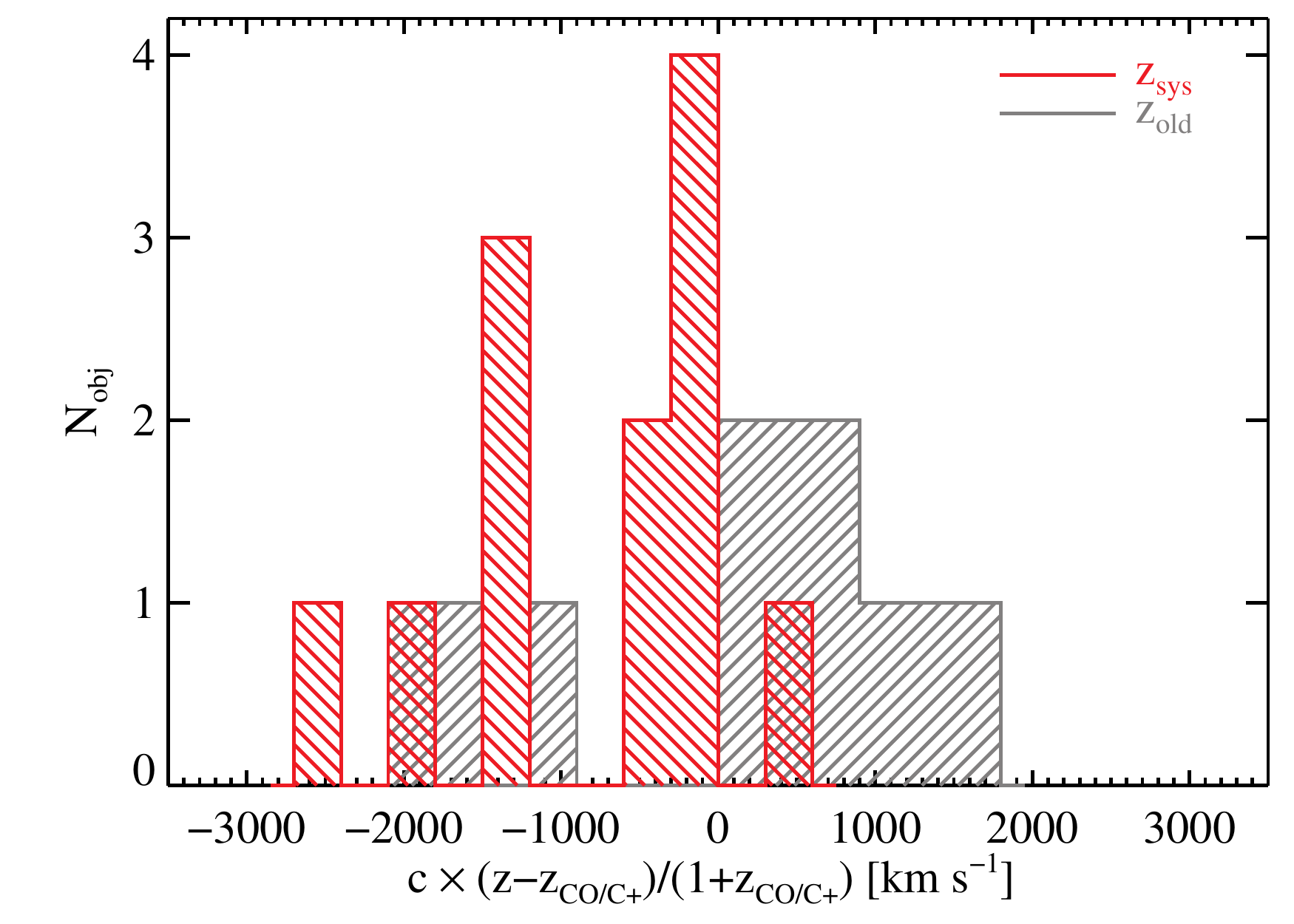}
     \caption{Comparison between the broad-line-based redshifts and those derived from CO or \CII\ molecular lines for a subset of 12 objects. The red histogram is for our improved redshifts based on near-IR spectra and the gray histogram is for the old redshifts. Our near-IR redshifts are in excellent agreement with the molecular line redshifts for half of the sample. There are two quasars for which our near-IR redshifts are lower than the molecular line redshifts by more than $1500\,\kms$, as discussed in the text. }
    \label{fig:zmm}
\end{figure}

We further compare our broad-line based redshifts with those measured for 12 quasars from [CII] and CO mm observations in Table \ref{tab:zcomp}. The molecular lines mostly trace the gas in the host galaxy of these quasars, and may have a velocity offset with respect to the broad-line region. Fig.\ \ref{fig:zmm} shows the velocity difference between the molecular line redshifts and our near-IR spectroscopic redshifts. There is excellent agreement between the two redshifts for half of the objects. There are two quasars for which our redshift estimates are lower than those based on molecular lines by more than $1500\,\kms$. These two quasars, J1429$+$5447 and J2310$+$1855, appear to have reasonably well measured broad emission lines. The UV broad lines in these objects may be more blueshifted from systemic than their lower-redshift counterparts with comparable luminosities, or we underestimated the systematic uncertainties of our broad-line-based redshifts. A larger sample of high-$z$ quasars with molecular line measurements will be useful to fully address this issue.

We compile our spectral measurements for our high-$z$ sample in an online FITS table, with the contents detailed in Table \ref{tab:format}. The calibrated GNIRS spectra are displayed in Fig.\ \ref{fig:spec} and its online extension set. 

\begin{table*}
\caption{Redshift Comparison}\label{tab:zcomp}
\centering
\begin{tabular}{lcccccc}
\hline\hline
ObjID  & $z_{\rm old}$  & $z_{\rm NIR}$ & $z_{\rm [CII]}$   & ref       & $z_{\rm CO}$ & ref \\
\hline
J0055+0146  & 6.02  & $6.017\pm 0.054$  & $6.0060\pm 0.0008$  & wil15   &      & \\
J0840+5624  & 5.85  & $5.816\pm0.010$  &                                  &            & $5.8441\pm0.0013$& wan10\\
J0842+1218  & 6.055  & $6.069\pm0.009$  & $6.0763\pm0.0005$ &dec18 &       &  \\
J0927+2001  & 5.79 & $5.770\pm0.013$ &	       	      	      	    &           & $5.7722\pm0.0006$ &	car07  \\
  ...                 &   &  &                                  &           & $5.7716\pm0.0012$ &	wan11a \\
J1044$-$0125 & 5.8 & $5.780\pm0.007$ & $5.7847\pm0.0007$ &wan13 & $5.7824\pm0.0007$ &	wan10 \\
J1148+5251 & 6.43 & $6.416\pm0.006$	& 6.419  	      	            & wal09b & $6.4189\pm	0.0006$ &	mai05 \\ 
J1207+0630 & 6.04	& $6.028\pm0.013$ & $6.0366\pm0.0009$ & dec18 & & \\
J1335+3533 & 5.93 & $5.870\pm0.020$  &	       	       	      	&            & $5.9012\pm0.0019$ &	wan10 \\
J1425+3254 & 5.85 & $5.862\pm0.006$  &	       	      	      	&           & $5.8918\pm0.0018$ &	wan10 \\
J1429+5447 & 6.21 & $6.119\pm0.008$ &	       	      	      	 &     	      &	  $6.1831\pm0.0007$ &	wan11a \\
J1623+3112 & 6.22 & $6.254\pm0.006$ & $6.2605\pm0.0005$ & wan11a & & \\
J2310+1855 & 6.04 & $5.956\pm0.011$ & $6.0031\pm0.0002$ & wan13  & $6.0025\pm0.0007$ & wan13\\
\hline
\hline\\
\end{tabular}
\begin{tablenotes} 
  \item NOTE. Comparison between our near-IR redshifts based on broad emission lines and those derived from \CII\ and CO. Reference keys are: wil15 -- \citet{Willott_etal_2015}; wan10 -- \citet{Wang_etal_2010}; dec18 -- \citet{Decarli_etal_2018}; car07 -- \citet{Carilli_etal_2007}; wan11a -- \citet{Wang_etal_2011a}; wan13 -- \citet{Wang_etal_2013}; wal09b --  \citet{Walter_etal_2009}.
\end{tablenotes}
\end{table*}

Among the 50 quasars in our sample, we visually identified 8 quasars with apparently strong broad \CIV\ absorption lines (J0008-0626, J0203+0012, J0353+0104, J0841+2905, J1044-0125, J1250+3130, J1427+3312, J2329-0403). This fraction ($\sim 16\%$) of broad-absorption line (BAL) quasars is roughly consistent with the apparent fraction for lower-redshift quasars \citep[e.g.,][]{Gibson_etal_2009} but notably smaller than the intrinsic fraction of broad absorption line quasars \citep[e.g.,][]{Dai_etal_2008, Allen_etal_2011}. {As a result of our qualitative identification scheme and limited SNR of the spectra, some weak BALs may have been missed}; on the other hand, we did not account for any luminosity differences between our sample and earlier samples. A more careful analysis of broad absorption lines in the high-$z$ quasar sample will be presented elsewhere. 

A few quasars have low SNR in our GNIRS spectra (e.g., J0055+0146, J0221$-$0802, and J0227$-$0605), which result in low-quality spectral fits. We included these objects in our sample since the measurement uncertainties in the measured quantities are still reasonably quantified. They also contribute to the construction of the composite spectrum in \S\ref{sec:results}. In addition, the extraction of J1427$+$3312 was complicated by a nearby star that falls in one of the AB positions during the observation, and at wavelengths greater than 2.3\,\micron\ the spectrum may be compromised by this complication.

We notice that one object, J0136+0226, shows apparently narrow profiles in \lya\ (${\rm FWHM}=1467\pm 25\,\kms$) and \CIV\ (${\rm FWHM}=3291\pm247\,\kms$). {This may be due to emission line contamination beyond the broad-line region (such as the narrow-line region)}, or a combination of absorption and intrinsically narrower broad lines due to a small black hole mass. Similar unobscured quasars (e.g., with detected quasar continuum) with narrow broad-lines have been seen in $z\sim 3$ SDSS quasars \citep{Alexandroff_etal_2013} and in low-luminosity $z>5.7$ quasars \citep[][]{Matsuoka_etal_2018}. 

The apparent radio-loud fraction of our GNIRS sample is $\sim 8\%$ (4/50), similar to the fraction reported in \citet{Banados_etal_2015}.

\begin{figure*}
\centering
    \includegraphics[width=0.85\textwidth]{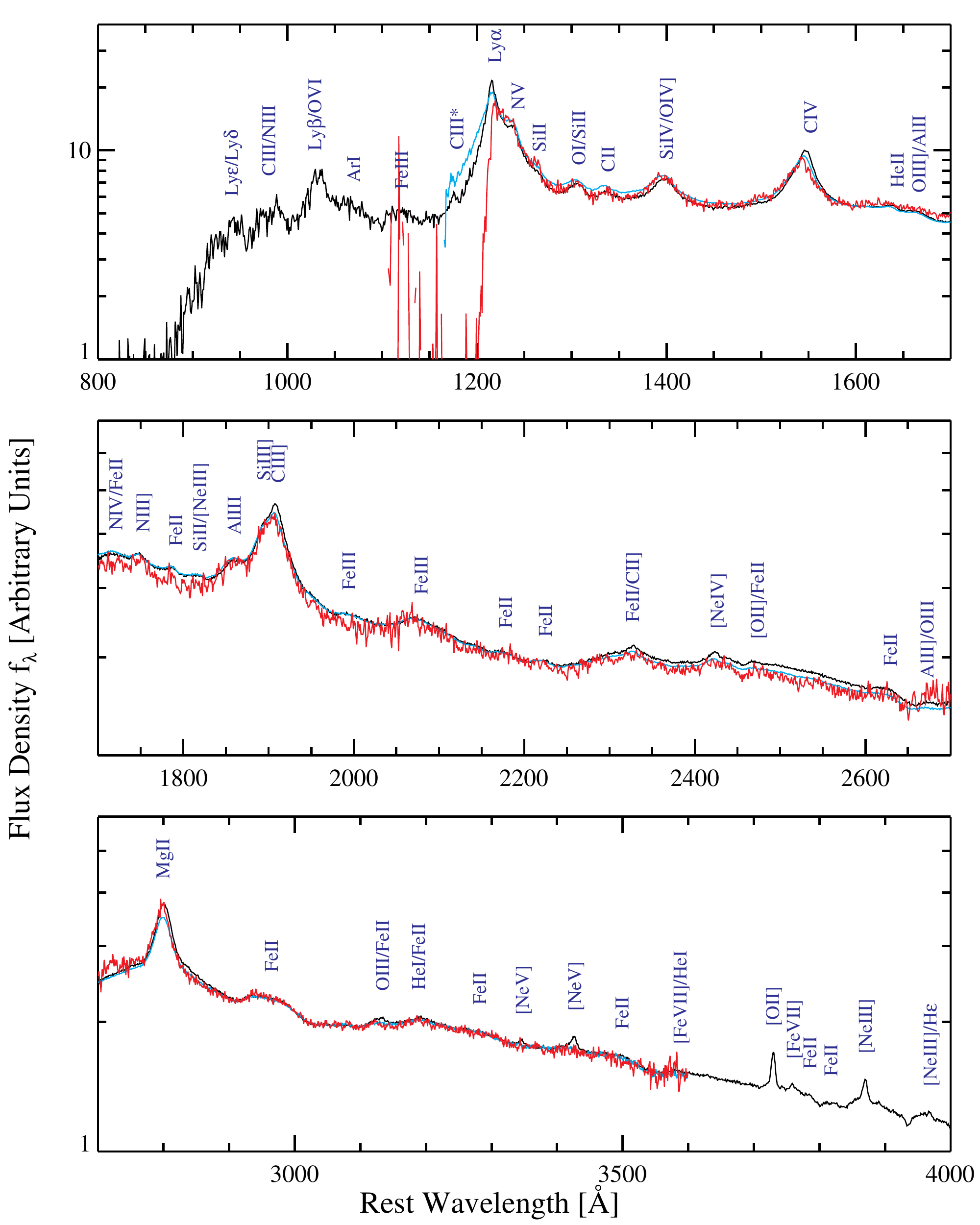}
     \caption{Rest-frame UV median composite spectra for our high-$z$ sample (red), the luminosity-matched control sample from SDSS DR7 (cyan) and the sample in \citet[][black]{Vandenberk_etal_2001}. The spectra are normalized around different continuum wavelengths (1600, 2200, 3000\,\AA) in the three panels to compensate for the slight (and insignificant) differences in the continuum slopes of these samples. The high-$z$ composite around 2700\,\AA\ is not well constrained due to the small number of contributing objects as this is around the telluric absorption band between $H$ and $K$. Other than the heavily absorbed \lya\ line, the high-$z$ sample has similar average UV spectral properties as the control sample matched in quasar luminosity. The composite spectrum for the high-$z$ sample is provided in Table \ref{tab:composite}.}
    \label{fig:coadd1}
\end{figure*}

\section{Rest-frame UV properties}\label{sec:results}

\begin{table}
\caption{Composite Spectrum}\label{tab:composite}
\centering
\scalebox{1.0}{
\begin{tabular}{lrcc}
\hline\hline
Rest Wavelength & $f_{\lambda}$ & $\sigma_{f_\lambda}$ & $N_{\rm obj}$ \\
(1) & (2) & (3) & (4) \\
\hline
1116.5 & 0.085  & 0.108 & 2 \\
1117.5 & 2.846  & 0.988 & 2 \\
... & ... & ... & ... \\
3738.5 & 0.460 & 0.002 & 2 \\
3739.5 & 0.497 & 0.113 & 2\\
\hline
\hline\\
\end{tabular}
}
\begin{tablenotes}
      \small
      \item NOTE. --- Median composite spectrum for our near-IR quasar sample. Wavelengths are in units of \AA. Flux density and flux density error units are arbitrary. The last column indicates how many objects contributed to the median composite at each wavelength pixel. 
\end{tablenotes}
\end{table}

We now proceed to study the rest-frame UV spectral properties of our GNIRS sample. In this work we only consider our GNIRS sample since all data are processed and analyzed in a uniform manner. We are in the process of collecting other existing near-IR spectroscopic data of $z\gtrsim5.7$ quasars and will jointly analyze the expanded sample with a uniform spectral analysis. Nevertheless, our GNIRS sample already represents the largest statistical sample to date for the study of rest-frame UV properties of $z\gtrsim 5.7$ quasars, in particular at the bright end.

To compare the properties of these high-$z$ quasars with their lower-$z$ counterparts, we create a control sample matched in restframe 1350\,\AA\, continuum luminosity. \footnote{We choose this specific monochromatic luminosity instead of a broad-band luminosity because the same quantity is explicitly used in earlier work in BH mass and bolometric luminosity calculations, and in the correlation between \CIV\ properties and quasar luminosity. We do not find any significant difference in our results if we adopt the monochromatic luminosity at a different wavelength (e.g., at 3000\,\AA) to construct our matched control sample. } It is crucial to match the luminosity of quasars, not only because luminosity is related to the accretion power, but also because many emission-line properties are functions of quasar luminosity. This is particularly true for high-ionization lines such as \CIV, where the line rest-frame equivalent width (REW) decreases with luminosity \citep[e.g., the Baldwin effect,][]{Baldwin_1977} and the line profile (velocity shift and asymmetry) changes with luminosity in a systematic manner \citep[e.g.,][]{Richards_etal_2002,Shen_etal_2008,Richards_etal_2011}.

We select control quasars from the SDSS DR7 quasar catalog in \citet[][]{Shen_etal_2011} with both \CIV\ and \MgII\ coverage, since this range contains most of the important broad emission lines that trace the properties of quasar accretion. Using a luminosity grid (with $\Delta \log L=0.2$) and for each bin that the high-$z$ quasars reside in, we randomly select 50 times more quasars from the DR7 catalog. Thus the resulting control sample has the same distribution in 1350\,\AA\ continuum luminosity as the high-$z$ sample. This control sample contains SDSS DR7 quasars at $z\sim 1.5-2.3$ that cover most of the major rest-frame UV lines from \lya\ to \MgII, providing a good reference sample to compare with our high-$z$ sample. We fit the control sample with the same global fitting recipe as for our GNIRS sample.\footnote{Since we used a different continuum fitting recipe for our high-$z$ sample than the one used in the \citet{Shen_etal_2011} catalog, we re-fit the control DR7 quasars with the same methodology. The new continuum fitting method mostly affected the REW of the broad \CIV\ line. We found that the global continuum fitting method adopted in this paper produces $\sim 18\%$ larger REWs of the \CIV\ line on average than those reported in the \citet[][]{Shen_etal_2011} catalog. On the other hand, other spectral properties, such as line peaks and widths, are all consistent within the errors with those compiled in the \citet{Shen_etal_2011} catalog. }

\begin{figure}
\centering
    \includegraphics[width=0.48\textwidth]{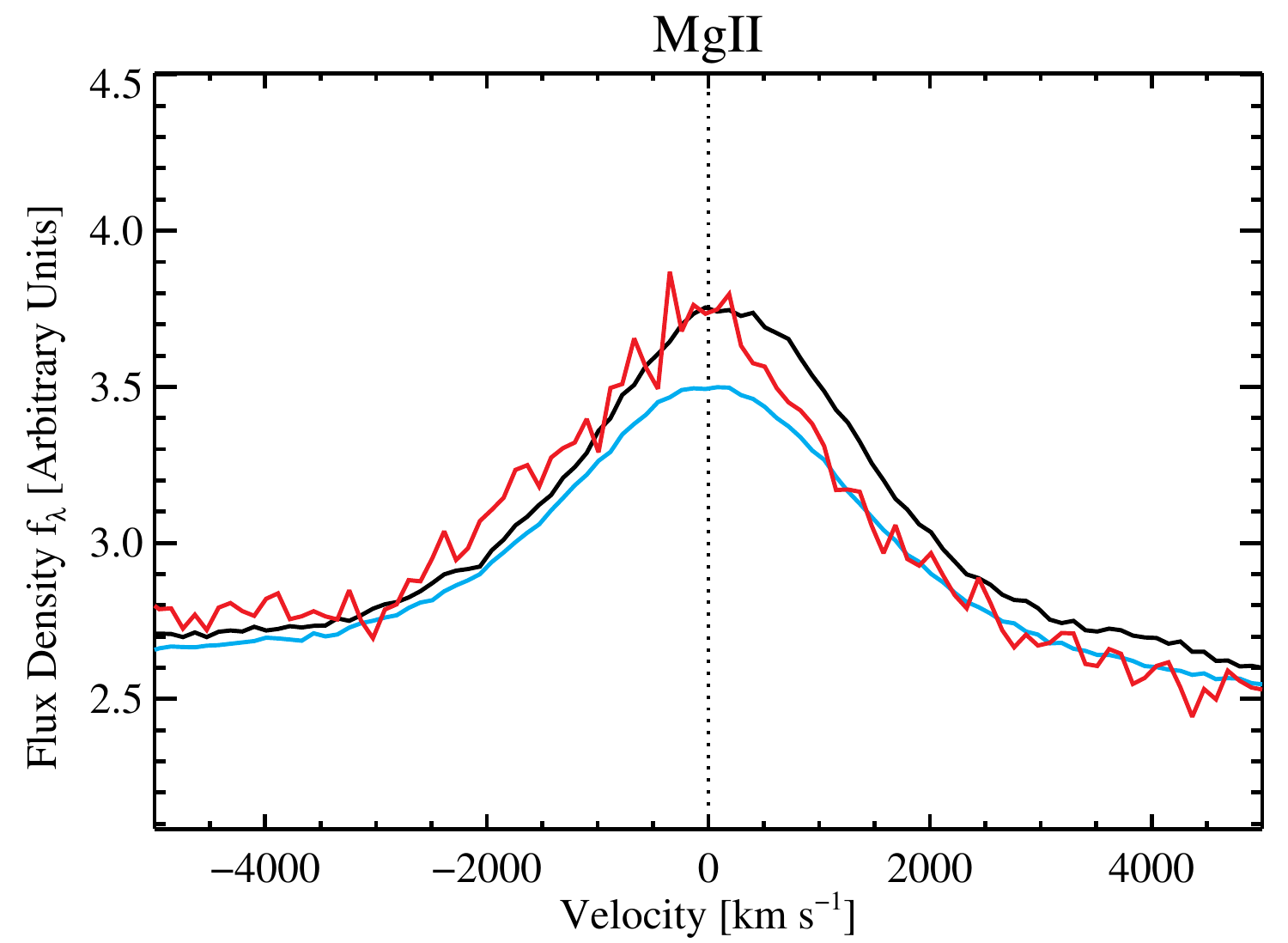}
    \includegraphics[width=0.48\textwidth]{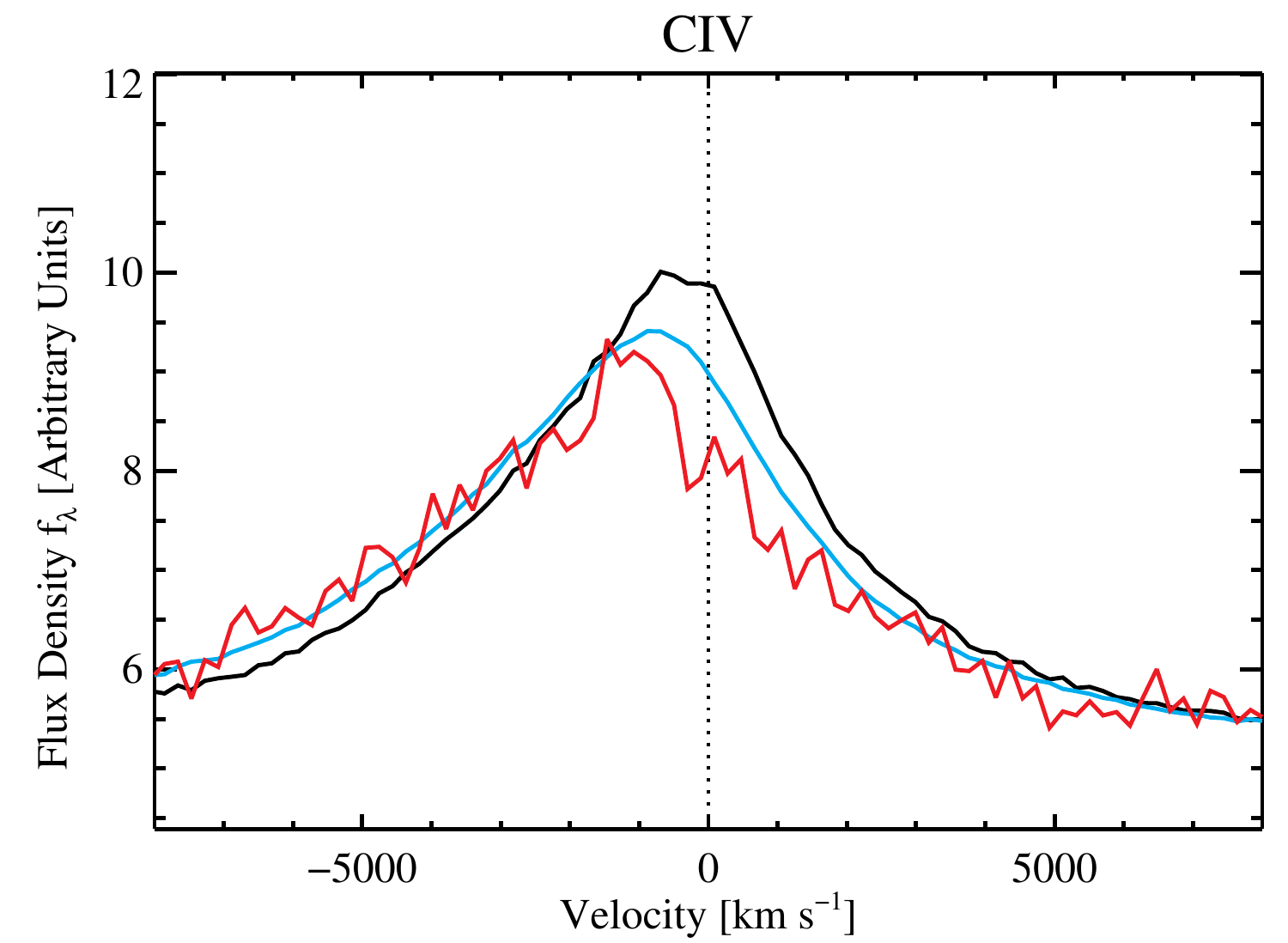}
     \caption{A detailed view of the median composite spectra as shown in Fig.\ \ref{fig:coadd1} around the \MgII\ (top) and \CIV\ (bottom) lines for the high-$z$ sample (red), the luminosity-matched control sample at lower-$z$ (cyan) and the sample from \citet{Vandenberk_etal_2001}. The improved redshifts for the control sample from \citet{Hewett_Wild_2010} and our best estimates for the high-$z$ sample consistently place the \MgII\ peak near zero velocity, while there is a small (but negligible) \MgII\ redshift for the \citet{Vandenberk_etal_2001} composite spectrum due to differences in the systemic redshift estimation in earlier work. On the other hand, \CIV\ is notably blueshifted from \MgII, with both the high-$z$ sample and the control sample having a larger blueshift than the \citet{Vandenberk_etal_2001} sample. This is because the \CIV\ blueshift increases with quasar luminosity \citep[e.g.,][]{Richards_etal_2002,Shen_etal_2016b}, and the \citet{Vandenberk_etal_2001} sample has a lower average luminosity than both the high-$z$ and the control samples. }
    \label{fig:coadd2}
\end{figure}

We create median composite spectra for the high-$z$ sample and the control sample following \citet{Vandenberk_etal_2001} to compare their average spectral properties. For the high-$z$ sample we adopt our best systemic redshift estimates, and for the control DR7 quasars we adopt the improved redshifts from \citet{Hewett_Wild_2010} during the coadding process. The \citet{Hewett_Wild_2010} redshifts are as good as those using our own redshift recipes in \citet{Shen_etal_2016b}, since both approaches take into account the velocity shifts of broad lines. As discussed in \citet{Vandenberk_etal_2001}, the median composite spectrum is more suitable for studying the relative line strength than are other spectral averaging methods. Fig.\ \ref{fig:coadd1} displays the median composite spectra for different samples. High-$z$ quasars have similar line strengths in most of the broad lines. In particular, the UV \FeII\ strength relative to broad \MgII\ is almost identical to that of low-$z$ quasars, as discovered in earlier small samples \citep[][]{Barth_etal_2003,DeRosa_etal_2011}. This line ratio can be used to measure the most precise alpha/iron abundance ratio for these high-$z$ quasars. However, \lya\ is apparently much weaker because of the much stronger absorption in the high-$z$ sample. Some of the narrow emission lines such as \NeV\ are also considerably weaker in the high-$z$ and control composite spectra than in the Vanden Berk et al. composite spectrum. This is likely a combination of the Baldwin effect and the fact that higher-luminosity quasars (in the high-$z$ and the control samples) have more massive host galaxies, leading to broader and thus less prominent narrow emission lines such as \NeV.

Both the high-$z$ sample and the control sample have an average \CIV\ profile that is more blueshifted from the low-ionization lines such as \MgII\, than the median composite spectrum from \citet{Vandenberk_etal_2001}, as further demonstrated in Fig.\ \ref{fig:coadd2}. The latter composite spectrum was generated using a sample of quasars that have lower luminosity than our high-$z$ sample and the matched control sample. The \CIV\ blueshift relative to \MgII\ increases with quasar luminosity \citep[e.g.,][]{Richards_etal_2002, Shen_etal_2016b}, so the \CIV\ blueshift is smaller in that composite than in our sample and the control sample.

\begin{figure}
\centering
    \includegraphics[width=0.48\textwidth]{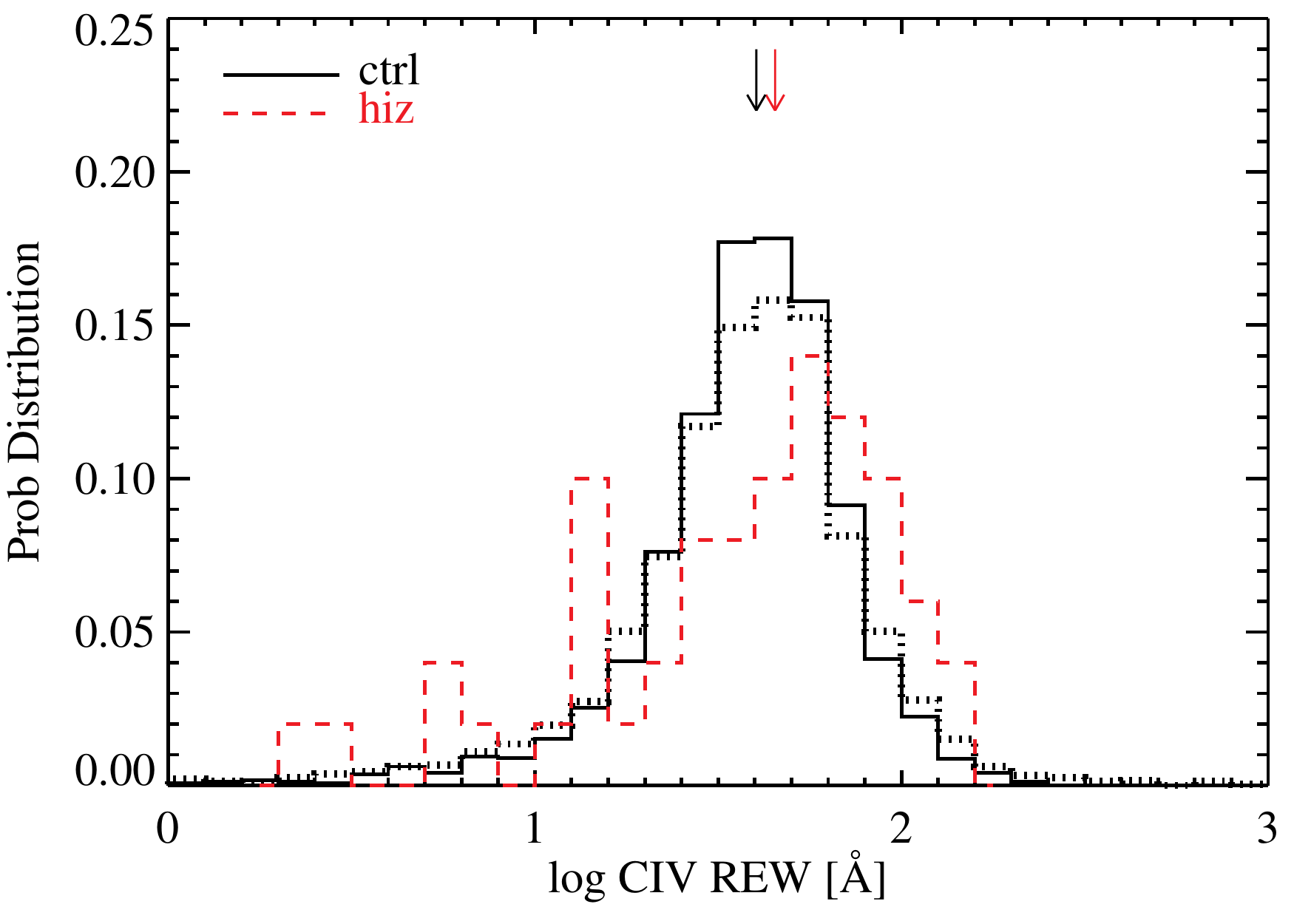}
    \includegraphics[width=0.48\textwidth]{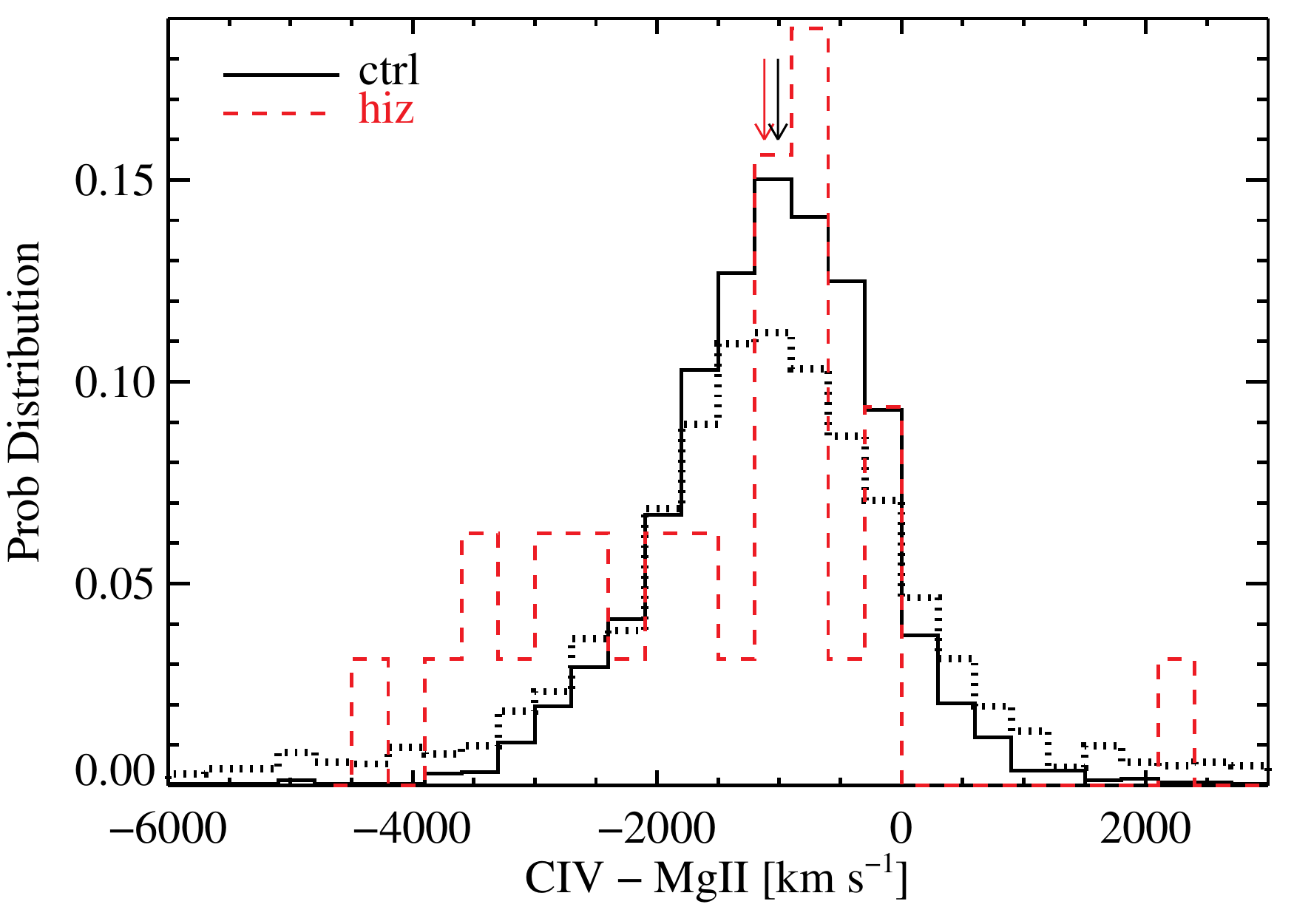}
     \caption{{Histograms of the \CIV\ rest equivalent width (top) and the \CIV\ blueshift relative to \MgII\ (bottom). The red dashed histograms are for the high-$z$ sample and the black solid histograms are for the control sample at lower redshifts matched in quasar luminosity. The median value of the distribution is indicated by an arrow at the top. The black dotted lines are the results of the control sample measurements randomly shuffled by the uncertainty distribution of the high-$z$ sample to compensate for the different SNR in the two samples. The median values remain almost the same, and KS tests on the distributions for the high-$z$ sample and the shuffled control sample suggest insignificant difference between the two distributions, with a null probability of $\sim 0.07$ and $\sim 0.4$ for the \CIV\ REW and blueshift distributions, respectively. Therefore we conclude that there is no significant difference in the plotted quantities between the high-$z$ sample and the control sample. }}
    \label{fig:civ_hist}
\end{figure}

We now examine the measurements for individual objects. Fig.\ \ref{fig:civ_hist} shows the distributions of the \CIV\ REWs and \CIV-\MgII\ blueshift for the high-$z$ sample and the control sample. Similar to our findings using the composite spectra, there is no significant difference in the typical \CIV\ properties when quasar luminosity is matched. The \CIV-\MgII\ velocity shift is independent of the systemic redshift estimate and both samples show a similar median \CIV-\MgII\ blueshift of $\sim 1000\,\kms$. Examinations of other broad lines covered in our GNIRS spectra also did not result in any significant differences between the two samples. Interestingly, \citet{Mazzucchelli_etal_2017} reported a significantly higher average \CIV\ blueshift relative to \MgII\ in 9 quasars at $z\gtrsim 6.5$ compared to low-$z$ SDSS quasars matched in luminosity. Mazzucchelli et al.\ fit a single Gaussian to the \CIV\ line, which will overestimate the peak blueshift given the blue-asymmetric profile (see Fig.\ \ref{fig:coadd2}), while we fit multiple Gaussians consistently to both the high-$z$ sample and the low-$z$ SDSS sample. Despite the small sample statistics and the difference in the fitting details, it is possible that the observed $z\gtrsim 6.5$ quasars have intrinsically different properties than their lower-$z$ counterpart. 

Although our control sample is matched in luminosity to our high-$z$ GNIRS sample, the SNR distributions of the two samples are different. This does not affect median values of the quantities we compare or the composite spectrum, but may impact the comparison of the full distribution. The median uncertainties in the measured spectral quantities in our GNIRS sample are only slightly larger than those for the SDSS control sample, but our GNIRS sample has a higher fraction of objects with large uncertainties. To test the impact of different SNR, we perturb the measured spectral quantities of the control sample using a Gaussian random variable with variance randomly assigned using the measurement uncertainties for the GNIRS sample (i.e., duplicated to match the number of the control sample). The results for the shuffled measurements for the control sample are shown as the black dotted lines in Fig.\ \ref{fig:civ_hist}. We found that for all the distributions studied here, the difference in SNR does not impact the results much. In fact, the shuffled distributions for the control sample better match those for the high-$z$ sample than the un-shuffled distributions in the tails that are mostly contributed by noisy measurements. 

\begin{figure}
\centering
    \includegraphics[width=0.48\textwidth]{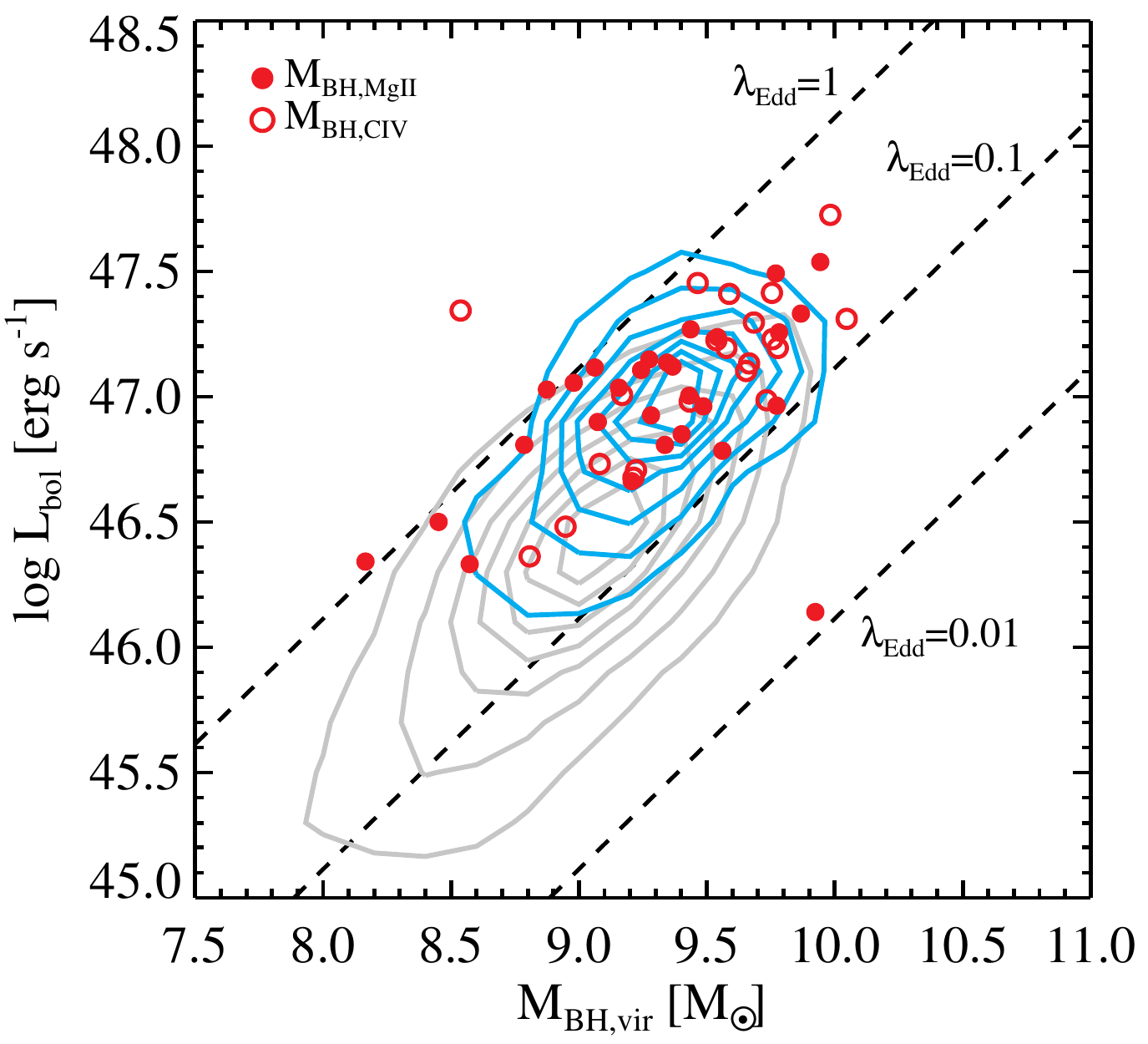}
    \includegraphics[width=0.48\textwidth]{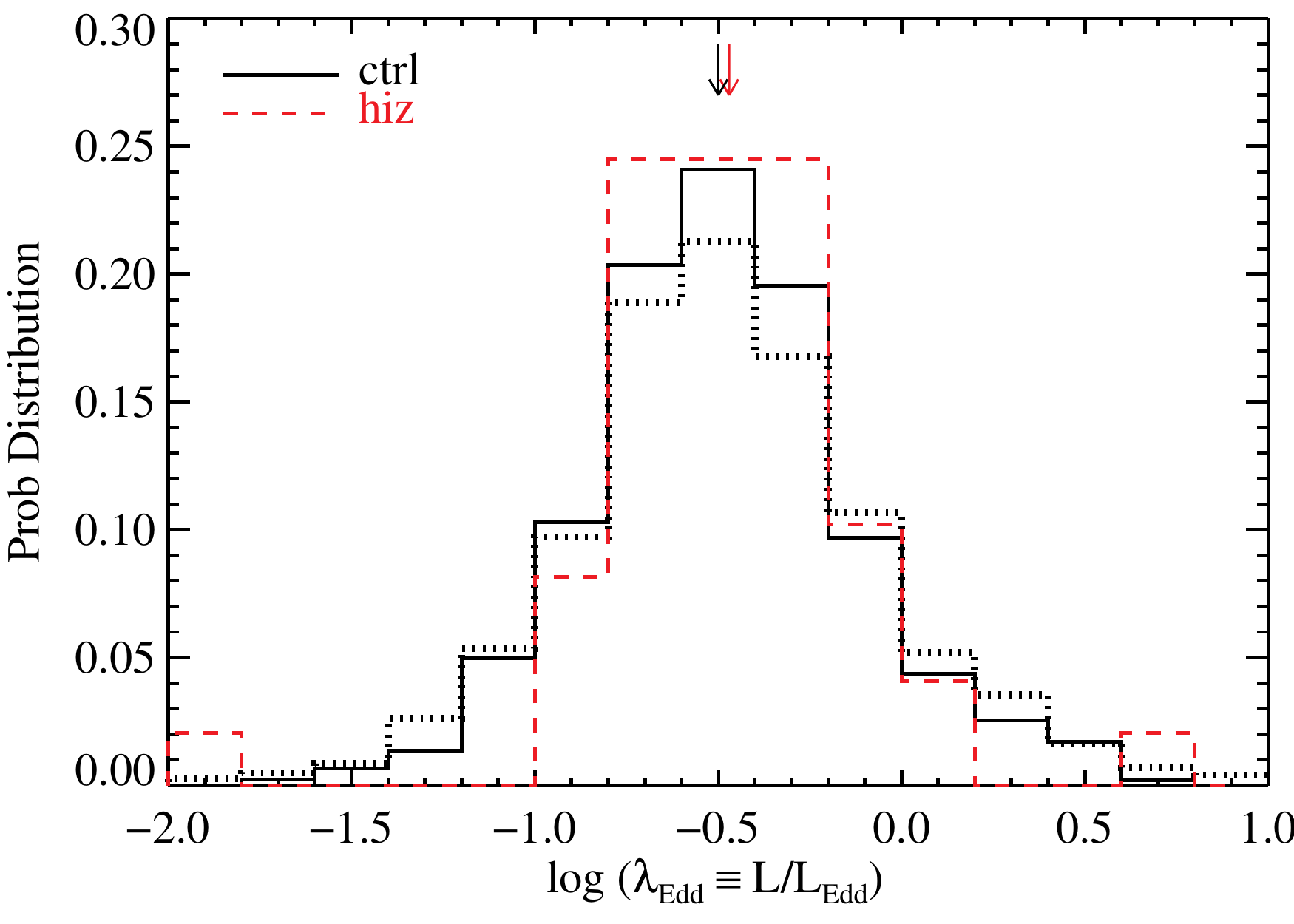}
     \caption{{{\it Top:} the BH mass-luminosity plane. The gray contours show the distribution for all SDSS DR7 quasars from the \citet{Shen_etal_2011} catalog, and the cyan contours show the distribution for the control sample matched in luminosity to our high-$z$ sample. The red points show our high-$z$ sample, with the \MgII-based virial BH masses shown in filled circles and the less reliable \CIV-based virial BH masses shown in open circles. We caution that individual virial BH masses could have a systematic uncertainty of $\sim 0.4$ dex \citep[e.g.,][]{Shen_2013} and individual bolometric luminosities could also be uncertain by a factor of up to 2 \citep[e.g.,][]{Richards_etal_2006b}. The one quasar with apparent super-Eddington accretion (J$0300-2232$) has its virial BH mass estimated from \CIV, whose fit is likely impacted by the absorption features adjacent to \CIV. Another apparent outlier with the lowest Eddington ratio (J$2356+0023$) is caused by a poor fit to the noisy \MgII\ line. {\it Bottom:} the distribution of Eddington ratio $L/L_{\rm Edd}$ for our high-$z$ sample (red dashed line) and the luminosity-matched control quasars at lower redshifts (black solid line). The median values of the distributions are marked by the arrows. The black dotted line shows the result for the control sample measurements randomly shuffled by the uncertainty distribution of the high-$z$ sample to compensate for the different SNR in the two samples. A KS test on the distributions for the high-$z$ sample and the shuffled control sample suggests an insignificant difference between the two distributions with a null probability of $\sim 0.05$.} }
    \label{fig:ML}
\end{figure}

We derive BH mass estimates for our high-$z$ sample using the so-called ``single-epoch virial BH masses'' that utilize the widths of the broad emission lines and the continuum luminosity measured from single-epoch spectroscopy. Despite the popularity of these virial BH mass estimates in recent years, there are substantial systematic uncertainties associated with them and differences among different broad-line estimators \citep[for a comprehensive review, see, e.g.,][]{Shen_2013}. Here we simply present these BH mass estimates, following the fiducial recipes summarized in \citet{Shen_etal_2011} based on \MgII\ and \CIV, and caution that there is substantial systematic uncertainty ($\sim 0.4$ dex) for individual estimates, which is typically much larger than the measurement uncertainty from spectral fits. These adopted fiducial mass recipes from \citet{Shen_etal_2011} were tested using SDSS quasars with a similar spectral fitting methodology as used in this work.

For a subset of 25 quasars we have BH mass estimates from both \MgII\ and \CIV\ lines. There has been extensive discussion in the literature on the reliability of \CIV\ as a virial BH mass estimator for high-redshift quasars \citep[e.g.,][and references therein]{Shen_2013}. Unlike the low-ionization broad lines such as \hbeta\ and \MgII, the \CIV\ line often displays a more asymmetric profile and a significant blueshift that correlate with the luminosity and Eddington ratio of the quasar \citep[e.g.,][]{Sulentic_etal_2000,Richards_etal_2011,Brotherton_etal_2015}. \CIV\ line width is poorly correlated with that of \MgII\ or \hbeta\ in high-luminosity quasars \citep[e.g.,][]{Baskin_Laor_2005,Shen_etal_2008,Shen_Liu_2012,Coatman_etal_2017}, and the difference correlates with quasar luminosity. It is likely that the \CIV\ line includes a significant non-virial component \citep[e.g.,][]{Denney_2012}, particularly in high-luminosity objects. While it is possible to calibrate the \CIV\ estimator to yield average BH masses unbiased relative to those based on \hbeta\ and \MgII\ \citep[e.g.,][]{Shen_etal_2008,Assef_etal_2011,Shen_Liu_2012,Runnoe_etal_2013,Coatman_etal_2017,Park_etal_2017,Mejia-Restrepo_etal_2018}, the scatter between the \CIV\ mass and \hbeta\ or \MgII\ mass is still considerably larger than that between \MgII\ and \hbeta. Given that most reverberation mapping (RM) measurements to date are on the \hbeta\ line \citep[e.g.,][]{Peterson_etal_2004} and that the current single-epoch mass estimators are all directly or indirectly based on the \hbeta\ RM results, it is reasonable to assume that \hbeta\ is the safest line from which to estimate a virial BH mass. Since \MgII\ width correlates with \hbeta\ width well \citep[e.g.,][]{Shen_etal_2008,Wang_etal_2009b}, it is also reasonable to assume that \MgII\ is relatively safe to use as a BH mass estimator. \CIV\ can still be used since on average it provides consistent BH masses with those from \MgII\ or \hbeta, albeit with a large intrinsic scatter. 

For this work we adopt fiducial BH mass estimates based on \MgII\ if available; otherwise we use \CIV-based masses. However, we have tested using alternative fiducial BH masses, e.g., using \CIV\ over \MgII\ masses or the average of the two masses \citep[as suggested by][]{Vestergaard_etal_2011}, and did not find any significant changes in our results. We require an additional criterion for BH mass estimation that the line must have a measured REW at $>1\sigma$ to avoid extremely noisy measurements, which leaves one object, J$0055+0146$, without a virial BH mass estimate.

Fig.\ \ref{fig:ML} (top panel) displays the distribution of quasars in the BH mass versus bolometric luminosity plane. The gray and cyan contours show the distributions for the entire SDSS DR7 quasar sample \citep{Shen_etal_2011} and the control sample. Our high-$z$ sample is shown in red points, where filled symbols use the \MgII-based BH masses and open symbols use the \CIV-based BH masses. The bottom panel compares the distribution of Eddington ratio, $\lambda_{\rm Edd}\equiv L_{\rm bol}/L_{\rm Edd}$ where $L_{\rm Edd}=1.3\times 10^{38}\,{\rm erg\,s^{-1}}(M_{\rm BH}/M_\odot)$ is the Eddington luminosity of the BH, between the high-$z$ sample and the luminosity-matched control sample. Our high-$z$ quasars have similar BH masses and Eddington ratios as the control sample, based on the virial BH mass estimates. The median Eddington ratio for the high-$z$ sample is $\log\lambda_{\rm Edd}\sim -0.5$, and we estimate a logarithmic dispersion of $\sim 0.3$ dex by fitting a Gaussian function to the distribution of $\log\lambda_{\rm Edd}$. {There is one quasar (J$0300-2232$) with apparent super-Eddington accretion, with the BH mass estimated from \CIV, which may be impacted by the absorption features adjacent to \CIV. Another object, J2356+0023, has a noisy \MgII\ line which resulted in a large uncertainty in the \MgII-based BH mass ($\sim 0.5$ dex), and it appears as an apparent outlier in Fig.\ \ref{fig:ML} (top panel). The Eddington ratio based on \CIV\ is $\log \lambda_{\rm Edd}\sim -1.2$ for this object.}

We note that our \MgII-based virial BH mass recipe was calibrated to agree with both \hbeta\ and \CIV-based masses on average with the \citet{Vestergaard_Peterson_2006} recipe using SDSS quasar samples that cover two lines at low and high redshift \citep{Shen_etal_2011}. Our \MgII\ mass recipe is almost identical to that derived by \citet{Trakhtenbrot_Netzer_2012}; both have the same slope on continuum luminosity as in \citet{McLure_Dunlop_2004} and similar zero points. Earlier work \citep[e.g.,][]{Kurk_etal_2007,Jiang_etal_2007,DeRosa_etal_2011} utilized the \MgII\ recipe from \citet{McLure_Dunlop_2004}, which has the same luminosity slope but a smaller zero point by $0.22$ dex, compared with our \MgII\ masses. We have checked the handful of objects\footnote{There are three objects, J$0836+0054$, J$1044-0125$ and J$1623+3112$, in common with \citet{Jiang_etal_2007}; one object, J$0836+0054$, in common with \citet{Kurk_etal_2007}; and three objects, J$0050+3445$, J$0055+0146$ and J$0221-0802$, in common with \citet{Willott_etal_2010b}. However, our GNIRS spectra for J$0055+0146$ and J$0221-0802$ are too noisy for a meaningful comparison with \citet{Willott_etal_2010b}. } in our sample that have NIR spectroscopy in earlier work and found reasonably good agreement in line width and continuum luminosity measurements considering the different epochs and spectral fitting recipes. Therefore the apparent difference in the reported Eddington ratios for the common objects is largely due to the difference in the adopted \MgII\ mass estimators. On the other hand, \citet{Willott_etal_2010b} reported Eddington ratios near unity for a sample of 9 low-luminosity quasars at $z\sim 6$. There is negligible overlap in the \citet{Willott_etal_2010b} sample and our GNIRS sample. But the distribution at the faint luminosity end in Fig.\ \ref{fig:ML} shows a few quasars closer to unity Eddington ratio than the rest of the sample, which is more in line with the typical Eddington ratios reported in \citet{Willott_etal_2010b} for low-luminosity high-$z$ quasars. Also,  the \MgII\ BH masses in \citet{Willott_etal_2010b} were based on the recipe from \citet{Vestergaard_Osmer_2009}, which has a shallower luminosity dependence than our recipe, and would yield on average smaller BH masses (higher Eddington ratios) than those based on our \MgII\ recipe for the luminosity regime of high-$z$ quasars. \citet{Willott_etal_2010b} also used a slightly larger bolometric correction, which further helps to explain the higher Eddington ratios reported in their work.

There are other existing \MgII\ based mass recipes that use slightly different zero points, luminosity and line width dependences \citep[e.g.,][]{Onken_Kollmeier_2008,Wang_etal_2009b,Woo_etal_2018}. Using different mass recipes introduces $\sim 0.2-0.3$ dex offset in the virial BH masses, which we consider tolerable given the $\sim 0.4$ dex systematic uncertainty in single-epoch virial BH masses \citep[e.g.,][]{Shen_2013}. Considering the large uncertainties in the BH mass estimates and bolometric corrections in individual objects, as well as the much smaller statistics in earlier work, we do not consider the factor of $\sim 2-3$ difference in the typical Eddington ratios between our work and earlier results significant.

%\red{Compare the BH mass and Eddington ratio results to previous work at similar or higher redshifts. }

\section{Discussion}\label{sec:dis}

\begin{figure}
\centering
    \includegraphics[width=0.48\textwidth]{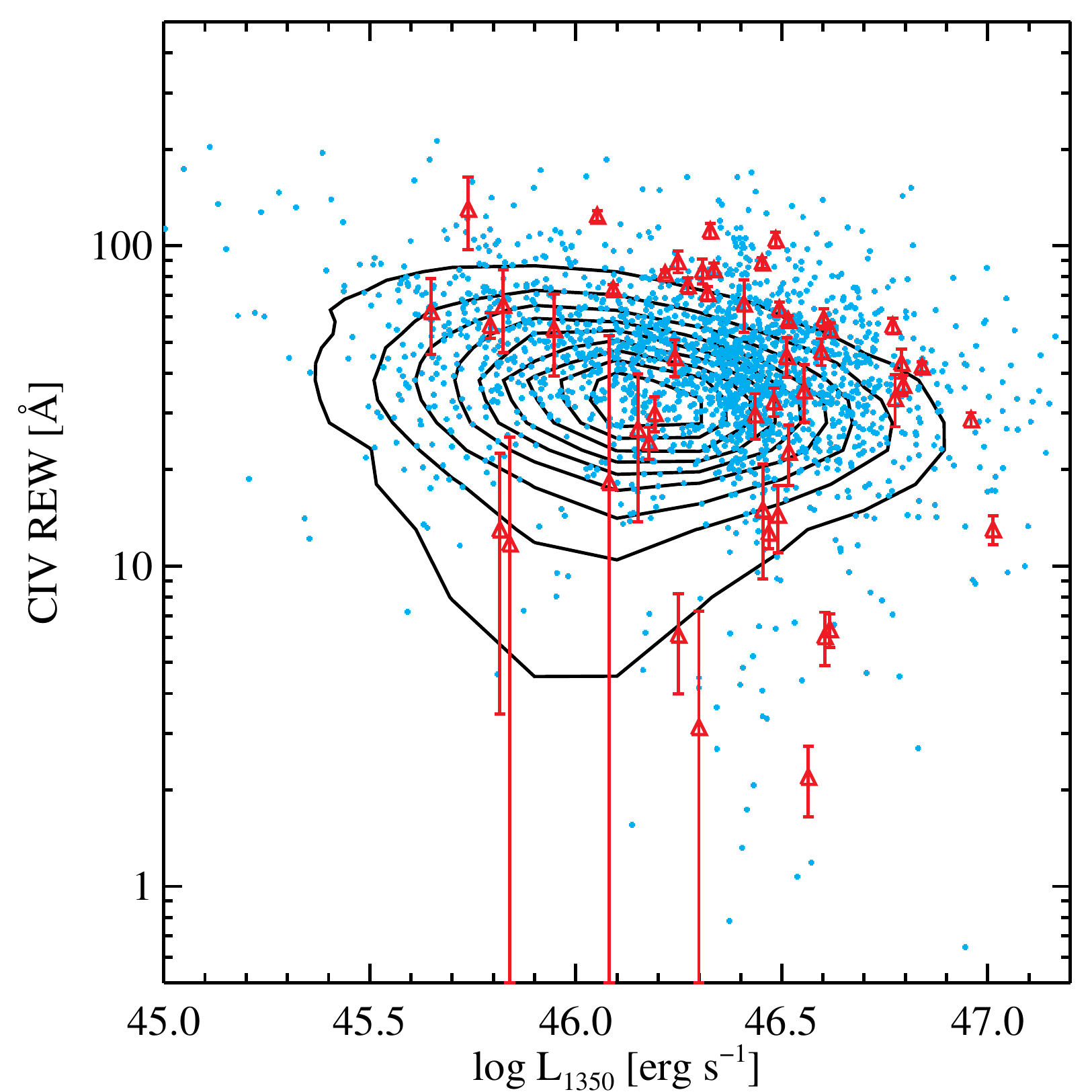}
    \includegraphics[width=0.48\textwidth]{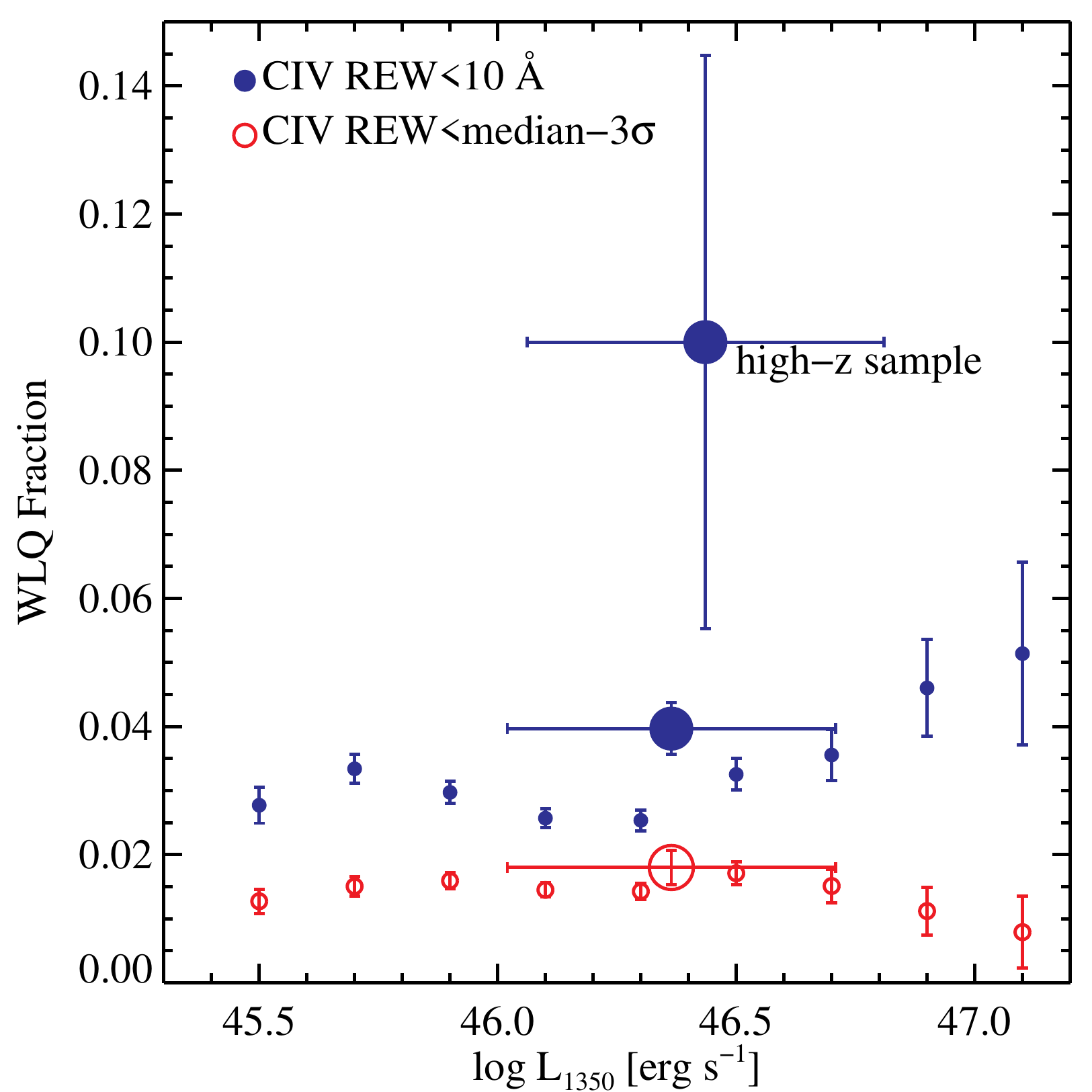}
     \caption{{{\it Top:} \CIV\ rest-frame EW as a function of luminosity. The black contours show all SDSS DR7 quasars from the \citet{Shen_etal_2011} catalog. The cyan points and the red triangles show the control sample and the high-$z$ sample, respectively. The general trend of decreasing \CIV\ REW with continuum luminosity is known as the Baldwin effect \citep[e.g.,][]{Baldwin_1977}. {\it Bottom:} the WLQ fraction as a function of continuum luminosity. The small symbols are computed using all SDSS DR7 quasars in \citet{Shen_etal_2011} binned by luminosity. The large symbols are for our high-$z$ sample (top blue filled circle) and the control sample (bottom blue filled circle and red open circle), computed for the entire sample. The two types of symbols represent the two definitions of WLQ (see text). The error bars on luminosity for the high-$z$ and control samples are the standard deviations in each sample, and the vertical error bars are estimated with Poisson counting uncertainties. For the high-$z$ sample we only use the first WLQ definition (\CIV\ ${\rm REW}<10$\,\AA) since the dispersion in the \CIV\ REW is not well defined due both to the smaller sample size and noisier measurements.} }
    \label{fig:wlq}
\end{figure}

\subsection{Weak-line quasars}

A small fraction of radio-quiet quasars have significantly weaker (small REW) broad emission lines (in particular the UV lines), dubbed ``weak line quasars'' \citep[WLQs, e.g., ][]{Fan_etal_1999, Plotkin_etal_2008, Plotkin_etal_2010, Diamond-Stanic_etal_2009, Shemmer_etal_2010,Shemmer_Lieber_2015}, where the weakness of the line is not apparently due to contamination from jet emission as in radio-loud quasars. WLQs sometimes have weaker X-ray emission than quasars with normal broad-line strength \citep[e.g.,][]{Wu_etal_2011, Luo_etal_2015}. While the exact cause of the weaker line strengths in WLQs is still unclear, one working scenario is that WLQs have a softer ionizing continuum than normal quasars such that the UV broad-line region sees less ionizing photons relative to the continuum underneath the line in WLQs, leading to reduced equivalent width \citep[e.g.,][]{Luo_etal_2015}. To produce a different ionizing spectrum, whether it is intrinsic or caused by filtering by some intermediate gas between the accretion disk and the broad-line region, requires that the accretion properties be different in WLQs. Thus studying the WLQ fraction at different redshifts can reveal possible changes in the accretion modes of SMBHs. Since the UV broad line strength also decreases with quasar luminosity \citep{Baldwin_1977}, one needs to study the WLQ fraction as a function of quasar luminosity. 

Since the definition of WLQs is somewhat arbitrary, we consider two definitions: 1) a quasar is a WLQ if its \CIV\ rest-frame EW is less than 10\,\AA\, \citep[e.g.,][]{Diamond-Stanic_etal_2009}, or 2) its \CIV\ REW is lower than $3\sigma$ from the median of a population (e.g., quasars in a certain luminosity bin), where $\sigma$ is the sample standard deviation. The second definition is meaningful when we consider the WLQ fraction as a function of quasar luminosity, where the median and standard deviation are calculated for each luminosity bin. We focus on the \CIV\ line for our study on WLQs since \lya\ is heavily absorbed and other broad lines are not as strong as \CIV\ in our sample. 

Fig.\ \ref{fig:wlq} (top) shows the Baldwin effect of \CIV\ for various samples. Our high-$z$ sample follows a similar trend as the control sample. In the bottom panel of Fig.\ \ref{fig:wlq} we show the WLQ fraction as a function of luminosity, using all SDSS DR7 quasars in \citet{Shen_etal_2011}. The blue filled symbols use the first WLQ definition with a constant threshold in \CIV\ REW. The WLQ fraction with this definition rises towards high luminosities due to the Baldwin effect. The red open symbols use the second WLQ definition, which is roughly flat with luminosity since the average Baldwin effect is taken out. Clearly, if one uses a constant cut in \CIV\ REW to define WLQs and compare two different samples, one must match the luminosity of these two samples for a fair comparison. 

For our high-$z$ sample, the sample dispersion in \CIV\ REW is not well defined given the smaller number of objects than the control sample, as well as substantial measurement uncertainties in individual objects. Therefore we only consider the first definition of WLQs, and compare to the control sample which is matched in luminosity to the high-$z$ sample. We identify 5 WLQ quasars in the high-$z$ sample with measured \CIV\ REW less than 10\,\AA: J0850+3246, J1257+6349, J1335+3533, J1429+5447, and J1621+5155. Inspection of their spectra suggests that these should be genuine WLQs. The WLQ fraction for the high-$z$ sample is thus $\sim 10\pm4\%$ (5 out of 50), compared to $4.0\pm0.4\%$ for the control sample. Thus the high-$z$ sample seems to have a mild excess in the WLQ fraction compared to their low-$z$ counterparts matched in luminosity, albeit with large uncertainties. This result is consistent with the distributions of \CIV\ REW shown in Fig.\ \ref{fig:civ_hist}, where the distribution of the high-$z$ sample has a more prominent tail at the low REW end. \citet{Banados_etal_2016} reported a WLQ fraction of $\sim 14\%$ (16/117) using \lya\ with a larger sample of $z>5.6$ quasars, which is fully consistent with our results here based on \CIV.

%This WLQ fraction among our high-$z$ quasars is smaller than the $\sim 25\%$ (2 out of 8) value reported in \citet{Banados_etal_2014} although the uncertainties are large enough for them to be formally consistent. 

There is one known radio-loud quasar (J1429$+$5447) among the 5 quasars with \CIV\ REW$<10$\,\AA, where the UV continuum could be contaminated by jet emission.

%Earlier studies which used \lya\ to identify WLQs may suffer more from absorption for these high-$z$ quasars than their low-$z$ counterparts. 

\subsection{Early formation of SMBHs}

Our results in \S\ref{sec:results} and Fig. \ref{fig:ML} reveal that our high-$z$ quasars are accreting at moderately high Eddington ratios, with an average value of $\lambda_{\rm Edd}\approx 0.3$, similar to their low-$z$ counterparts with matched luminosities. This typical Eddington ratio is very similar to the value of $\sim 0.4$ for a sample of 15 $z\gtrsim 6.5$ quasars studied in \citet{Mazzucchelli_etal_2017}. Interestingly, these objects appear to be well bounded by the Eddington limit, even considering the systematic uncertainties in the estimation of BH masses and bolometric luminosities. 

The observed high-$z$ quasars are drawn from an underlying quasar population with an intrinsic distribution of BH masses accreting at a range of Eddington ratios. The inevitable flux limit will preferentially select the most luminous objects from the intrinsic distribution. As we naively expect that there are more abundant lower-mass SMBHs than higher-mass ones, those lower-mass BHs with higher-than-average Eddington ratios may scatter above the flux limit and bias the observed Eddington ratios high. At the same time, massive BHs accreting at lower-than-average Eddington ratios may be missed from the sample due to the flux limit. Thus our observed apparent Eddington ratios suggest that the majority of the unobserved quasars should be accreting at even lower Eddington ratios, as argued with robust statistical approaches using lower-redshift quasar samples \citep[e.g.,][]{Kelly_etal_2010,Shen_Kelly_2012,Kelly_Shen_2013}.

The $e$-folding time of BH growth at constant Eddington ratio is $t_e\approx 4.5\times 10^{8}\epsilon/[\lambda_{\rm Edd}(1-\epsilon)]\,$yr, where $\epsilon$ is the radiative efficiency. The typical radiative efficiency for $z\gtrsim 5.7$ quasars is constrained to be $\epsilon\approx 0.1$ \citep[e.g.,][]{Trakhtenbrot_etal_2017}, using the observed luminosity and BH mass, as well as scaling relations derived from accretion disk theories \citep[e.g.,][]{Davis_Laor_2011}. It would have been difficult \citep[although not impossible, e.g.,][]{Li_etal_2007} for these SMBHs to grow to their current mass from small ($\sim 100\,M_\odot$) seeds at these sub-Eddington levels of accretion rates, given the limited time available for their growth from seed BHs. Episodes of higher accretion rates \citep[e.g.,][]{Pezzulli_etal_2016} or heavier seed BHs \citep[e.g.,][]{Latif_etal_2013} are possible solutions to this early growth problem \citep[for recent reviews on this topic, see, e.g.,][]{Volonteri_2010,Haiman_2013,Latif_Schleicher_2018}. If the early BH growth invokes much higher accretion rates, this phase is most likely obscured in the UV-optical \citep[e.g.,][]{Sanders_Mirabel_1996,Hopkins_etal_2008a} and/or has much lower radiative efficiency \citep[e.g.,][]{Pezzulli_etal_2016}, otherwise we would have observed much higher Eddington ratios among the current high-$z$ quasar sample. These heavily obscured quasars can only be probed at other wavelengths \citep[e.g.,][and references therein]{Glikman_etal_2018,Hickox_Alexander_2018}.

%\red{Expand the discussion: 3) more references on hypermassive black holes. }

We also do not find evidence for a population of excessively large ($>10^{10}\,M_\odot$) black holes at these highest redshifts. The one object (J0203+0012) with a \CIV-only virial BH mass slightly greater than $10^{10}\,M_\odot$ is a broad absorption line quasar, where the \CIV\ line fit is likely impacted by the broad absorption, and the individual \CIV-based mass may not be as reliable as that based on the \MgII\ line as discussed earlier. If such a population of hyper-massive SMBHs exist in large numbers, they must be accreting at substantially lower Eddington ratios, or are obscured, and missed from our sample. It is possible that even more luminous quasars discovered at $z\gtrsim 5.7$ could harbor $>10^{10}\,M_\odot$ BHs \citep[e.g.,][]{Wu_etal_2015} but such objects are increasingly rare. In any case, we caution that individual quasars with reported virial BH mass greater than $10^{10}\,M_\odot$ in some past studies could be due to systematic uncertainties in their BH mass estimation \citep[see discussions in, e.g.,][]{Vestergaard_2004,Shen_2013}. Using a robust statistical approach on the lower-redshift SDSS quasar samples, \citet{Kelly_etal_2010} and \citet{Kelly_Shen_2013} constrained the maximum BH mass in quasars that can be observed in a thoretical all-sky survey without a flux limit to be several times $10^{10}\,M_\odot$, albeit with considerable systematic uncertainties. The absence of such hyper-massive BHs in our sample is consistent with the constraints for lower-$z$ quasars. From a theoretical perspective, some papers have argued for an upper limit of SMBH mass of $\sim 10^{10-11}\,M_\odot$ \citep[e.g.,][]{Natarajan_Treister_2009,King_2016}. But even if these hyper-massive SMBHs do exist, they would be extremely difficult to find given their apparent rareness and potential low activity (or possible obscuration). 

On the other hand, the distribution of high-$z$ quasars in the mass-luminosity plane of Fig.\ \ref{fig:ML} suffers from the flux limit of the search, lacking objects that are too faint to be included in the sample. Thus at fixed BH mass, the sample is incomplete towards lower luminosities, or lower Eddington ratios. While we observe these quasars accreting at moderately high Eddington ratios, it is possible that the majority of high-$z$ SMBHs are accreting at substantially lower Eddington ratios, as mentioned earlier. In order to constrain the intrinsic distribution of quasars in the mass-luminosity plane, one needs to take into account the selection function, which is based on flux (not BH mass), as well as the uncertainties of the BH mass estimates. The best approach is a forward modeling as detailed in earlier work \citep[e.g.,][]{Kelly_etal_2010,Shen_Kelly_2012,Kelly_Shen_2013}, which will be one of the follow-up studies based on our sample.  

\section{Summary}\label{sec:sum}

We have presented near-IR (simultaneous $YJHK$) spectra for a sample of 50 quasars at $z\gtrsim 5.7$ from our large Gemini GNIRS program. Based on these near-IR spectra, we measured the rest-frame UV spectral properties of these quasars, including broad emission line properties and derived quantities such as bolometric luminosities and virial BH mass estimates. 

We compared the UV spectral properties of these high-$z$ quasars with a control sample of quasars from SDSS at $z=1.5-2.3$ matched in continuum luminosity. We found that our high-$z$ quasars have a mild excess of weak-line quasars (with \CIV\ REW less than 10\,\AA) compared to the luminosity-matched control sample at lower redshifts. But other than that, the UV spectral properties are very similar between the high-$z$ and the low-$z$ quasars. This similarity between the high-$z$ and low-$z$ quasars also extends to X-ray properties when luminosity is matched \citep[e.g.,][]{Nanni_etal_2017,Banados_etal_2018b}. These results suggest that the same physical mechanisms are operating for these accreting SMBHs at all redshifts. Using broad \MgII\ and \CIV\ lines covered in the near-IR spectra, we estimated the BH masses and Eddington ratios of these high-$z$ quasars using the single-epoch virial mass method. We found that these high-$z$ quasars are accreting at moderately high Eddington ratios with a median value of $L/L_{\rm Edd}\sim 0.3$, similar to their low-$z$ counterparts at comparable luminosities as well as quasars at even higher redshifts \citep[e.g.,][]{Mazzucchelli_etal_2017}. We did not find evidence of a population of hyper-massive ($>10^{10}\,M_\odot$) BHs in our sample. Such objects are either too rare to be included in our sample, or are accreting at substantially lower Eddington ratios (or are heavily obscured) and are missed from current high-$z$ quasar searches. Our results are consistent with earlier studies of high-$z$ quasars with near-IR spectroscopy but with substantially better statistics. 

The data and spectral measurements presented in this work can form the basis of additional follow-up studies on $z\gtrsim 5.7$ quasars, which include the measurements of their demography in the mass-luminosity plane, detailed analysis of narrow and broad absorption lines, chemical abundances of gas in the broad-line region, etc. Moreover, our new data add to the increasingly larger multi-wavelength database to study these earliest SMBHs, and strongly motivate new spectroscopic observations to extend the wavelength coverage, e.g., mid-IR observations with JWST to study the rest-frame optical properties of these high-$z$ quasars \citep[][]{Kalirai_2018}. 

\acknowledgements

We thank the Gemini Observatory staff for their help with our observing program, and the anonymous referee for comments that improved the manuscript. J.W., L.J., and L.C.H. acknowledge support from the National Key R\&D Program of China (2016YFA0400702, 2016YFA0400703), and from the National Science Foundation of China (grant 11473002, 11533001, 11721303). Y.S. acknowledges support from an Alfred P. Sloan Research Fellowship and National Science Foundation grant AST-1715579. D.R. acknowledges support from the National Science Foundation under grant number AST-1614213. M.V. gratefully acknowledges support from the Independent Research Fund Denmark via grant numbers DFF 4002-00275 and 8021-00130.

%\appendix

\bibliography{/Users/yshen/Research/refs,toShen}

\begin{thebibliography}{}
\expandafter\ifx\csname natexlab\endcsname\relax\def\natexlab#1{#1}\fi

\bibitem[{{Alexandroff} {et~al.}(2013){Alexandroff}, {Strauss}, {Greene},
  {Zakamska}, {Ross}, {Brandt}, {Liu}, {Smith}, {Ge}, {Hamann}, {Myers},
  {Petitjean}, {Schneider}, {Yesuf}, \& {York}}]{Alexandroff_etal_2013}
{Alexandroff}, R., {Strauss}, M.~A., {Greene}, J.~E., {et~al.} 2013, \mnras,
  435, 3306

\bibitem[{{Allen} {et~al.}(2011){Allen}, {Hewett}, {Maddox}, {Richards}, \&
  {Belokurov}}]{Allen_etal_2011}
{Allen}, J.~T., {Hewett}, P.~C., {Maddox}, N., {Richards}, G.~T., \&
  {Belokurov}, V. 2011, \mnras, 410, 860

\bibitem[{{Assef} {et~al.}(2011){Assef}, {Denney}, {Kochanek}, {Peterson},
  {Koz{\l}owski}, {Ageorges}, {Barrows}, {Buschkamp}, {Dietrich}, {Falco},
  {Feiz}, {Gemperlein}, {Germeroth}, {Grier}, {Hofmann}, {Juette}, {Khan},
  {Kilic}, {Knierim}, {Laun}, {Lederer}, {Lehmitz}, {Lenzen}, {Mall}, {Madsen},
  {Mandel}, {Martini}, {Mathur}, {Mogren}, {Mueller}, {Naranjo}, {Pasquali},
  {Polsterer}, {Pogge}, {Quirrenbach}, {Seifert}, {Stern}, {Shappee}, {Storz},
  {Van Saders}, {Weiser}, \& {Zhang}}]{Assef_etal_2011}
{Assef}, R.~J., {Denney}, K.~D., {Kochanek}, C.~S., {et~al.} 2011, \apj, 742,
  93

\bibitem[{{Ba{\~n}ados} {et~al.}(2014){Ba{\~n}ados}, {Venemans}, {Morganson},
  {Decarli}, {Walter}, {Chambers}, {Rix}, {Farina}, {Fan}, {Jiang}, {McGreer},
  {De Rosa}, {Simcoe}, {Wei{\ss}}, {Price}, {Morgan}, {Burgett}, {Greiner},
  {Kaiser}, {Kudritzki}, {Magnier}, {Metcalfe}, {Stubbs}, {Sweeney}, {Tonry},
  {Wainscoat}, \& {Waters}}]{Banados_etal_2014}
{Ba{\~n}ados}, E., {Venemans}, B.~P., {Morganson}, E., {et~al.} 2014, \aj, 148,
  14

\bibitem[{{Ba{\~n}ados} {et~al.}(2015){Ba{\~n}ados}, {Venemans}, {Morganson},
  {Hodge}, {Decarli}, {Walter}, {Stern}, {Schlafly}, {Farina}, {Greiner},
  {Chambers}, {Fan}, {Rix}, {Burgett}, {Draper}, {Flewelling}, {Kaiser},
  {Metcalfe}, {Morgan}, {Tonry}, \& {Wainscoat}}]{Banados_etal_2015}
---. 2015, \apj, 804, 118

\bibitem[{{Ba{\~n}ados} {et~al.}(2016){Ba{\~n}ados}, {Venemans}, {Decarli},
  {Farina}, {Mazzucchelli}, {Walter}, {Fan}, {Stern}, {Schlafly}, {Chambers},
  {Rix}, {Jiang}, {McGreer}, {Simcoe}, {Wang}, {Yang}, {Morganson}, {De Rosa},
  {Greiner}, {Balokovi{\'c}}, {Burgett}, {Cooper}, {Draper}, {Flewelling},
  {Hodapp}, {Jun}, {Kaiser}, {Kudritzki}, {Magnier}, {Metcalfe}, {Miller},
  {Schindler}, {Tonry}, {Wainscoat}, {Waters}, \& {Yang}}]{Banados_etal_2016}
{Ba{\~n}ados}, E., {Venemans}, B.~P., {Decarli}, R., {et~al.} 2016, \apjs, 227,
  11

\bibitem[{{Ba{\~n}ados} {et~al.}(2018{\natexlab{a}}){Ba{\~n}ados}, {Venemans},
  {Mazzucchelli}, {Farina}, {Walter}, {Wang}, {Decarli}, {Stern}, {Fan},
  {Davies}, {Hennawi}, {Simcoe}, {Turner}, {Rix}, {Yang}, {Kelson}, {Rudie}, \&
  {Winters}}]{Banados_etal_2018}
{Ba{\~n}ados}, E., {Venemans}, B.~P., {Mazzucchelli}, C., {et~al.}
  2018{\natexlab{a}}, \nat, 553, 473

\bibitem[{{Ba{\~n}ados} {et~al.}(2018{\natexlab{b}}){Ba{\~n}ados}, {Connor},
  {Stern}, {Mulchaey}, {Fan}, {Decarli}, {Farina}, {Mazzucchelli}, {Venemans},
  {Walter}, {Wang}, \& {Yang}}]{Banados_etal_2018b}
{Ba{\~n}ados}, E., {Connor}, T., {Stern}, D., {et~al.} 2018{\natexlab{b}},
  \apjl, 856, L25

\bibitem[{{Baldwin}(1977)}]{Baldwin_1977}
{Baldwin}, J.~A. 1977, \apj, 214, 679

\bibitem[{{Barth} {et~al.}(2003){Barth}, {Martini}, {Nelson}, \&
  {Ho}}]{Barth_etal_2003}
{Barth}, A.~J., {Martini}, P., {Nelson}, C.~H., \& {Ho}, L.~C. 2003, \apjl,
  594, L95

\bibitem[{{Baskin} \& {Laor}(2005)}]{Baskin_Laor_2005}
{Baskin}, A., \& {Laor}, A. 2005, \mnras, 356, 1029

\bibitem[{{Brotherton} {et~al.}(2015){Brotherton}, {Runnoe}, {Shang}, \&
  {DiPompeo}}]{Brotherton_etal_2015}
{Brotherton}, M.~S., {Runnoe}, J.~C., {Shang}, Z., \& {DiPompeo}, M.~A. 2015,
  \mnras, 451, 1290

\bibitem[{{Carilli} {et~al.}(2007){Carilli}, {Neri}, {Wang}, {Cox}, {Bertoldi},
  {Walter}, {Fan}, {Menten}, {Wagg}, {Maiolino}, {Omont}, {Strauss},
  {Riechers}, {Lo}, {Bolatto}, \& {Scoville}}]{Carilli_etal_2007}
{Carilli}, C.~L., {Neri}, R., {Wang}, R., {et~al.} 2007, \apjl, 666, L9

\bibitem[{{Coatman} {et~al.}(2017){Coatman}, {Hewett}, {Banerji}, {Richards},
  {Hennawi}, \& {Prochaska}}]{Coatman_etal_2017}
{Coatman}, L., {Hewett}, P.~C., {Banerji}, M., {et~al.} 2017, \mnras, 465, 2120

\bibitem[{{Cool} {et~al.}(2006){Cool}, {Kochanek}, {Eisenstein}, {Stern},
  {Brand}, {Brown}, {Dey}, {Eisenhardt}, {Fan}, {Gonzalez}, {Green}, {Jannuzi},
  {McKenzie}, {Rieke}, {Rieke}, {Soifer}, {Spinrad}, \&
  {Elston}}]{Cool_etal_2006}
{Cool}, R.~J., {Kochanek}, C.~S., {Eisenstein}, D.~J., {et~al.} 2006, \aj, 132,
  823

\bibitem[{{Dai} {et~al.}(2008){Dai}, {Shankar}, \& {Sivakoff}}]{Dai_etal_2008}
{Dai}, X., {Shankar}, F., \& {Sivakoff}, G.~R. 2008, \apj, 672, 108

\bibitem[{{Davis} \& {Laor}(2011)}]{Davis_Laor_2011}
{Davis}, S.~W., \& {Laor}, A. 2011, \apj, 728, 98

\bibitem[{{De Rosa} {et~al.}(2011){De Rosa}, {Decarli}, {Walter}, {Fan},
  {Jiang}, {Kurk}, {Pasquali}, \& {Rix}}]{DeRosa_etal_2011}
{De Rosa}, G., {Decarli}, R., {Walter}, F., {et~al.} 2011, \apj, 739, 56

\bibitem[{{De Rosa} {et~al.}(2014){De Rosa}, {Venemans}, {Decarli}, {Gennaro},
  {Simcoe}, {Dietrich}, {Peterson}, {Walter}, {Frank}, {McMahon}, {Hewett},
  {Mortlock}, \& {Simpson}}]{DeRosa_etal_2014}
{De Rosa}, G., {Venemans}, B.~P., {Decarli}, R., {et~al.} 2014, \apj, 790, 145

\bibitem[{{Decarli} {et~al.}(2017){Decarli}, {Walter}, {Venemans},
  {Ba{\~n}ados}, {Bertoldi}, {Carilli}, {Fan}, {Farina}, {Mazzucchelli},
  {Riechers}, {Rix}, {Strauss}, {Wang}, \& {Yang}}]{Decarli_etal_2017}
{Decarli}, R., {Walter}, F., {Venemans}, B.~P., {et~al.} 2017, \nat, 545, 457

\bibitem[{{Decarli} {et~al.}(2018){Decarli}, {Walter}, {Venemans},
  {Ba{\~n}ados}, {Bertoldi}, {Carilli}, {Fan}, {Farina}, {Mazzucchelli},
  {Riechers}, {Rix}, {Strauss}, {Wang}, \& {Yang}}]{Decarli_etal_2018}
---. 2018, \apj, 854, 97

\bibitem[{{Denney}(2012)}]{Denney_2012}
{Denney}, K.~D. 2012, \apj, 759, 44

\bibitem[{{Diamond-Stanic} {et~al.}(2009){Diamond-Stanic}, {Fan}, {Brandt},
  {Shemmer}, {Strauss}, {Anderson}, {Carilli}, {Gibson}, {Jiang}, {Kim},
  {Richards}, {Schmidt}, {Schneider}, {Shen}, {Smith}, {Vestergaard}, \&
  {Young}}]{Diamond-Stanic_etal_2009}
{Diamond-Stanic}, A.~M., {Fan}, X., {Brandt}, W.~N., {et~al.} 2009, \apj, 699,
  782

\bibitem[{{Dye} {et~al.}(2018){Dye}, {Lawrence}, {Read}, {Fan}, {Kerr},
  {Varricatt}, {Furnell}, {Edge}, {Irwin}, {Hambly}, {Lucas}, {Almaini},
  {Chambers}, {Green}, {Hewett}, {Liu}, {McGreer}, {Best}, {Zhang}, {Sutorius},
  {Froebrich}, {Magnier}, {Hasinger}, {Lederer}, {Bold}, \&
  {Tedds}}]{Dye_etal_2018}
{Dye}, S., {Lawrence}, A., {Read}, M.~A., {et~al.} 2018, \mnras, 473, 5113

\bibitem[{{Edge} {et~al.}(2013){Edge}, {Sutherland}, {Kuijken}, {Driver},
  {McMahon}, {Eales}, \& {Emerson}}]{Edge_etal_2013}
{Edge}, A., {Sutherland}, W., {Kuijken}, K., {et~al.} 2013, The Messenger, 154,
  32

\bibitem[{{Fan} {et~al.}(2006{\natexlab{a}}){Fan}, {Carilli}, \&
  {Keating}}]{Fan_etal_2006araa}
{Fan}, X., {Carilli}, C.~L., \& {Keating}, B. 2006{\natexlab{a}}, \araa, 44,
  415

\bibitem[{{Fan} {et~al.}(1999){Fan}, {Strauss}, {Gunn}, {Lupton}, {Carilli},
  {Rupen}, {Schmidt}, {Moustakas}, {Davis}, {Annis}, {Bahcall}, {Brinkmann},
  {Brunner}, {Csabai}, {Doi}, {Fukugita}, {Heckman}, {Hennessy}, {Hindsley},
  {Ivezi{\'c} }, {Knapp}, {Lamb}, {Munn}, {Pauls}, {Pier}, {Rockosi},
  {Schneider}, {Szalay}, {Tucker}, \& {York}}]{Fan_etal_1999}
{Fan}, X., {Strauss}, M.~A., {Gunn}, J.~E., {et~al.} 1999, \apjl, 526, L57

\bibitem[{{Fan} {et~al.}(2000){Fan}, {White}, {Davis}, {Becker}, {Strauss},
  {Haiman}, {Schneider}, {Gregg}, {Gunn}, {Knapp}, {Lupton}, {Anderson},
  {Anderson}, {Annis}, {Bahcall}, {Boroski}, {Brunner}, {Chen}, {Connolly},
  {Csabai}, {Doi}, {Fukugita}, {Hennessy}, {Hindsley}, {Ichikawa},
  {Ivezi{\'c}}, {Loveday}, {Meiksin}, {McKay}, {Munn}, {Newberg}, {Nichol},
  {Okamura}, {Pier}, {Sekiguchi}, {Shimasaku}, {Stoughton}, {Szalay},
  {Szokoly}, {Thakar}, {Vogeley}, \& {York}}]{Fan_etal_2000}
{Fan}, X., {White}, R.~L., {Davis}, M., {et~al.} 2000, \aj, 120, 1167

\bibitem[{{Fan} {et~al.}(2001){Fan}, {Narayanan}, {Lupton}, {Strauss}, {Knapp},
  {Becker}, {White}, {Pentericci}, {Leggett}, {Haiman}, {Gunn}, {Ivezi{\'c}},
  {Schneider}, {Anderson}, {Brinkmann}, {Bahcall}, {Connolly}, {Csabai}, {Doi},
  {Fukugita}, {Geballe}, {Grebel}, {Harbeck}, {Hennessy}, {Lamb}, {Miknaitis},
  {Munn}, {Nichol}, {Okamura}, {Pier}, {Prada}, {Richards}, {Szalay}, \&
  {York}}]{Fan_etal_2001}
{Fan}, X., {Narayanan}, V.~K., {Lupton}, R.~H., {et~al.} 2001, \aj, 122, 2833

\bibitem[{{Fan} {et~al.}(2003){Fan}, {Strauss}, {Schneider}, {Becker}, {White},
  {Haiman}, {Gregg}, {Pentericci}, {Grebel}, {Narayanan}, {Loh}, {Richards},
  {Gunn}, {Lupton}, {Knapp}, {Ivezi{\'c}}, {Brandt}, {Collinge}, {Hao},
  {Harbeck}, {Prada}, {Schaye}, {Strateva}, {Zakamska}, {Anderson},
  {Brinkmann}, {Bahcall}, {Lamb}, {Okamura}, {Szalay}, \&
  {York}}]{Fan_etal_2003}
{Fan}, X., {Strauss}, M.~A., {Schneider}, D.~P., {et~al.} 2003, \aj, 125, 1649

\bibitem[{{Fan} {et~al.}(2004){Fan}, {Hennawi}, {Richards}, {Strauss},
  {Schneider}, {Donley}, {Young}, {Annis}, {Lin}, {Lampeitl}, {Lupton}, {Gunn},
  {Knapp}, {Brandt}, {Anderson}, {Bahcall}, {Brinkmann}, {Brunner}, {Fukugita},
  {Szalay}, {Szokoly}, \& {York}}]{Fan_etal_2004}
{Fan}, X., {Hennawi}, J.~F., {Richards}, G.~T., {et~al.} 2004, \aj, 128, 515

\bibitem[{{Fan} {et~al.}(2006{\natexlab{b}}){Fan}, {Strauss}, {Becker},
  {White}, {Gunn}, {Knapp}, {Richards}, {Schneider}, {Brinkmann}, \&
  {Fukugita}}]{Fan_etal_2006b}
{Fan}, X., {Strauss}, M.~A., {Becker}, R.~H., {et~al.} 2006{\natexlab{b}}, \aj,
  132, 117

\bibitem[{{Gibson} {et~al.}(2009){Gibson}, {Jiang}, {Brandt}, {Hall}, {Shen},
  {Wu}, {Anderson}, {Schneider}, {Vanden Berk}, {Gallagher}, {Fan}, \&
  {York}}]{Gibson_etal_2009}
{Gibson}, R.~R., {Jiang}, L., {Brandt}, W.~N., {et~al.} 2009, \apj, 692, 758

\bibitem[{{Glikman} {et~al.}(2018){Glikman}, {Lacy}, {LaMassa}, {Stern},
  {Djorgovski}, {Graham}, {Urrutia}, {Lovdal}, {Crnogorcevic}, {Daniels-Koch},
  {Hundal}, {Urry}, {Gates}, \& {Murray}}]{Glikman_etal_2018}
{Glikman}, E., {Lacy}, M., {LaMassa}, S., {et~al.} 2018, \apj, 861, 37

\bibitem[{{Goto}(2006)}]{Goto_2006}
{Goto}, T. 2006, \mnras, 371, 769

\bibitem[{{Haiman}(2013)}]{Haiman_2013}
{Haiman}, Z. 2013, in Astrophysics and Space Science Library, Vol. 396, The
  First Galaxies, ed. T.~{Wiklind}, B.~{Mobasher}, \& V.~{Bromm}, 293

\bibitem[{{Hewett} \& {Wild}(2010)}]{Hewett_Wild_2010}
{Hewett}, P.~C., \& {Wild}, V. 2010, \mnras, 405, 2302

\bibitem[{{Hickox} \& {Alexander}(2018)}]{Hickox_Alexander_2018}
{Hickox}, R.~C., \& {Alexander}, D.~M. 2018, ArXiv e-prints, arXiv:1806.04680

\bibitem[{{Hopkins} {et~al.}(2008){Hopkins}, {Hernquist}, {Cox}, \& {Kere{\v
  s}}}]{Hopkins_etal_2008a}
{Hopkins}, P.~F., {Hernquist}, L., {Cox}, T.~J., \& {Kere{\v s}}, D. 2008,
  \apjs, 175, 356

\bibitem[{{Horne}(1986)}]{Horne_1986}
{Horne}, K. 1986, \pasp, 98, 609

\bibitem[{{Jiang} {et~al.}(2007){Jiang}, {Fan}, {Vestergaard}, {Kurk},
  {Walter}, {Kelly}, \& {Strauss}}]{Jiang_etal_2007}
{Jiang}, L., {Fan}, X., {Vestergaard}, M., {et~al.} 2007, \aj, 134, 1150

\bibitem[{{Jiang} {et~al.}(2015){Jiang}, {McGreer}, {Fan}, {Bian}, {Cai},
  {Cl{\'e}ment}, {Wang}, \& {Fan}}]{Jiang_etal_2015}
{Jiang}, L., {McGreer}, I.~D., {Fan}, X., {et~al.} 2015, \aj, 149, 188

\bibitem[{{Jiang} {et~al.}(2008){Jiang}, {Fan}, {Annis}, {Becker}, {White},
  {Chiu}, {Lin}, {Lupton}, {Richards}, {Strauss}, {Jester}, \&
  {Schneider}}]{Jiang_etal_2008}
{Jiang}, L., {Fan}, X., {Annis}, J., {et~al.} 2008, \aj, 135, 1057

\bibitem[{{Jiang} {et~al.}(2009){Jiang}, {Fan}, {Bian}, {Annis}, {Chiu},
  {Jester}, {Lin}, {Lupton}, {Richards}, {Strauss}, {Malanushenko},
  {Malanushenko}, \& {Schneider}}]{Jiang_etal_2009}
{Jiang}, L., {Fan}, X., {Bian}, F., {et~al.} 2009, \aj, 138, 305

\bibitem[{{Jiang} {et~al.}(2016){Jiang}, {McGreer}, {Fan}, {Strauss},
  {Ba{\~n}ados}, {Becker}, {Bian}, {Farnsworth}, {Shen}, {Wang}, {Wang},
  {Wang}, {White}, {Wu}, {Wu}, {Yang}, \& {Yang}}]{Jiang_etal_2016}
{Jiang}, L., {McGreer}, I.~D., {Fan}, X., {et~al.} 2016, \apj, 833, 222

\bibitem[{{Kalirai}(2018)}]{Kalirai_2018}
{Kalirai}, J. 2018, Contemporary Physics, 59, 251

\bibitem[{{Kelly} \& {Shen}(2013)}]{Kelly_Shen_2013}
{Kelly}, B.~C., \& {Shen}, Y. 2013, \apj, 764, 45

\bibitem[{{Kelly} {et~al.}(2010){Kelly}, {Vestergaard}, {Fan}, {Hopkins},
  {Hernquist}, \& {Siemiginowska}}]{Kelly_etal_2010}
{Kelly}, B.~C., {Vestergaard}, M., {Fan}, X., {et~al.} 2010, \apj, 719, 1315

\bibitem[{Kelson(2003)}]{Kelson2003BSpline}
Kelson, D.~D. 2003, \pasp, 115, 688

\bibitem[{{Kim} {et~al.}(2018){Kim}, {Im}, {Jeon}, {Kim}, {Hyun}, {Kim}, {Kim},
  {Taak}, {Yoon}, {Choi}, {Hong}, {Jun}, {Karouzos}, {Kim}, {Kim}, {Lee},
  {Pak}, \& {Park}}]{Kim_etal_2018}
{Kim}, Y., {Im}, M., {Jeon}, Y., {et~al.} 2018, \apj, 855, 138

\bibitem[{{King}(2016)}]{King_2016}
{King}, A. 2016, \mnras, 456, L109

\bibitem[{{Kurk} {et~al.}(2007){Kurk}, {Walter}, {Fan}, {Jiang}, {Riechers},
  {Rix}, {Pentericci}, {Strauss}, {Carilli}, \& {Wagner}}]{Kurk_etal_2007}
{Kurk}, J.~D., {Walter}, F., {Fan}, X., {et~al.} 2007, \apj, 669, 32

\bibitem[{{Latif} \& {Schleicher}(2019)}]{Latif_Schleicher_2018}
{Latif}, M., \& {Schleicher}, D.~R.~G. 2019, {Formation of the First Black
  Holes}, ed. {{Latif}, M.\& {Schleicher}, D.~R.~G.}

\bibitem[{{Latif} {et~al.}(2013){Latif}, {Schleicher}, {Schmidt}, \&
  {Niemeyer}}]{Latif_etal_2013}
{Latif}, M.~A., {Schleicher}, D.~R.~G., {Schmidt}, W., \& {Niemeyer}, J. 2013,
  \mnras, 433, 1607

\bibitem[{{Lawrence} {et~al.}(2007){Lawrence}, {Warren}, {Almaini}, {Edge},
  {Hambly}, {Jameson}, {Lucas}, {Casali}, {Adamson}, {Dye}, {Emerson},
  {Foucaud}, {Hewett}, {Hirst}, {Hodgkin}, {Irwin}, {Lodieu}, {McMahon},
  {Simpson}, {Smail}, {Mortlock}, \& {Folger}}]{Lawrence_etal_2007}
{Lawrence}, A., {Warren}, S.~J., {Almaini}, O., {et~al.} 2007, \mnras, 379,
  1599

\bibitem[{{Li} {et~al.}(2007){Li}, {Hernquist}, {Robertson}, {Cox}, {Hopkins},
  {Springel}, {Gao}, {Di Matteo}, {Zentner}, {Jenkins}, \&
  {Yoshida}}]{Li_etal_2007}
{Li}, Y., {Hernquist}, L., {Robertson}, B., {et~al.} 2007, \apj, 665, 187

\bibitem[{{Luo} {et~al.}(2015){Luo}, {Brandt}, {Hall}, {Wu}, {Anderson},
  {Garmire}, {Gibson}, {Plotkin}, {Richards}, {Schneider}, {Shemmer}, \&
  {Shen}}]{Luo_etal_2015}
{Luo}, B., {Brandt}, W.~N., {Hall}, P.~B., {et~al.} 2015, \apj, 805, 122

\bibitem[{Mason {et~al.}(2015)Mason, Rodr{\'\i}guez-Ardila, Martins, Riffel,
  Gonz{\'a}lez~Mart{\'\i}n, Ramos~Almeida, Ruschel~Dutra, Ho, Thanjavur,
  Flohic, Alonso-Herrero, Lira, McDermid, Riffel, Schiavon, Winge, Hoenig, \&
  Perlman}]{Mason2015XDGNIRS}
Mason, R.~E., Rodr{\'\i}guez-Ardila, A., Martins, L., {et~al.} 2015, \apjs,
  217, 13

\bibitem[{{Matsuoka} {et~al.}(2016){Matsuoka}, {Onoue}, {Kashikawa}, {Iwasawa},
  {Strauss}, {Nagao}, {Imanishi}, {Niida}, {Toba}, {Akiyama}, {Asami}, {Bosch},
  {Foucaud}, {Furusawa}, {Goto}, {Gunn}, {Harikane}, {Ikeda}, {Kawaguchi},
  {Kikuta}, {Komiyama}, {Lupton}, {Minezaki}, {Miyazaki}, {Morokuma},
  {Murayama}, {Nishizawa}, {Ono}, {Ouchi}, {Price}, {Sameshima}, {Silverman},
  {Sugiyama}, {Tait}, {Takada}, {Takata}, {Tanaka}, {Tang}, \&
  {Utsumi}}]{Matsuoka_etal_2016}
{Matsuoka}, Y., {Onoue}, M., {Kashikawa}, N., {et~al.} 2016, \apj, 828, 26

\bibitem[{{Matsuoka} {et~al.}(2018{\natexlab{a}}){Matsuoka}, {Onoue},
  {Kashikawa}, {Iwasawa}, {Strauss}, {Nagao}, {Imanishi}, {Lee}, {Akiyama},
  {Asami}, {Bosch}, {Foucaud}, {Furusawa}, {Goto}, {Gunn}, {Harikane}, {Ikeda},
  {Izumi}, {Kawaguchi}, {Kikuta}, {Kohno}, {Komiyama}, {Lupton}, {Minezaki},
  {Miyazaki}, {Morokuma}, {Murayama}, {Niida}, {Nishizawa}, {Oguri}, {Ono},
  {Ouchi}, {Price}, {Sameshima}, {Schulze}, {Shirakata}, {Silverman},
  {Sugiyama}, {Tait}, {Takada}, {Takata}, {Tanaka}, {Tang}, {Toba}, {Utsumi},
  \& {Wang}}]{Matsuoka_etal_2018a}
---. 2018{\natexlab{a}}, \pasj, 70, S35

\bibitem[{{Matsuoka} {et~al.}(2018{\natexlab{b}}){Matsuoka}, {Iwasawa},
  {Onoue}, {Kashikawa}, {Strauss}, {Lee}, {Imanishi}, {Nagao}, {Akiyama},
  {Asami}, {Bosch}, {Furusawa}, {Goto}, {Gunn}, {Harikane}, {Ikeda}, {Izumi},
  {Kawaguchi}, {Kato}, {Kikuta}, {Kohno}, {Komiyama}, {Lupton}, {Minezaki},
  {Miyazaki}, {Morokuma}, {Murayama}, {Niida}, {Nishizawa}, {Oguri}, {Ono},
  {Ouchi}, {Price}, {Sameshima}, {Schulze}, {Shirakata}, {Silverman},
  {Sugiyama}, {Tait}, {Takada}, {Takata}, {Tanaka}, {Tang}, {Toba}, {Utsumi},
  {Wang}, \& {Yamashita}}]{Matsuoka_etal_2018}
{Matsuoka}, Y., {Iwasawa}, K., {Onoue}, M., {et~al.} 2018{\natexlab{b}}, \apjs,
  237, 5

\bibitem[{{Mazzucchelli} {et~al.}(2017){Mazzucchelli}, {Ba{\~n}ados},
  {Venemans}, {Decarli}, {Farina}, {Walter}, {Eilers}, {Rix}, {Simcoe},
  {Stern}, {Fan}, {Schlafly}, {De Rosa}, {Hennawi}, {Chambers}, {Greiner},
  {Burgett}, {Draper}, {Kaiser}, {Kudritzki}, {Magnier}, {Metcalfe}, {Waters},
  \& {Wainscoat}}]{Mazzucchelli_etal_2017}
{Mazzucchelli}, C., {Ba{\~n}ados}, E., {Venemans}, B.~P., {et~al.} 2017, \apj,
  849, 91

\bibitem[{{McGreer} {et~al.}(2006){McGreer}, {Becker}, {Helfand}, \&
  {White}}]{McGreer_etal_2006}
{McGreer}, I.~D., {Becker}, R.~H., {Helfand}, D.~J., \& {White}, R.~L. 2006,
  \apj, 652, 157

\bibitem[{{McLure} \& {Dunlop}(2004)}]{McLure_Dunlop_2004}
{McLure}, R.~J., \& {Dunlop}, J.~S. 2004, \mnras, 352, 1390

\bibitem[{{McMahon} {et~al.}(2013){McMahon}, {Banerji}, {Gonzalez}, {Koposov},
  {Bejar}, {Lodieu}, {Rebolo}, \& {VHS Collaboration}}]{McMahon_et_al_2013}
{McMahon}, R.~G., {Banerji}, M., {Gonzalez}, E., {et~al.} 2013, The Messenger,
  154, 35

\bibitem[{{Mej{\'{\i}}a-Restrepo} {et~al.}(2018){Mej{\'{\i}}a-Restrepo},
  {Trakhtenbrot}, {Lira}, \& {Netzer}}]{Mejia-Restrepo_etal_2018}
{Mej{\'{\i}}a-Restrepo}, J.~E., {Trakhtenbrot}, B., {Lira}, P., \& {Netzer}, H.
  2018, \mnras, 478, 1929

\bibitem[{{Morganson} {et~al.}(2012){Morganson}, {De Rosa}, {Decarli},
  {Walter}, {Chambers}, {McGreer}, {Fan}, {Burgett}, {Flewelling}, {Greiner},
  {Hodapp}, {Kaiser}, {Magnier}, {Price}, {Rix}, {Sweeney}, \&
  {Waters}}]{Morganson_etal_2012}
{Morganson}, E., {De Rosa}, G., {Decarli}, R., {et~al.} 2012, \aj, 143, 142

\bibitem[{{Mortlock} {et~al.}(2009){Mortlock}, {Patel}, {Warren}, {Venemans},
  {McMahon}, {Hewett}, {Simpson}, {Sharp}, {Burningham}, {Dye}, {Ellis},
  {Gonzales-Solares}, \& {Hu{\'e}lamo}}]{Mortlock_etal_2009}
{Mortlock}, D.~J., {Patel}, M., {Warren}, S.~J., {et~al.} 2009, \aap, 505, 97

\bibitem[{{Mortlock} {et~al.}(2011){Mortlock}, {Warren}, {Venemans}, {Patel},
  {Hewett}, {McMahon}, {Simpson}, {Theuns}, {Gonz{\'a}les-Solares}, {Adamson},
  {Dye}, {Hambly}, {Hirst}, {Irwin}, {Kuiper}, {Lawrence}, \&
  {R{\"o}ttgering}}]{Mortlock_etal_2011}
{Mortlock}, D.~J., {Warren}, S.~J., {Venemans}, B.~P., {et~al.} 2011, \nat,
  474, 616

\bibitem[{{Nanni} {et~al.}(2017){Nanni}, {Vignali}, {Gilli}, {Moretti}, \&
  {Brandt}}]{Nanni_etal_2017}
{Nanni}, R., {Vignali}, C., {Gilli}, R., {Moretti}, A., \& {Brandt}, W.~N.
  2017, \aap, 603, A128

\bibitem[{{Natarajan} \& {Treister}(2009)}]{Natarajan_Treister_2009}
{Natarajan}, P., \& {Treister}, E. 2009, \mnras, 393, 838

\bibitem[{{Onken} \& {Kollmeier}(2008)}]{Onken_Kollmeier_2008}
{Onken}, C.~A., \& {Kollmeier}, J.~A. 2008, \apjl, 689, L13

\bibitem[{{Park} {et~al.}(2017){Park}, {Barth}, {Woo}, {Malkan}, {Treu},
  {Bennert}, {Assef}, \& {Pancoast}}]{Park_etal_2017}
{Park}, D., {Barth}, A.~J., {Woo}, J.-H., {et~al.} 2017, \apj, 839, 93

\bibitem[{{Peterson} {et~al.}(2004){Peterson}, {Ferrarese}, {Gilbert}, {Kaspi},
  {Malkan}, {Maoz}, {Merritt}, {Netzer}, {Onken}, {Pogge}, {Vestergaard}, \&
  {Wandel}}]{Peterson_etal_2004}
{Peterson}, B.~M., {Ferrarese}, L., {Gilbert}, K.~M., {et~al.} 2004, \apj, 613,
  682

\bibitem[{{Pezzulli} {et~al.}(2016){Pezzulli}, {Valiante}, \&
  {Schneider}}]{Pezzulli_etal_2016}
{Pezzulli}, E., {Valiante}, R., \& {Schneider}, R. 2016, \mnras, 458, 3047

\bibitem[{{Plotkin} {et~al.}(2010){Plotkin}, {Anderson}, {Brandt},
  {Diamond-Stanic}, {Fan}, {MacLeod}, {Schneider}, \&
  {Shemmer}}]{Plotkin_etal_2010}
{Plotkin}, R.~M., {Anderson}, S.~F., {Brandt}, W.~N., {et~al.} 2010, \apj, 721,
  562

\bibitem[{{Plotkin} {et~al.}(2008){Plotkin}, {Anderson}, {Hall}, {Margon},
  {Voges}, {Schneider}, {Stinson}, \& {York}}]{Plotkin_etal_2008}
{Plotkin}, R.~M., {Anderson}, S.~F., {Hall}, P.~B., {et~al.} 2008, \aj, 135,
  2453

\bibitem[{{Reed} {et~al.}(2017){Reed}, {McMahon}, {Martini}, {Banerji},
  {Auger}, {Hewett}, {Koposov}, {Gibbons}, {Gonzalez-Solares}, {Ostrovski},
  {Tie}, {Abdalla}, {Allam}, {Benoit-L{\'e}vy}, {Bertin}, {Brooks},
  {Buckley-Geer}, {Burke}, {Carnero Rosell}, {Carrasco Kind}, {Carretero}, {da
  Costa}, {DePoy}, {Desai}, {Diehl}, {Doel}, {Evrard}, {Finley}, {Flaugher},
  {Fosalba}, {Frieman}, {Garc{\'{\i}}a-Bellido}, {Gaztanaga}, {Goldstein},
  {Gruen}, {Gruendl}, {Gutierrez}, {James}, {Kuehn}, {Kuropatkin}, {Lahav},
  {Lima}, {Maia}, {Marshall}, {Melchior}, {Miller}, {Miquel}, {Nord}, {Ogando},
  {Plazas}, {Romer}, {Sanchez}, {Scarpine}, {Schubnell}, {Sevilla-Noarbe},
  {Smith}, {Sobreira}, {Suchyta}, {Swanson}, {Tarle}, {Tucker}, {Walker}, \&
  {Wester}}]{Reed_etal_2017}
{Reed}, S.~L., {McMahon}, R.~G., {Martini}, P., {et~al.} 2017, \mnras, 468,
  4702

\bibitem[{{Richards} {et~al.}(2002){Richards}, {Fan}, {Newberg}, {Strauss},
  {Vanden Berk}, {Schneider}, {Yanny}, {Boucher}, {Burles}, {Frieman}, {Gunn},
  {Hall}, {Ivezi{\'c}}, {Kent}, {Loveday}, {Lupton}, {Rockosi}, {Schlegel},
  {Stoughton}, {SubbaRao}, \& {York}}]{Richards_etal_2002}
{Richards}, G.~T., {Fan}, X., {Newberg}, H.~J., {et~al.} 2002, \aj, 123, 2945

\bibitem[{{Richards} {et~al.}(2006){Richards}, {Lacy}, {Storrie-Lombardi},
  {Hall}, {Gallagher}, {Hines}, {Fan}, {Papovich}, {Vanden Berk}, {Trammell},
  {Schneider}, {Vestergaard}, {York}, {Jester}, {Anderson}, {Budav{\'a}ri}, \&
  {Szalay}}]{Richards_etal_2006b}
{Richards}, G.~T., {Lacy}, M., {Storrie-Lombardi}, L.~J., {et~al.} 2006, \apjs,
  166, 470

\bibitem[{{Richards} {et~al.}(2011){Richards}, {Kruczek}, {Gallagher}, {Hall},
  {Hewett}, {Leighly}, {Deo}, {Kratzer}, \& {Shen}}]{Richards_etal_2011}
{Richards}, G.~T., {Kruczek}, N.~E., {Gallagher}, S.~C., {et~al.} 2011, \aj,
  141, 167

\bibitem[{{Riechers} {et~al.}(2009){Riechers}, {Walter}, {Bertoldi}, {Carilli},
  {Aravena}, {Neri}, {Cox}, {Wei{\ss}}, \& {Menten}}]{Riechers_etal_2009}
{Riechers}, D.~A., {Walter}, F., {Bertoldi}, F., {et~al.} 2009, \apj, 703, 1338

\bibitem[{{Runnoe} {et~al.}(2013){Runnoe}, {Brotherton}, {Shang}, \&
  {DiPompeo}}]{Runnoe_etal_2013}
{Runnoe}, J.~C., {Brotherton}, M.~S., {Shang}, Z., \& {DiPompeo}, M.~A. 2013,
  \mnras, 434, 848

\bibitem[{{Ryan-Weber} {et~al.}(2009){Ryan-Weber}, {Pettini}, {Madau}, \&
  {Zych}}]{Ryan-Weber_etal_2009}
{Ryan-Weber}, E.~V., {Pettini}, M., {Madau}, P., \& {Zych}, B.~J. 2009, \mnras,
  395, 1476

\bibitem[{{Sanders} \& {Mirabel}(1996)}]{Sanders_Mirabel_1996}
{Sanders}, D.~B., \& {Mirabel}, I.~F. 1996, \araa, 34, 749

\bibitem[{{Shemmer} \& {Lieber}(2015)}]{Shemmer_Lieber_2015}
{Shemmer}, O., \& {Lieber}, S. 2015, \apj, 805, 124

\bibitem[{{Shemmer} {et~al.}(2010){Shemmer}, {Trakhtenbrot}, {Anderson},
  {Brandt}, {Diamond-Stanic}, {Fan}, {Lira}, {Netzer}, {Plotkin}, {Richards},
  {Schneider}, \& {Strauss}}]{Shemmer_etal_2010}
{Shemmer}, O., {Trakhtenbrot}, B., {Anderson}, S.~F., {et~al.} 2010, \apjl,
  722, L152

\bibitem[{{Shen}(2013)}]{Shen_2013}
{Shen}, Y. 2013, Bulletin of the Astronomical Society of India, 41, 61

\bibitem[{{Shen}(2016)}]{Shen_2016}
---. 2016, \apj, 817, 55

\bibitem[{{Shen} {et~al.}(2008){Shen}, {Greene}, {Strauss}, {Richards}, \&
  {Schneider}}]{Shen_etal_2008}
{Shen}, Y., {Greene}, J.~E., {Strauss}, M.~A., {Richards}, G.~T., \&
  {Schneider}, D.~P. 2008, \apj, 680, 169

\bibitem[{{Shen} \& {Kelly}(2012)}]{Shen_Kelly_2012}
{Shen}, Y., \& {Kelly}, B.~C. 2012, \apj, 746, 169

\bibitem[{{Shen} \& {Liu}(2012)}]{Shen_Liu_2012}
{Shen}, Y., \& {Liu}, X. 2012, \apj, 753, 125

\bibitem[{{Shen} {et~al.}(2011){Shen}, {Richards}, {Strauss}, {Hall},
  {Schneider}, {Snedden}, {Bizyaev}, {Brewington}, {Malanushenko},
  {Malanushenko}, {Oravetz}, {Pan}, \& {Simmons}}]{Shen_etal_2011}
{Shen}, Y., {Richards}, G.~T., {Strauss}, M.~A., {et~al.} 2011, \apjs, 194, 45

\bibitem[{{Shen} {et~al.}(2016){Shen}, {Brandt}, {Richards}, {Denney},
  {Greene}, {Grier}, {Ho}, {Peterson}, {Petitjean}, {Schneider}, {Tao}, \&
  {Trump}}]{Shen_etal_2016b}
{Shen}, Y., {Brandt}, W.~N., {Richards}, G.~T., {et~al.} 2016, \apj, 831, 7

\bibitem[{{Shen} {et~al.}(2018){Shen}, {Hall}, {Horne}, {Zhu}, {McGreer},
  {Simm}, {Trump}, {Kinemuchi}, {Brandt}, {Green}, {Grier}, {Guo}, {Ho},
  {Homayouni}, {Jiang}, {I-Hsiu Li}, {Morganson}, {Petitjean}, {Richards},
  {Schneider}, {Starkey}, {Wang}, {Chambers}, {Kaiser}, {Kudritzki}, {Magnier},
  \& {Waters}}]{Shen_etal_2018b}
{Shen}, Y., {Hall}, P.~B., {Horne}, K., {et~al.} 2018, ArXiv e-prints,
  arXiv:1810.01447

\bibitem[{{Sulentic} {et~al.}(2000){Sulentic}, {Zwitter}, {Marziani}, \&
  {Dultzin-Hacyan}}]{Sulentic_etal_2000}
{Sulentic}, J.~W., {Zwitter}, T., {Marziani}, P., \& {Dultzin-Hacyan}, D. 2000,
  \apjl, 536, L5

\bibitem[{{Tang} {et~al.}(2017){Tang}, {Goto}, {Ohyama}, {Chen}, {Walter},
  {Venemans}, {Chambers}, {Ba{\~n}ados}, {Decarli}, {Fan}, {Farina},
  {Mazzucchelli}, {Kaiser}, \& {Magnier}}]{Tang_etal_2017}
{Tang}, J.-J., {Goto}, T., {Ohyama}, Y., {et~al.} 2017, \mnras, 466, 4568

\bibitem[{{Trakhtenbrot} \& {Netzer}(2012)}]{Trakhtenbrot_Netzer_2012}
{Trakhtenbrot}, B., \& {Netzer}, H. 2012, \mnras, 427, 3081

\bibitem[{{Trakhtenbrot} {et~al.}(2017){Trakhtenbrot}, {Volonteri}, \&
  {Natarajan}}]{Trakhtenbrot_etal_2017}
{Trakhtenbrot}, B., {Volonteri}, M., \& {Natarajan}, P. 2017, \apjl, 836, L1

\bibitem[{{van Dokkum}(2001)}]{Dokkum2001LACosmic}
{van Dokkum}, P.~G. 2001, \pasp, 113, 1420

\bibitem[{{Vanden Berk} {et~al.}(2001){Vanden Berk}, {Richards}, {Bauer},
  {Strauss}, {Schneider}, {Heckman}, {York}, {Hall}, {Fan}, {Knapp},
  {Anderson}, {Annis}, {Bahcall}, {Bernardi}, {Briggs}, {Brinkmann}, {Brunner},
  {Burles}, {Carey}, {Castander}, {Connolly}, {Crocker}, {Csabai}, {Doi},
  {Finkbeiner}, {Friedman}, {Frieman}, {Fukugita}, {Gunn}, {Hennessy},
  {Ivezi{\'c}}, {Kent}, {Kunszt}, {Lamb}, {Leger}, {Long}, {Loveday}, {Lupton},
  {Meiksin}, {Merelli}, {Munn}, {Newberg}, {Newcomb}, {Nichol}, {Owen}, {Pier},
  {Pope}, {Rockosi}, {Schlegel}, {Siegmund}, {Smee}, {Snir}, {Stoughton},
  {Stubbs}, {SubbaRao}, {Szalay}, {Szokoly}, {Tremonti}, {Uomoto}, {Waddell},
  {Yanny}, \& {Zheng}}]{Vandenberk_etal_2001}
{Vanden Berk}, D.~E., {Richards}, G.~T., {Bauer}, A., {et~al.} 2001, \aj, 122,
  549

\bibitem[{{Venemans} {et~al.}(2007){Venemans}, {McMahon}, {Warren},
  {Gonzalez-Solares}, {Hewett}, {Mortlock}, {Dye}, \&
  {Sharp}}]{Venemans_etal_2007}
{Venemans}, B.~P., {McMahon}, R.~G., {Warren}, S.~J., {et~al.} 2007, \mnras,
  376, L76

\bibitem[{{Venemans} {et~al.}(2016){Venemans}, {Walter}, {Zschaechner},
  {Decarli}, {De Rosa}, {Findlay}, {McMahon}, \&
  {Sutherland}}]{Venemans_etal_2016}
{Venemans}, B.~P., {Walter}, F., {Zschaechner}, L., {et~al.} 2016, \apj, 816,
  37

\bibitem[{{Venemans} {et~al.}(2015){Venemans}, {Ba{\~n}ados}, {Decarli},
  {Farina}, {Walter}, {Chambers}, {Fan}, {Rix}, {Schlafly}, {McMahon},
  {Simcoe}, {Stern}, {Burgett}, {Draper}, {Flewelling}, {Hodapp}, {Kaiser},
  {Magnier}, {Metcalfe}, {Morgan}, {Price}, {Tonry}, {Waters}, {AlSayyad},
  {Banerji}, {Chen}, {Gonz{\'a}lez-Solares}, {Greiner}, {Mazzucchelli},
  {McGreer}, {Miller}, {Reed}, \& {Sullivan}}]{Venemans_etal_2015}
{Venemans}, B.~P., {Ba{\~n}ados}, E., {Decarli}, R., {et~al.} 2015, \apjl, 801,
  L11

\bibitem[{{Vestergaard}(2004)}]{Vestergaard_2004}
{Vestergaard}, M. 2004, \apj, 601, 676

\bibitem[{{Vestergaard} {et~al.}(2011){Vestergaard}, {Denney}, {Fan}, {Jensen},
  {Kelly}, {Osmer}, {Peterson}, \& {Tremonti}}]{Vestergaard_etal_2011}
{Vestergaard}, M., {Denney}, K., {Fan}, X., {et~al.} 2011, in Narrow-Line
  Seyfert 1 Galaxies and their Place in the Universe, 38

\bibitem[{{Vestergaard} \& {Osmer}(2009)}]{Vestergaard_Osmer_2009}
{Vestergaard}, M., \& {Osmer}, P.~S. 2009, \apj, 699, 800

\bibitem[{{Vestergaard} \& {Peterson}(2006)}]{Vestergaard_Peterson_2006}
{Vestergaard}, M., \& {Peterson}, B.~M. 2006, \apj, 641, 689

\bibitem[{{Vestergaard} \& {Wilkes}(2001)}]{Vestergaard_Wilkes_2001}
{Vestergaard}, M., \& {Wilkes}, B.~J. 2001, \apjs, 134, 1

\bibitem[{{Volonteri}(2010)}]{Volonteri_2010}
{Volonteri}, M. 2010, A\&A Rev, 18, 279

\bibitem[{{Walter} {et~al.}(2009){Walter}, {Riechers}, {Cox}, {Neri},
  {Carilli}, {Bertoldi}, {Weiss}, \& {Maiolino}}]{Walter_etal_2009}
{Walter}, F., {Riechers}, D., {Cox}, P., {et~al.} 2009, \nat, 457, 699

\bibitem[{{Walter} {et~al.}(2003){Walter}, {Bertoldi}, {Carilli}, {Cox}, {Lo},
  {Neri}, {Fan}, {Omont}, {Strauss}, \& {Menten}}]{Walter_etal_2003}
{Walter}, F., {Bertoldi}, F., {Carilli}, C., {et~al.} 2003, \nat, 424, 406

\bibitem[{{Wang} {et~al.}(2016){Wang}, {Wu}, {Fan}, {Yang}, {Yi}, {Bian},
  {McGreer}, {Yang}, {Ai}, {Dong}, {Zuo}, {Jiang}, {Green}, {Wang}, {Cai},
  {Wang}, \& {Yue}}]{Wang_etal_2016}
{Wang}, F., {Wu}, X.-B., {Fan}, X., {et~al.} 2016, \apj, 819, 24

\bibitem[{{Wang} {et~al.}(2009){Wang}, {Dong}, {Wang}, {Ho}, {Yuan}, {Wang},
  {Zhang}, {Zhang}, \& {Zhou}}]{Wang_etal_2009b}
{Wang}, J.-G., {Dong}, X.-B., {Wang}, T.-G., {et~al.} 2009, \apj, 707, 1334

\bibitem[{{Wang} {et~al.}(2010){Wang}, {Carilli}, {Neri}, {Riechers}, {Wagg},
  {Walter}, {Bertoldi}, {Menten}, {Omont}, {Cox}, \& {Fan}}]{Wang_etal_2010}
{Wang}, R., {Carilli}, C.~L., {Neri}, R., {et~al.} 2010, \apj, 714, 699

\bibitem[{{Wang} {et~al.}(2011){Wang}, {Wagg}, {Carilli}, {Walter}, {Riechers},
  {Willott}, {Bertoldi}, {Omont}, {Beelen}, {Cox}, {Strauss}, {Bergeron},
  {Forveille}, {Menten}, \& {Fan}}]{Wang_etal_2011a}
{Wang}, R., {Wagg}, J., {Carilli}, C.~L., {et~al.} 2011, \apjl, 739, L34

\bibitem[{{Wang} {et~al.}(2013){Wang}, {Wagg}, {Carilli}, {Walter}, {Lentati},
  {Fan}, {Riechers}, {Bertoldi}, {Narayanan}, {Strauss}, {Cox}, {Omont},
  {Menten}, {Knudsen}, {Neri}, \& {Jiang}}]{Wang_etal_2013}
---. 2013, \apj, 773, 44

\bibitem[{{Willott} {et~al.}(2015){Willott}, {Bergeron}, \&
  {Omont}}]{Willott_etal_2015}
{Willott}, C.~J., {Bergeron}, J., \& {Omont}, A. 2015, \apj, 801, 123

\bibitem[{{Willott} {et~al.}(2017){Willott}, {Bergeron}, \&
  {Omont}}]{Willott_etal_2017}
---. 2017, \apj, 850, 108

\bibitem[{{Willott} {et~al.}(2007){Willott}, {Delorme}, {Omont}, {Bergeron},
  {Delfosse}, {Forveille}, {Albert}, {Reyl{\'e}}, {Hill}, {Gully-Santiago},
  {Vinten}, {Crampton}, {Hutchings}, {Schade}, {Simard}, {Sawicki}, {Beelen},
  \& {Cox}}]{Willott_etal_2007}
{Willott}, C.~J., {Delorme}, P., {Omont}, A., {et~al.} 2007, \aj, 134, 2435

\bibitem[{{Willott} {et~al.}(2009){Willott}, {Delorme}, {Reyl{\'e}}, {Albert},
  {Bergeron}, {Crampton}, {Delfosse}, {Forveille}, {Hutchings}, {McLure},
  {Omont}, \& {Schade}}]{Willott_etal_2009}
{Willott}, C.~J., {Delorme}, P., {Reyl{\'e}}, C., {et~al.} 2009, \aj, 137, 3541

\bibitem[{{Willott} {et~al.}(2010{\natexlab{a}}){Willott}, {Albert},
  {Arzoumanian}, {Bergeron}, {Crampton}, {Delorme}, {Hutchings}, {Omont},
  {Reyl{\'e}}, \& {Schade}}]{Willott_etal_2010b}
{Willott}, C.~J., {Albert}, L., {Arzoumanian}, D., {et~al.} 2010{\natexlab{a}},
  \aj, 140, 546

\bibitem[{{Willott} {et~al.}(2010{\natexlab{b}}){Willott}, {Delorme},
  {Reyl{\'e}}, {Albert}, {Bergeron}, {Crampton}, {Delfosse}, {Forveille},
  {Hutchings}, {McLure}, {Omont}, \& {Schade}}]{Willott_etal_2010a}
{Willott}, C.~J., {Delorme}, P., {Reyl{\'e}}, C., {et~al.} 2010{\natexlab{b}},
  \aj, 139, 906

\bibitem[{{Woo} {et~al.}(2018){Woo}, {Le}, {Karouzos}, {Park}, {Park},
  {Malkan}, {Treu}, \& {Bennert}}]{Woo_etal_2018}
{Woo}, J.-H., {Le}, H.~A.~N., {Karouzos}, M., {et~al.} 2018, ArXiv e-prints,
  arXiv:1804.02798

\bibitem[{{Wu} {et~al.}(2011){Wu}, {Brandt}, {Hall}, {Gibson}, {Richards},
  {Schneider}, {Shemmer}, {Just}, \& {Schmidt}}]{Wu_etal_2011}
{Wu}, J., {Brandt}, W.~N., {Hall}, P.~B., {et~al.} 2011, \apj, 736, 28

\bibitem[{{Wu} {et~al.}(2015){Wu}, {Wang}, {Fan}, {Yi}, {Zuo}, {Bian}, {Jiang},
  {McGreer}, {Wang}, {Yang}, {Yang}, {Thompson}, \& {Beletsky}}]{Wu_etal_2015}
{Wu}, X.-B., {Wang}, F., {Fan}, X., {et~al.} 2015, \nat, 518, 512

\end{thebibliography}

\clearpage
\begin{figure}
\centering
\includegraphics[width=1.0\columnwidth]{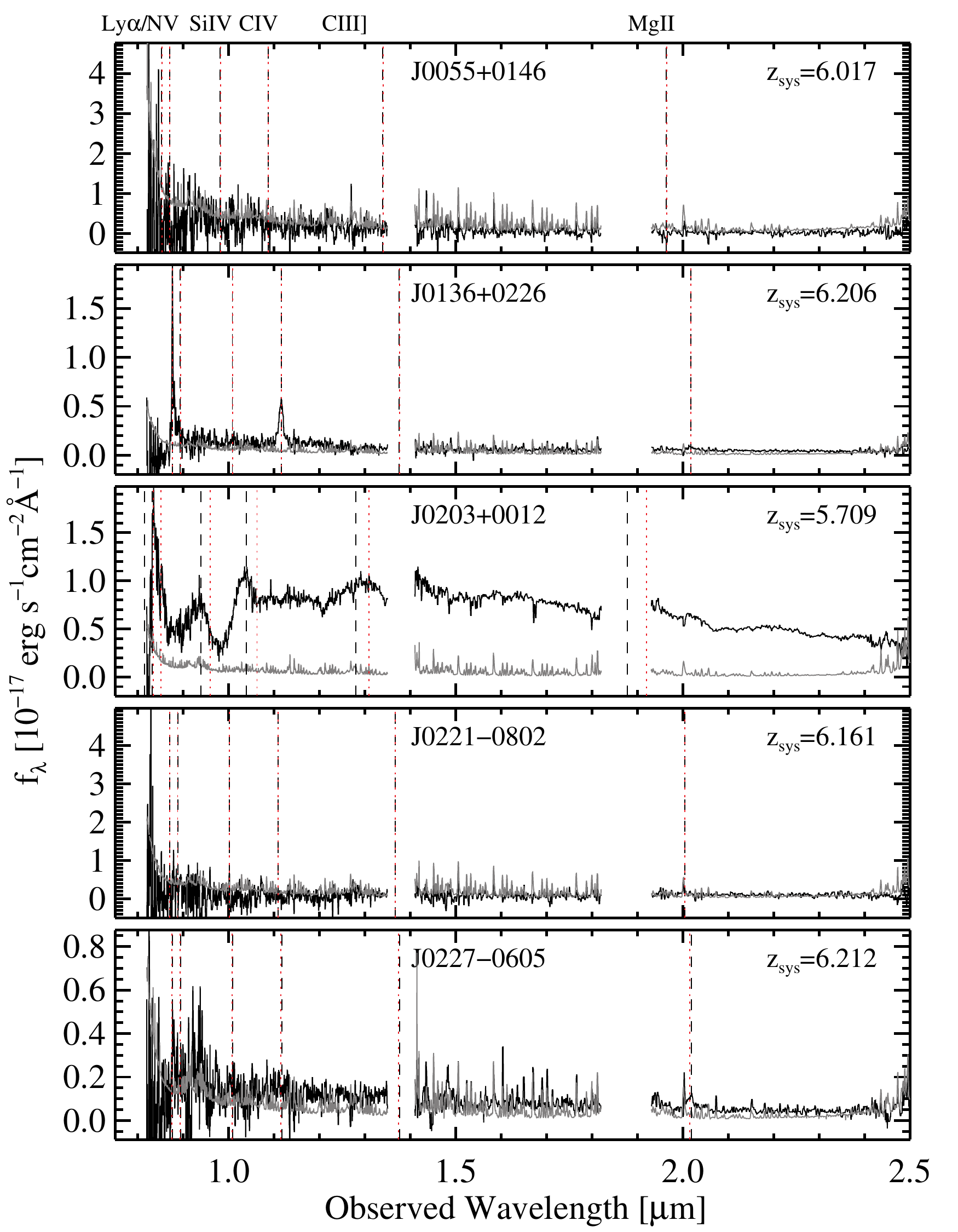}
\label{set1}
\caption{Another set of GNIRS spectra.}
%\phantomcaption  % <=== This line added to compensate for the missing \caption
\end{figure}
\begin{figure}
\centering
\includegraphics[width=1.0\columnwidth]{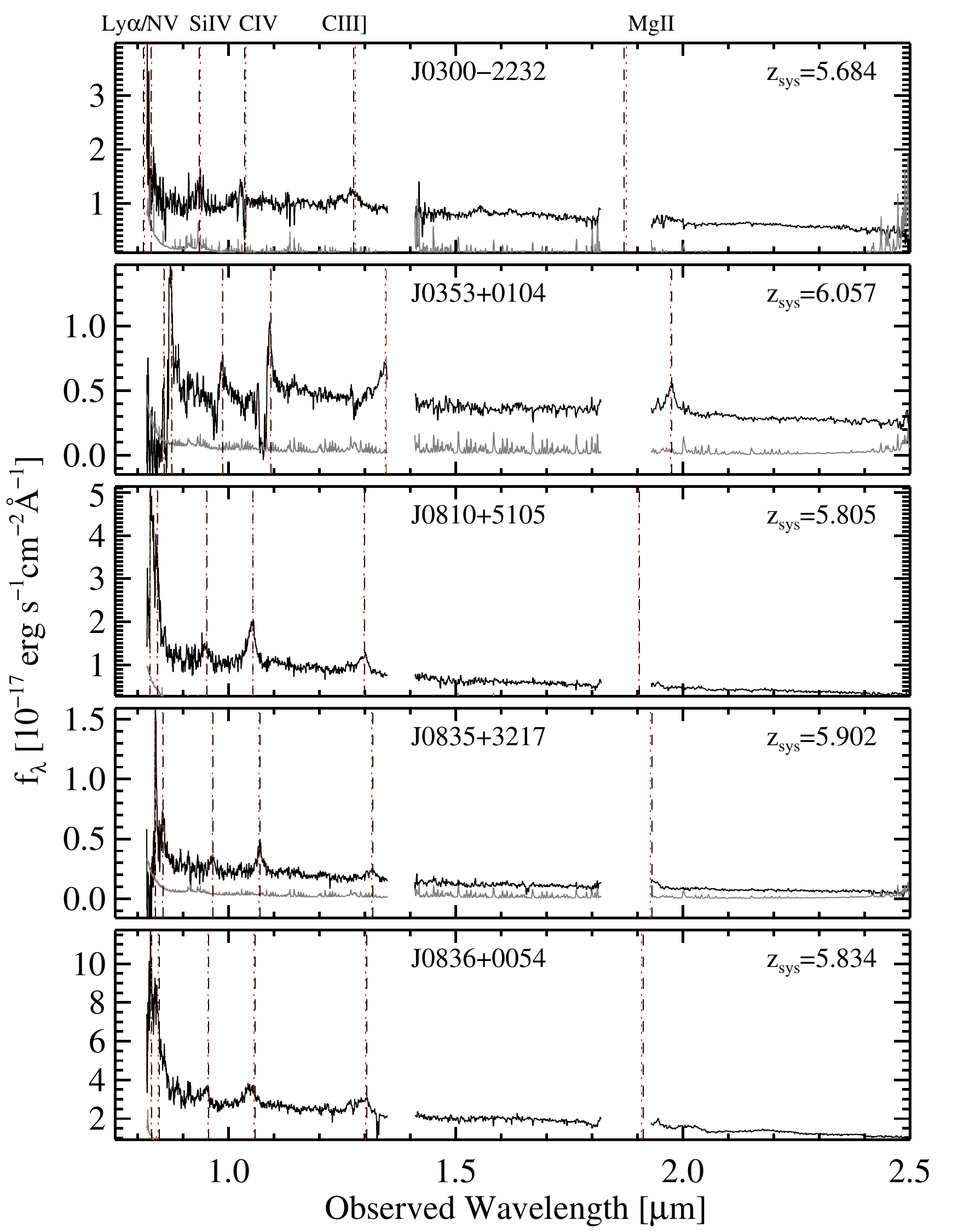} \, 
\label{set2} %
\caption{Another set of GNIRS spectra.}
\end{figure}
\begin{figure}
\centering
\includegraphics[width=1.0\columnwidth]{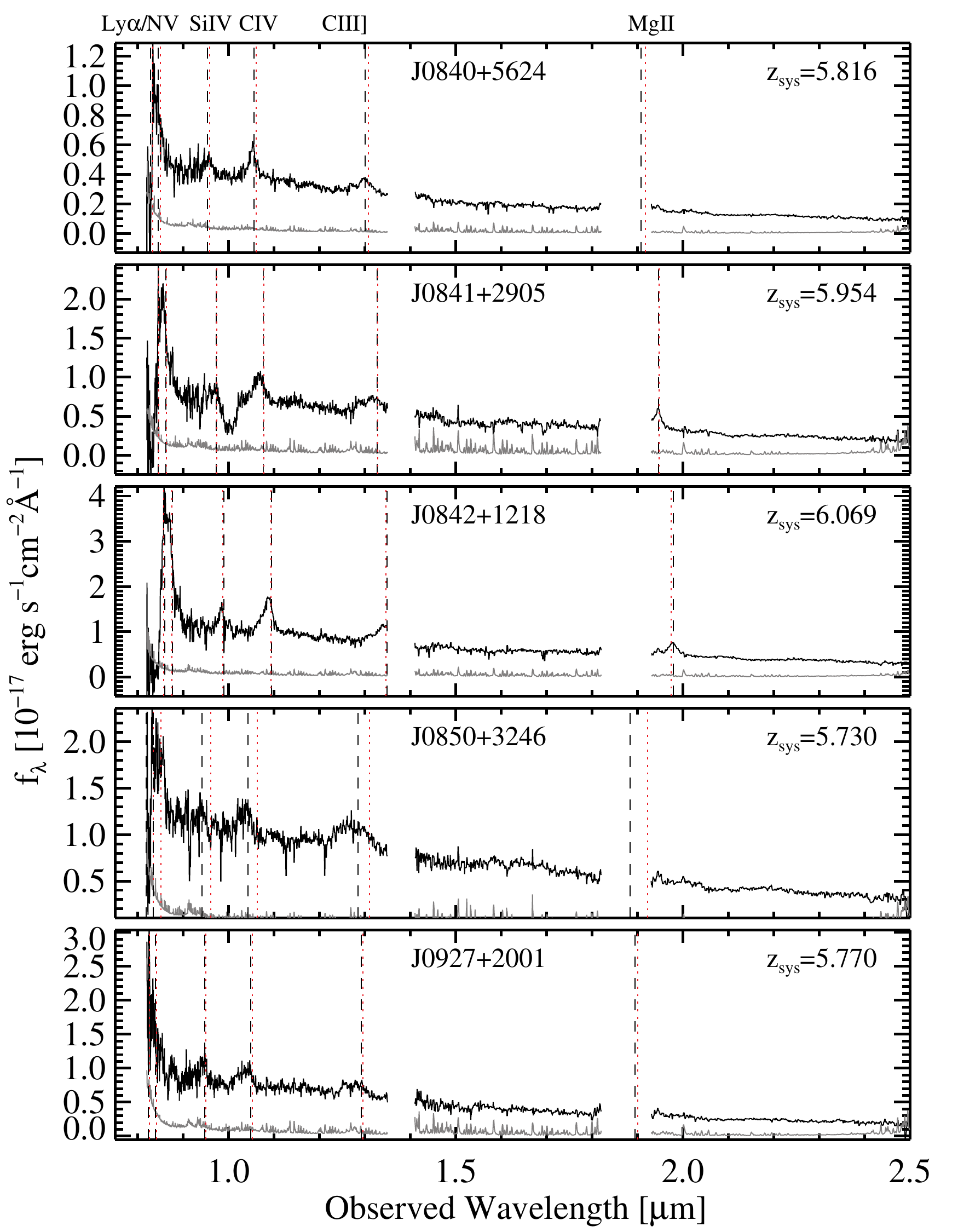} \, 
\label{set2} %
\caption{Another set of GNIRS spectra.}
\end{figure}
\begin{figure}
\centering
\includegraphics[width=1.0\columnwidth]{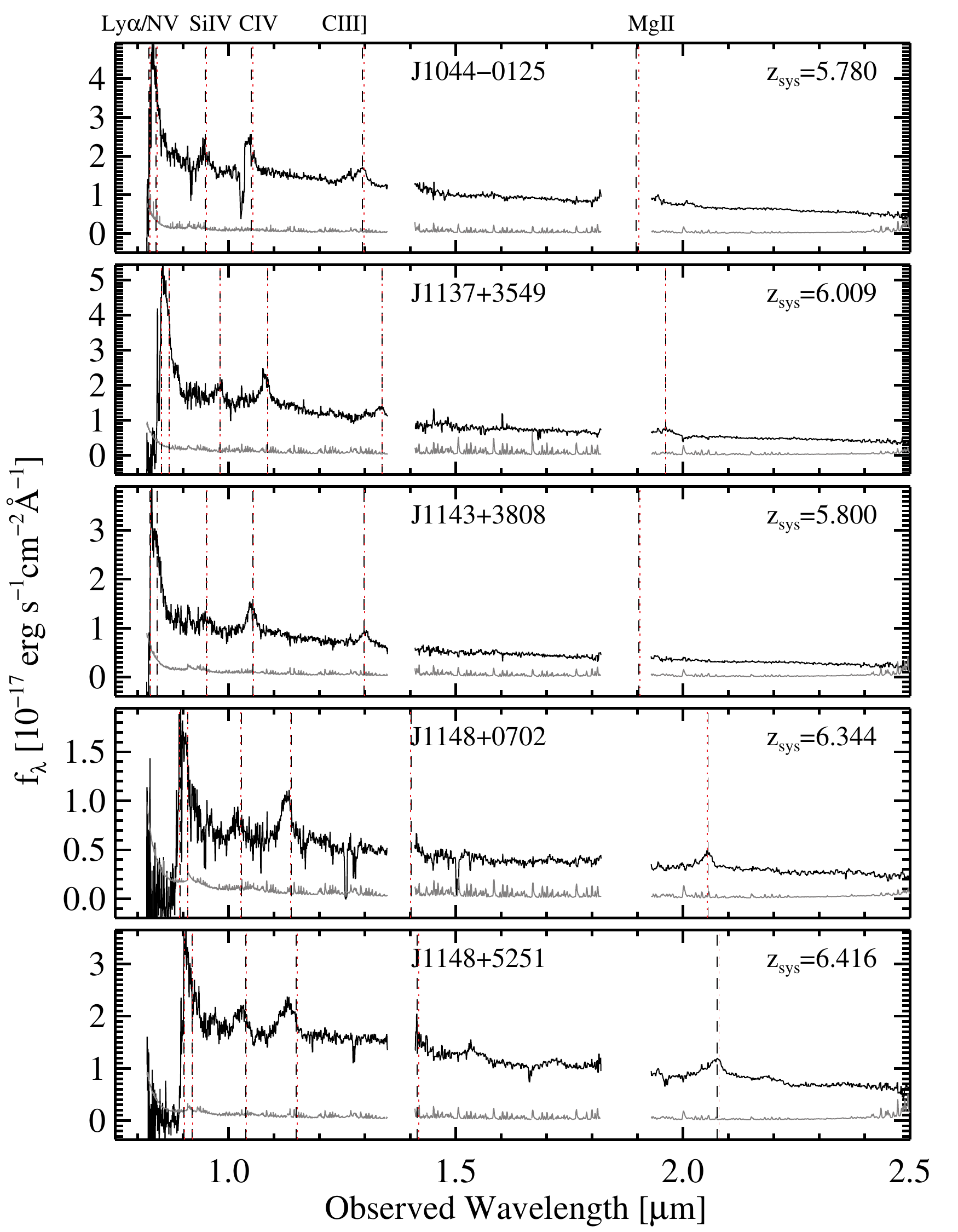} \, 
\label{set2} %
\caption{Another set of GNIRS spectra.}
\end{figure}
\begin{figure}
\centering
\includegraphics[width=1.0\columnwidth]{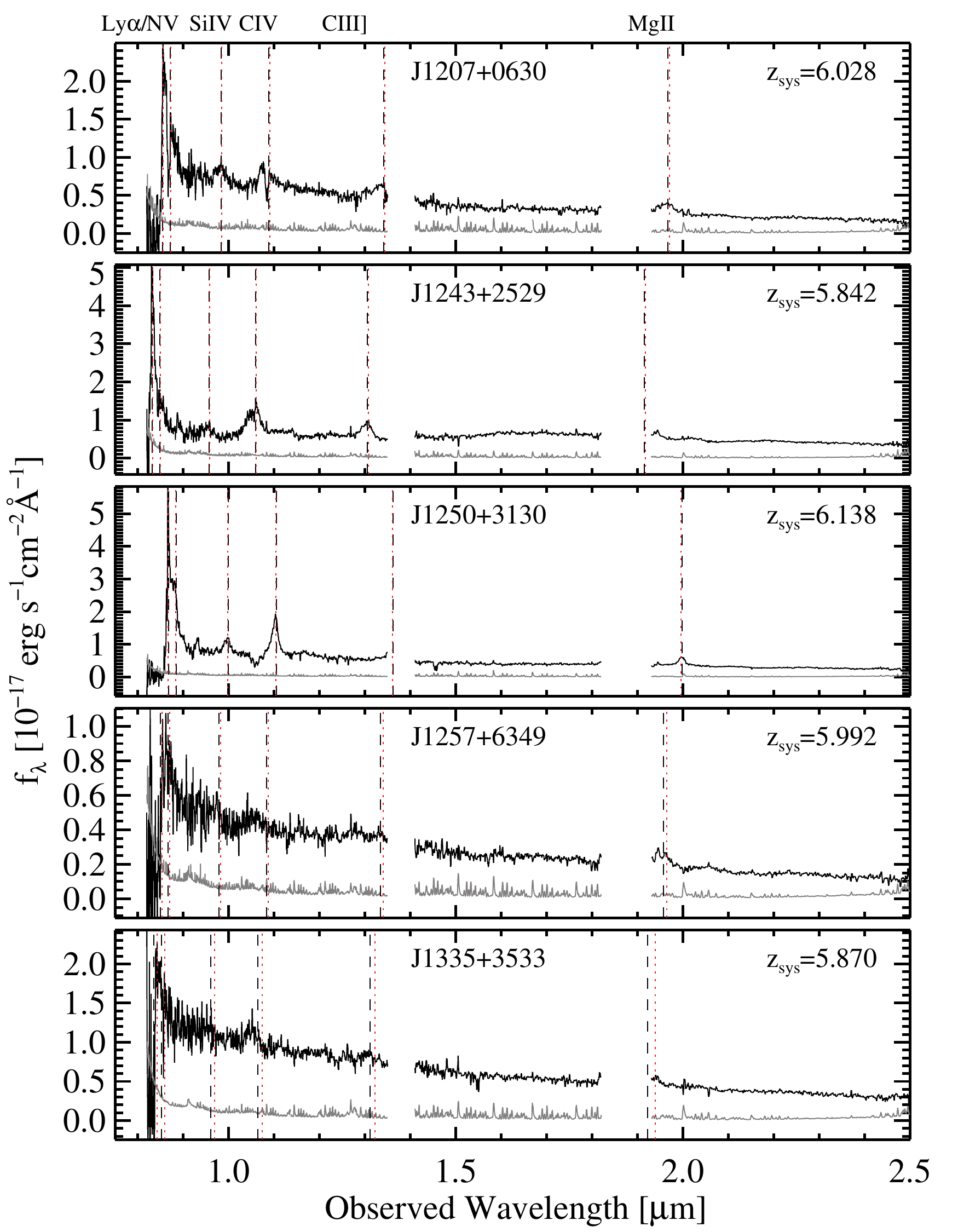} \, 
\label{set2} %
\caption{Another set of GNIRS spectra.}
\end{figure}
\begin{figure}
\centering
\includegraphics[width=1.0\columnwidth]{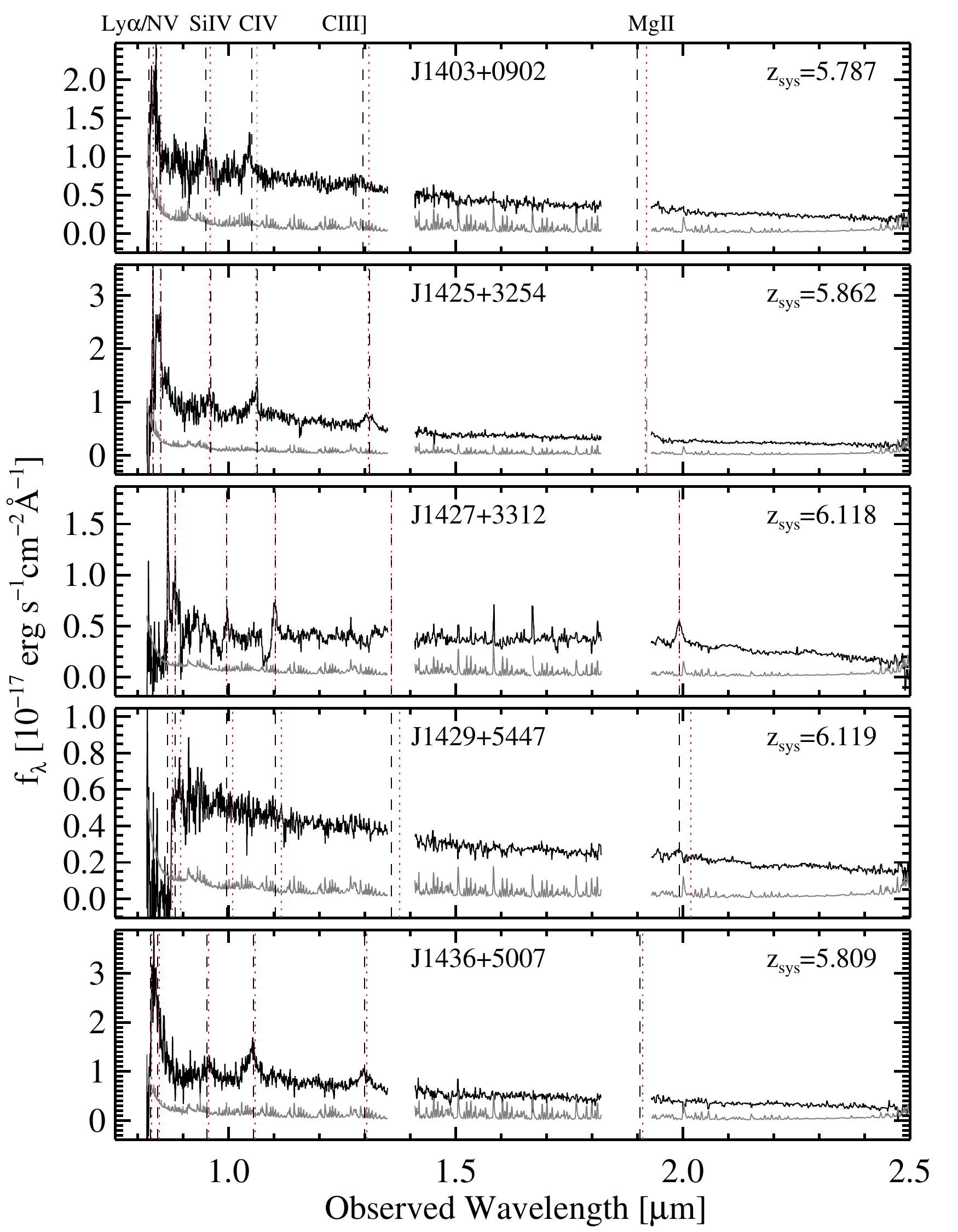} \, 
\label{set2} %
\caption{Another set of GNIRS spectra.}
\end{figure}
\begin{figure}
\centering
\includegraphics[width=1.0\columnwidth]{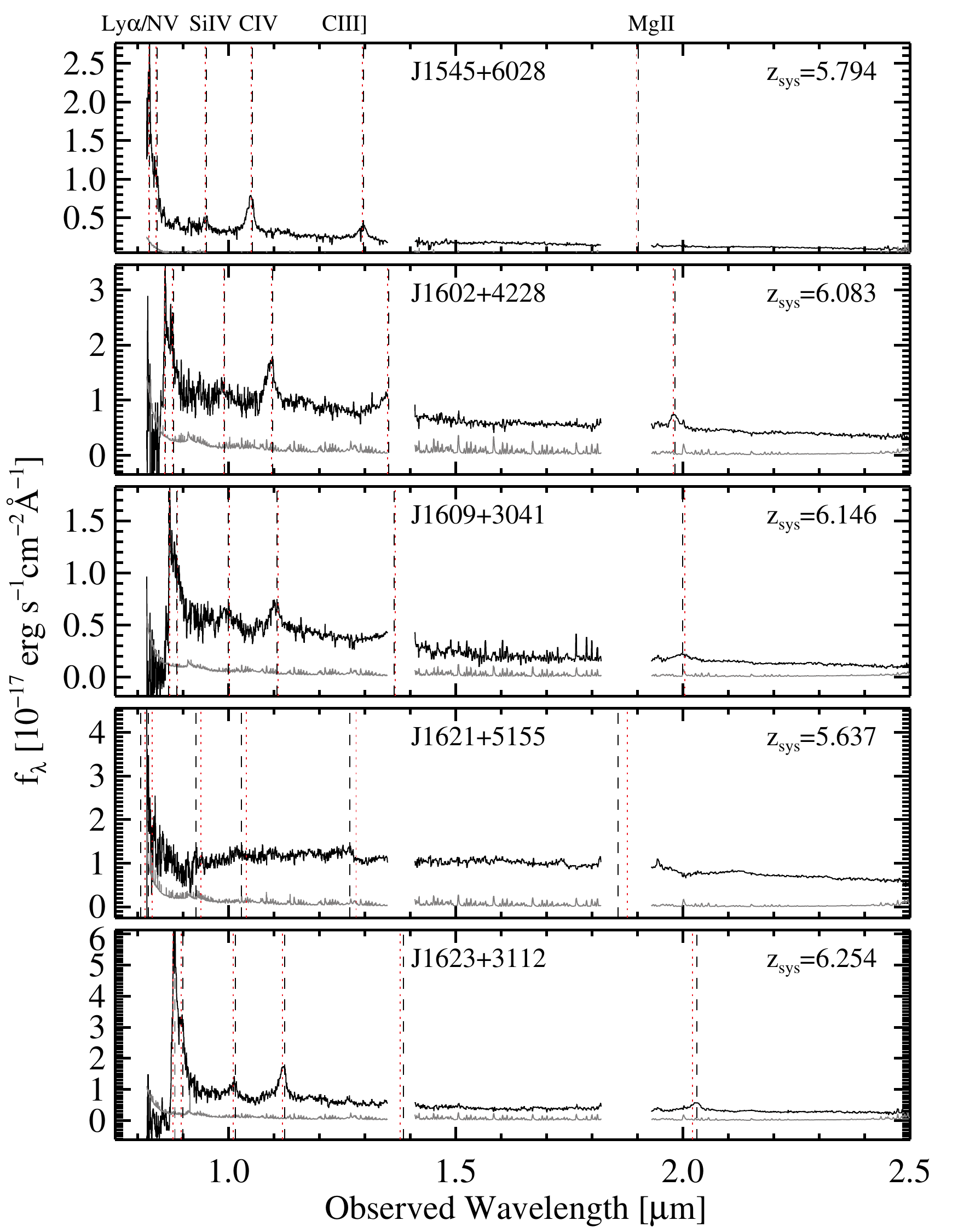} \, 
\label{set2} %
\caption{Another set of GNIRS spectra.}
\end{figure}
\begin{figure}
\centering
\includegraphics[width=1.0\columnwidth]{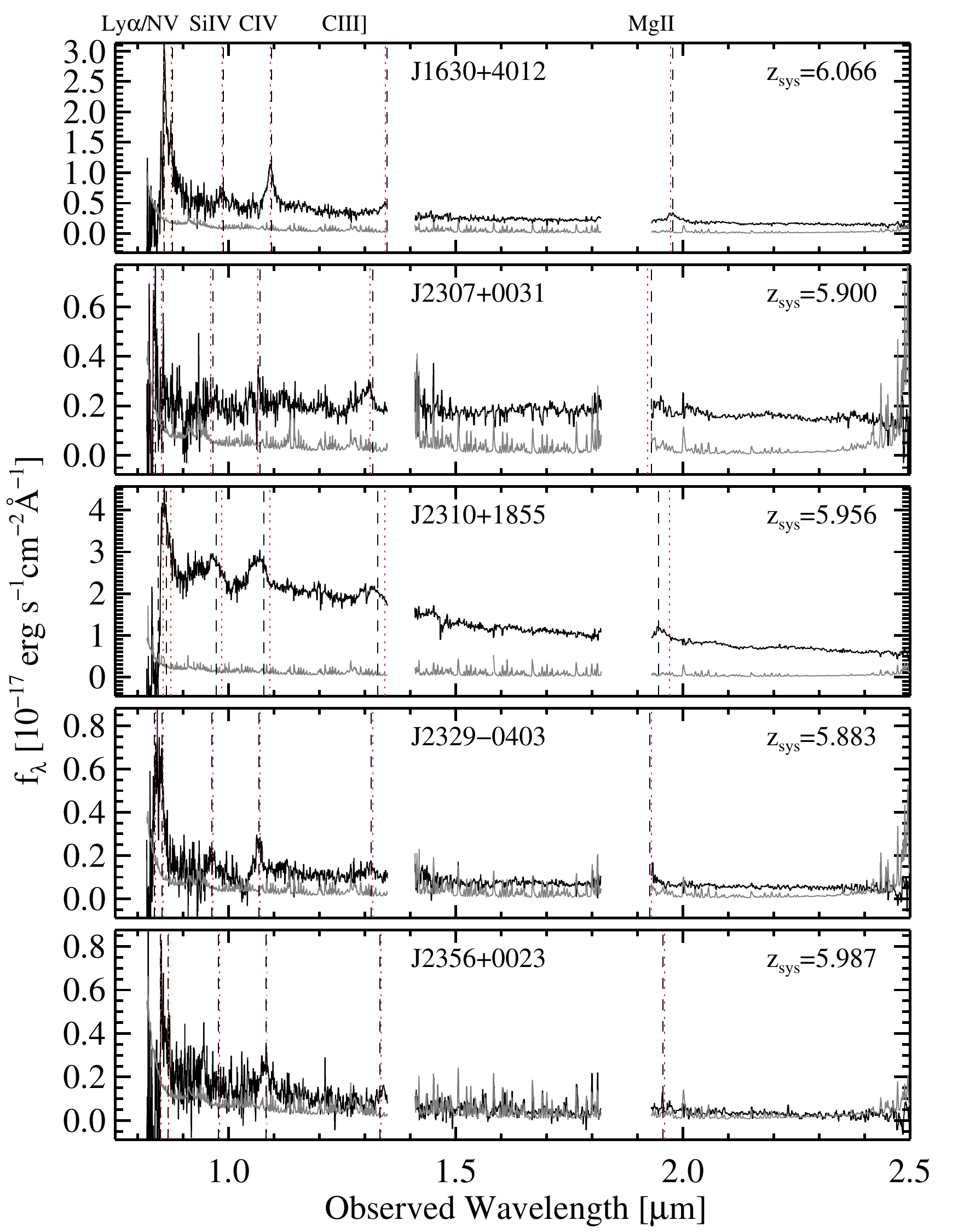} \, 
\label{set2} %
\caption{Another set of GNIRS spectra.}
\end{figure}
\begin{figure}
\centering
\includegraphics[width=1.0\columnwidth]{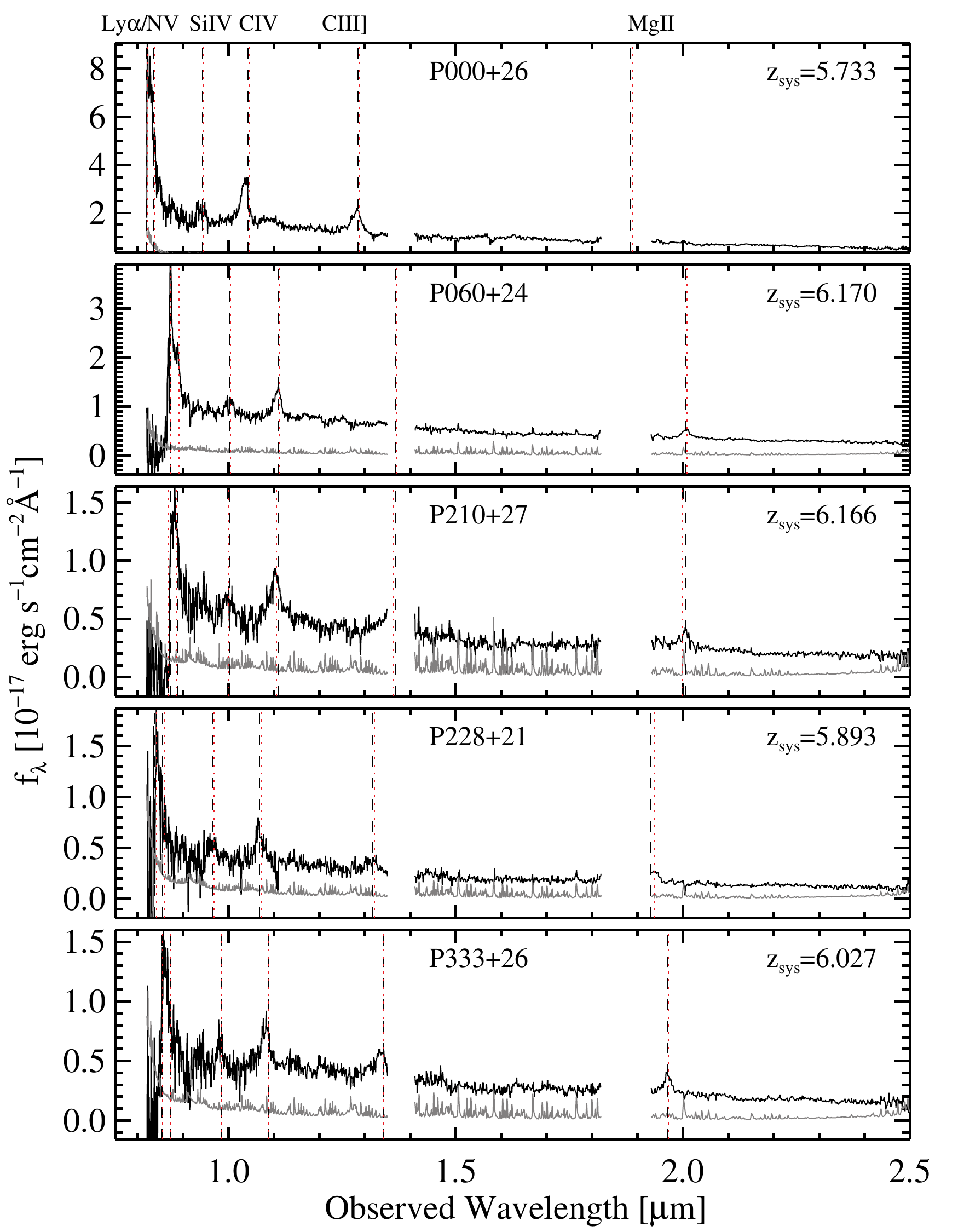} \, 
\label{set2} %
\caption{Another set of GNIRS spectra.}
\end{figure}

\end{document}